\documentclass[aps,showpacs,superscriptaddress,amsmath,amssymb,nofootinbib]{revtex4-1}

\newcommand{\overbar}[1]{\mkern 1.5mu\overline{\mkern-1.5mu#1\mkern-1.5mu}\mkern 1.5mu}

\usepackage{color}
\usepackage{graphicx}
\usepackage{epstopdf}
 \usepackage{verbatim}
\usepackage{url}

\usepackage{multirow}
\usepackage{graphicx}
\usepackage{dcolumn}
\usepackage{bm}
\usepackage{color}
\usepackage{txfonts}
\usepackage{soul}
\usepackage{amssymb}
\usepackage[utf8x]{inputenc}
\usepackage{amsmath}
\usepackage{titlesec}
\usepackage{booktabs}
\usepackage{rotating}
\usepackage{grffile}
\usepackage{epstopdf}
\usepackage{verbatim}
\usepackage{url}
\usepackage{mathtools}
\usepackage{url}
\usepackage{enumerate}
\usepackage{setspace}

\begin{document}

\title{Null Models for Community Detection in Spatially-Embedded, Temporal Networks}
\author{Marta Sarzynska}
\email[Corresponding author: ]{sarzynska@maths.ox.ac.uk}
\affiliation{Oxford Centre for Industrial and Applied Mathematics, Mathematical Institute, University of Oxford, Oxford, OX2 6GG, UK}
\author{Elizabeth A. Leicht}
\affiliation{CABDyN Complexity Centre, University of Oxford, Oxford, OX1 1HP, UK}
\author{Gerardo Chowell}
\affiliation{Mathematical, Computational \& Modeling Sciences Center, School of Human Evolution and Social Change, Arizona State University, 900 S. Cady Mall, Tempe 85287-2402, Arizona, USA}
\affiliation{Division of International Epidemiology and Population Studies, Fogarty International Center, National Institutes of Health, Bethesda, MD, USA}
\author{Mason A. Porter}
\affiliation{Oxford Centre for Industrial and Applied Mathematics, Mathematical Institute, University of Oxford, Oxford, OX2 6GG, UK}
\affiliation{CABDyN Complexity Centre, University of Oxford, Oxford, OX1 1HP, UK}




\begin{abstract}
{In the study of networks, it is often insightful to use algorithms to determine mesoscale features such as ``community structure'', in which densely connected sets of nodes constitute ``communities'' that have sparse connections to other communities.  The most popular way of detecting communities algorithmically is to optimize the quality function known as modularity.  When optimizing modularity, one compares the actual connections in a (static or time-dependent) network to the connections obtained from a random-graph ensemble that acts as a null model.  The communities are then the sets of nodes that are connected to each other densely relative to what is expected from the null model.  Clearly, the process of community detection depends fundamentally on the choice of null model, so it is important to develop and analyze novel null models that take into account appropriate features of the system under study.  In this paper, we investigate the effects of using null models that incorporate spatial information, and we propose a novel null model based on the radiation model of population spread.  We also develop novel synthetic spatial benchmark networks in which the connections between entities are based on distance or flux between nodes, and we compare the performance of both static and time-dependent radiation null model to the standard (``Newman-Girvan'') null model for modularity optimization and a recently-proposed gravity null model. In our comparisons, we use both the above synthetic benchmarks and time-dependent correlation networks that we construct using countrywide dengue fever incidence data for Peru. We also evaluate a recently-proposed correlation null model, which was developed specifically for correlation networks that are constructed from time series, on the epidemic-correlation data. Our findings underscore the need to use appropriate generative models for the development of spatial null models for community detection. }
{Community detection, spatial null model}\\
{87.19.Xx,89.20.-a,89.75.Fb,05.45.Tp}
\end{abstract}

\maketitle





\section{Introduction}

A network formalism is often very useful for describing complex systems of interacting entities~\cite{Newman2010,Wasserman1994Social}. Scholars in a diverse set of disciplines have studied networks for many decades, and network science has experienced particularly explosive growth during the past 20 years~\cite{Newman2010}. 

The most traditional network representation is a static graph, in which nodes represent entities and edges represent pairwise connections between nodes.  However, many networks are time-dependent \cite{holme12,holme13} or multiplex (include multiple types of connections between nodes) \cite{mikko-review,bocca-review}.  Moreover, network structure is influenced profoundly by spatial effects \cite{Barthelemy2011}. To avoid discarding potentially important information, which can lead to misleading results, it is thus crucial to develop methods that incorporate features such as time-dependence, multiplexity, and spatial embeddedness in a context-dependent manner \cite{holme12,mikko-review,Barthelemy2011}.  Because of the newfound wealth of rich data, it has now become possible to validate increasingly complicated network structures and methods using empirical data.

In the present paper, we study a mesoscale network structure known as \emph{community structure}.  A ``community'' is a set of nodes with dense connections among themselves, and with only sparse connections to other communities in a network~\cite{Fortunato2010,PorterMasonA2009}. Communities arise in numerous applications. For example, social networks typically include dense sets of nodes with common interests or other characteristics \cite{traud2012}, networks of legislators often contain dense sets of individuals who vote in similar ways \cite{waugh2009}, and protein-protein interaction networks include dense sets of nodes that constitute functional units \cite{anna2010}.  The algorithmic detection of communities and the subsequent investigation of both their aggregate properties and the properties of their component members can provide novel insights into the relationship between network structure and function (e.g., functional groupings of newly discovered proteins~\cite{Spirin2003}).

Myriad community detection methods have been developed~\cite{PorterMasonA2009,Fortunato2010}. The most popular family of methods entails the optimization of a quality function known as \emph{modularity} \cite{Newman2004,newman2006pre}.  To optimize modularity, one compares the actual network structure to some \emph{null model}, which quantifies what it means for a pair of nodes to be connected ``at random''. Traditionally, most studies have randomized only network structure (while preserving some structural properties) and not incorporated other features (such as spatial or other information). The standard null model for modularity optimization is the ``Newman-Girvan'' (NG) null model, in which one randomizes edge weights such that the expected strength distribution is preserved \cite{Newman2004,newman2006pre}.  It is thus related to the classical configuration model \cite{Newman2010}.  It has become very popular due to its simplicity and effectiveness, and it has been derived systematically through the consideration of Laplacian dynamics on networks \cite{lambiotte2008}.  However, it is also a naive choice, as it does not incorporate domain-specific information. 

The choice of a null model is an important consideration because (1) it can have a significant effect on the community structure obtained via optimization of a quality function, and (2) it changes the interpretation of communities \cite{good2010,MacMahon2013arXiv,bazzi2014}. The best choice for a null model depends on both one's data set and scientific question.  In the present paper, we explore the issue of null model choice in detail in the context of spatially embedded and temporal networks. 

Most existing research on community detection does not incorporate metadata about nodes (or edges) or information about the timing and location of interactions between nodes. However, with the increasing wealth of space-resolved and time-resolved data sets, it is important to develop community detection techniques that take advantage of the additional spatial and temporal information (and of domain-specific information, such as generative models for human interactions \cite{Expert2011}). Indeed, community detection in temporal networks has become increasingly popular~\cite{BERGERWOLF_2006,Fenn2009,Chan2009,Mucha2010,Kawadia2012,Chen2013, Bassett2013}, but the majority of methods use networks that are constructed from either static snapshots of data or aggregations of data over time windows.  Few investigations of community structure in temporal networks have used methods that exploit temporal structure (see, e.g., \cite{Mucha2010,Bassett2013}).  There is also starting to be more work on the influence of space on community structure~\cite{Cerina2012,Expert2011,Hannigan2013,Shakarian2013,bassett-granular2014}, but much more research is necessary. 

In the present paper, we use modularity maximization to study communities in spatially embedded and time-dependent networks. We compare the results of community detection using two different spatial null models --- a \emph{gravity null model}~\cite{Expert2011} and a new \emph{radiation null model} --- to the standard NG null model using novel synthetic benchmark networks that incorporate spatial effects via distance decay or disease flux as well as temporal correlation networks that we constructed using time-series data of recurrent epidemic outbreaks in Peru. We also evaluate a recently-proposed \emph{correlation null model}, which was developed specifically for correlation networks that are constructed from time series~\cite{MacMahon2013arXiv}, on the epidemic-correlation data.

Our direct analysis of disease data in the present paper provides a complementary (e.g., more systemic) approach to the majority of studies using network science methodology in this field, which focus on the importance of interpersonal contact networks on the disease spread on an individual level. These types of network methods have become increasingly prevalent in the modeling of infectious diseases~\cite{Barrat2008}. Our work also complements other approaches, such as large-scale compartmental models that incorporate transportation networks to link local populations. Such models have been used to study large-scale spatial disease spread (e.g., to examine the influence of features such as spatial location, climate, and facility of transportation  on phenomena such as disease persistence and synchronization of disease spread) \cite{Barthelemy2011,Balcan2009,Colizza2007,Xia04}.

The rest of the present paper is organized as follows. In Section \ref{Section:networks}, we give an overview of networks and community detection.  We also discuss the gravity null model and introduce a new radiation null model. We give our results for synthetic spatial networks in Section \ref{Section:benchmarks}, and we give our results for correlation networks that we construct from disease data in Section \ref{Section:dengue}.  We summarize our results in Section \ref{Section:conclusions}.  In appendices, we include the results of additional numerical experiments from varying parameters in the benchmark networks.  We also include an additional examination of the similarity between network partitions for the benchmarks and the dengue fever correlation networks.
%
%
\section{Networks and Community Structure}
\label{Section:networks}
A network describes a set of entities (called \emph{nodes}) that are connected by pairwise relationships (called \emph{edges}). In the present paper, we study weighted networks which are \emph{spatially embedded}: each node represents a location in space. One can represent a weighted network with $N$ nodes as an $N \times N$ adjacency matrix $W$, where an edge $W_{ij}$ represents the strength of the relationship between nodes $i$ and $j$. We seek to find \emph{communities}, which are sets of nodes that are densely connected to each other but sparsely connected to other dense sets in a network~\cite{Fortunato2010,PorterMasonA2009}.

We wish to study the evolution of network structure through time. The simplest way to represent temporal data is through an ordered set of \emph{static networks}, which can arise either as snapshots at different points in time or as a sequence of aggregations over consecutive time windows (which one can take either as overlapping or nonoverlapping).  

Static networks provide a good starting point for the development and investigation of new methods --- which, in our case, entails how to incorporate spatial information into null models for community detection via modularity maximization.  However, they do not take full advantage of temporal information in data that changes in time.  For example, it can be hard to track the identity of communities in temporal sequences of networks \cite{Mucha2010}.

To mitigate the community-tracking problem, we also use a type of \emph{multilayer network}~\cite{mikko-review,bocca-review} known as a multislice network \cite{Mucha2010}.  This gives an $N \times N \times m$ adjacency tensor $\overbar{W}$ that has $m$ layers and $N$ nodes in each layer, where each layer has a copy each node $i$. The intralayer edges in the network are exactly the same as they were for the sequence of static networks: the tensor element $\overbar{W}_{ijs}$ gives the weight of an intralayer edge between nodes $i$ and $j$ in layer $s$.  Additionally, each node is connected to copies of itself in consecutive layers $s$ and $r$ using interlayer edges of weight $C_{isr}$.  In this paper, we will suppose for simplicity that  $C_{isr} = \omega \in [0, \infty)$, but one can also consider more general situations \cite{mikko-review,DeDomenico2013}. A multislice network can have up to ($N \times m$) \emph{multilayer nodes} (i.e., node-layer tuples), each of which corresponds to a specific (node, time) pair. Hence, this structure makes it possible to detect temporally evolving communities in a natural way.

For our computations of community structure, we flatten the $N \times N \times m$ adjacency tensor into a $(N \times m) \times (N \times m)$ adjacency matrix, such that the intralayer connections are on the main block diagonal and the interlayer connections occur on the off-block-diagonal entries. We detect communities by maximizing modularity, which we use to describe the ``quality'' of a particular network partition into communities in terms of its departure from a null model~\cite{Newman2004}. The null model amounts to a prior belief regarding influences on network structure, so it is important to carefully  consider the choice of null model \cite{Expert2011,Bassett2013,MacMahon2013arXiv}.

For a weighted static network $W$, modularity is~\cite{Newman2004weight} 
\begin{equation}
\label{modularity-weighted}
	Q = \frac{1}{2w} \sum_{ij}{ (W_{ij} - \gamma P_{ij} ) \delta (c_i , c_j ) }\,,
\end{equation}
where $2w = \sum_{ij} W_{ij}$ is the total edge weight, $c_i$ denotes the community that contains node $i$, the function $\delta$ is the Kronecker delta, and $P_{ij}$ is the $ij$-th element of the null model matrix. One can examine different scales of community structure by incorporating a resolution parameter $\gamma$~\cite{Reichardt2006,Onnela2012}. Smaller values of $\gamma$ tend to yield larger communities and vice versa.

For multislice networks, modularity is given by
\begin{equation}
\label{modularity-multislice}
	\overbar{Q} = \frac{1}{2\overbar{w}} \sum_{ijsr} {\left[\left(\overbar{W}_{ijs} - \gamma \overbar{P}_{ijs} \right) \delta_{sr} + \delta_{ij} C_{jsr} \right] \delta (\overbar{c}_{is} , \overbar{c}_{jr})}\,,
\end{equation}
where $2\overbar{w} = \sum_{ijs} \overbar{W}_{ijs}$, the quantity $\overbar{c}_{is}$ denotes the community that contains node $i$ in layer $s$, and $\overbar{P}_{ijs}$ is the $ij$-th element of the null model tensor in layer $s$~\cite{Mucha2010}. 

To detect communities via modularity maximization, one searches the possible network partitions for the one with the highest modularity score. Because exhaustive search over all possible partitions is computationally intractable \cite{Brandes2008}, practical algorithms invariably use approximate optimization methods (e.g., greedy algorithms, simulated annealing, or spectral optimization), and different approaches offer different balances between speed and accuracy~\cite{PorterMasonA2009,Fortunato2010}. 

In the present paper, we optimize modularity using a two-phase iterative procedure similar to the Louvain method~\cite{Blondel2008}. However, rather than using the adjacency matrix $W$, we work with the modularity matrix $B$ with elements $B_{ij} = W_{ij} - \gamma P_{ij}$ for static networks and with the modularity tensor with elements $\overbar{B}_{ijs} = \overbar{W}_{ijs} - \gamma \overbar{P}_{ijs}$ for multislice networks~\cite{Netwiki}. 

The employed Louvain-like algorithm~\cite{Netwiki} is stochastic, and a modularity landscape for empirical networks typically includes a very large number of nearly-optimal partitions~\cite{good2010}.  For each of our numerical experiments, we thus apply the computational heuristic 100 times to obtain a \emph{consensus community structure}~\cite{Lancichinetti2012} by constructing an \emph{association matrix} $A^{\mathrm{rep}}$ (where the entries $A^{\mathrm{rep}}_{ij}$ represent the fraction of times that nodes $i$ and $j$ are classified together in the 100 partitions) and performing community detection on $A^{\mathrm{rep}}$ using the uniform null model $P_{ij}^U = 2w / [N(N-1)]$~\cite{Bassett2013}. We choose the uniform null model in order to detect the strongest community structure in the association matrix (i.e., one that is often detected by the original optimization process).  


For multislice networks, we perform community detection and then consensus clustering using the same basic procedure.  This yields an assignment of each multilayer node (i.e., node-layer tuple) to a community.  We are also sometimes interested in community assignments of the original entities (i.e., a partition of the set of nodes regardless of what layer they are in). For example, we might wish to compare the result of algorithmic community detection to known partitions, such as grouping a node (i.e., province) by climate, population, administrative region, etc. To do this, we perform what we call \emph{province-level community detection}, which proceeds in two rounds: (1) we detect communities in a multislice network using any method and null model of choice; 
(2) we use this partition to construct an $N \times N$ province-level association matrix (i.e., a matrix $A^{\mathrm{province}}$ where entries $A^{\mathrm{province}}_{ij}$ represent the fraction of times that nodes $i$ and $j$ are classified together in all layers), and we detect province-level communities by maximizing modularity on this association matrix using a uniform null model. We choose the uniform null model to detect the most temporally persistent community structure in the association matrix (i.e., one that is often detected in multiple layers). We can then follow this with consensus community detection across 100 repeat province-level structures. 
\subsection{Null Models for Community Detection}
\label{Section:commdetect}
The choice of null model is vital for the detection of communities using modularity maximization~\cite{good2010,Bassett2013,MacMahon2013arXiv}. The most common choice is the Newman-Girvan (NG) null model, which randomizes a network such that the expected strength sequence of nodes is preserved~\cite{Newman2006,Newman2006pre}. For static, weighted networks, the NG null model is given by
\begin{equation}
	P^{\mathrm{NG}}_{ij}=\frac{k_{i} k_{j}}{2w}\,, 
\end{equation}
where $k_i = \sum_j W_{ij}$ is the strength (i.e., weighted degree) of node $i$ and $2w = \sum_{ij} W_{ij}$ is the total edge weight in the network. 

For multislice networks, the NG null model is~\cite{Mucha2010}
\begin{equation}
	\overbar{P}^{\mathrm{NG}}_{ijs}=\frac{\overbar{k}_{is} \overbar{k}_{js}}{2\overbar{w}}\,, 
\end{equation}
where $\overbar{k}_{is} = \sum_j \overbar{W}_{ijs}$ is the intralayer strength of node $i$ in layer $s$ and $2\overbar{w} = \sum_{ijs} \overbar{W}_{ijs}$. 

Despite its popularity and demonstrated effectiveness in many situations, the NG null model is naive in the sense that only takes node strengths into account and it does not incorporate problem-specific information (such as spatial embeddedness).  It is often important to incorporate additional (domain-specific or even problem-specific) information, and what one considers to be connected ``at random'' depends fundamentally on the research question of interest. Consequently, NG null model is not suitable for all applications.


\subsubsection{Spatial Null Models: Gravity Model}

In many spatially embedded networks, proximity has a strong effect on the connections between nodes, as (all else held equal) neighboring nodes are more likely to be connected to each other (and their connections are likely to have to have larger weights) than nodes that are far away \cite{Barthelemy2011,Expert2011}.  Moreover, proximity can mask other underlying influences. Consequently, incorporating the expected influence of proximity on edge weights into null models for community detection should make it possible to discover new and important types of structures.

Expert et al.~\cite{Expert2011} proposed a spatial null model that was inspired by the ``gravity model'' of human mobility \cite{zipf1946,stewart1947,stewart1958,wilson1967}. A gravity model assumes that the interaction between two locations is proportional to their importance (e.g., population), but it decays with distance.

In the standard gravity model, the interaction between locations $i$ and $j$ with respective populations $n_i$ and $n_j$ that are a distance $d_{ij}$ apart is
\begin{equation}
\label{eq:gravitymodel}
	G_{ij} = n_i^\alpha n_j^\beta f(d_{ij})\,,
\end{equation}
where the ``deterrence function'' $f(d)$ describes the effect of space on node interactions. Common choices for the deterrence function include inverse proportionality to distance (i.e., $f(d_{ij}) = 1/d_{ij}$), inverse proportionality to squared distance (i.e., $f(d_{ij}) = 1/d_{ij}^2$), exponential decay (i.e., $f(d_{ij}) = e^{-d_{ij}}$), and other interactions of the form $f(d_{ij}) = d_{ij}^\kappa$~\cite{Barthelemy2011}.  It is common to estimate the parameters $\alpha$, $\beta$, and $ \kappa$ using regression.
Gravity models have been employed successfully during the past half century to model spatial interactions such as population migration~\cite{Barthelemy2011,Balcan2009a,shl-marriage}, trade~\cite{Disdier2008}, and disease spread~\cite{Xia04}. 

The simplest form of a gravity-like interaction in Eq.~(\ref{eq:gravitymodel}), with $\alpha = \beta = 1$ and $\kappa = -1$, was incorporated into a \emph{gravity null model}~\cite{Expert2011}, to give
\begin{equation}
	P^{\mathrm{grav}}_{ij} = I_i I_j f(d_{ij})\,, 
\label{eq:gravity}
\end{equation}
where $I_i$ is the importance of node $i$. One estimates the ``deterrence function'' 
\begin{equation}
	f(d) = \frac{\sum_{\{k,l|d_{kl}=d\}}{W_{kl}}}{\sum_{\{k,l|d_{kl}=d\}}{(I_{k}I_l})}\,,
\label{eq:gravityfd}
\end{equation}
from data for all nodes at distance $d$ between them in a data set. Expert et al.~\cite{Expert2011} used 
$n_i$, the population of province $i$, as their measure of node importance. After briefly experimenting with variations, such as using population density or a logarithm of the population (i.e., $I_i = \log(n_i)$) and observing no significant differences in performance, we will follow their lead. Another simple choice is node strength (i.e., $I_i = k_i = \sum_j W_{ij}$), though the null model then becomes very similar to the usual NG null model~\cite{Expert2011}.  Moreover, if $f(d)$ does not depend on distance, then the null model becomes exactly the NG null model in that case.

In most data sets, distances are continuous, so one needs to bin distance data to obtain enough nodes in each distance bin to construct a meaningful deterrence function $f(d)$ in Eq.~(\ref{eq:gravityfd}). Possible binning methods include binning into equal-distance bins (e.g., every $b$ km) and equal-sized bins (e.g., each bin containing $c$ elements). After testing the choice of binning procedure on the benchmark networks and applying the null model to empirical data and observing no qualitative differences in null model performance, we selected the equal-distance method for the rest of the paper, while choosing a bin size large enough that there are always more than 5 elements in each bin. Additionally, for the benchmark networks we can test the influence of bin sizes on similarity of algorithmic partitions to the planted community structure. We will give the specific bin sizes for spatial benchmark and dengue correlation networks in their respective Sections.

Combining Eqs.~(\ref{eq:gravity}) and (\ref{eq:gravityfd}) allows us to write the gravity null model as
\begin{equation}
	P^{\mathrm{grav}}_{ij} = I_i I_j \frac{\sum_{\{k,l|d_{kl}=d_{ij}\}}{W_{kl}}}{\sum_{\{k,l|d_{kl}=d_{ij}\}}{(I_{k}I_l})}\,.
\label{eq:gravitynm}
\end{equation}
Expert et al. used the null model (\ref{eq:gravitynm}) to uncover a linguistic partition of a network of Belgian mobile phone calls into the French and Flemish speaking parts of Belgium. This partition was obscured by geographical communities when using the NG null model~\cite{Expert2011}. 

In the present paper, we generalize the gravity null model to a multislice setting by calculating a separate gravity null model for each layer $s$. The resulting multislice gravity null model is
\begin{equation}
	\overbar{P}^{\mathrm{grav}_{ijs}} = I_i I_j \frac{\sum_{\{k,l|d_{kl}=d_{ij}\}}{\overbar{W}_{kls}}}{\sum_{\{k,l|d_{kl}=d_{ij}\}}{(I_{k}I_l})}\,,
\label{eq:gravitym}
\end{equation}
where we have assumed that the population stays constant across time. If one has reliable information about changes in population with time, one can incorporate such information into the null model (\ref{eq:gravitym}) by substituting $I_i$ with an analogous quantity $I_{is}$ that depends both on the node $i$ and on the layer $s$.


\subsubsection{Spatial Null Models: Radiation Model}

Gravity models include multiple parameters that one needs to either choose arbitrarily or estimate from data.  Moreover, by their design, gravity models are unable to predict different fluxes between locations that are the same distance apart but which have regions with different population densities between them. For example, one would expect a higher flux of infectious disease between two locations that are separated by a space with high population density than between locations that are separated by a space with low population density (because of the higher availability of susceptible hosts in the latter case)~\cite{Jones2008}. By contrast, one would expect a smaller commuting flux between such locations in the latter case due to higher availability of nearby jobs, as this reduces peoples' willingness to commute for longer distances~\cite{Simini2012}. 

The radiation model~\cite{Simini2012} was developed to attempt to address these issues. It was designed for population flows and has subsequently been applied successfully in several situations~\cite{Goh2012,Masucci2013}. Because the radiation model is designed to capture human mobility between populations, and the long-distance spread of many infectious diseases --- including dengue --- is believed to be largely due to long-distance mobility ~\cite{Stoddard2009}, the radiation model might provide a useful but simplified description for the spread of disease across space. In this section, we use it to construct a new spatial null model for community detection that we believe might be well-suited for studying the long-distance spread of dengue.

The mean commuting flux predicted by the radiation model for locations $i$ and $j$ with populations $n_i$ and $n_j$ is
\begin{equation}
 	T_{ij} = T_i \frac{n_i n_j}{(n_i + r_{ij})(n_i + n_j + r_{ij})}\,,
\label{eq:radiation}
\end{equation}
where $r_{ij}$ is the population between locations $i$ and $j$, and $T_i$ is the number of commuters in location $i$.  A simple way to calculate $r_{ij}$ is to use the population $q_{ij}$ in the circle of radius $d_{ij}$ centered at $i$ and subtract the total of the populations at the origin and destination.  That is, $r_{ij} = q_{ij} - (n_i + n_j)$.
Although the radiation model is relatively recent~\cite{Simini2012}, several modifications to it have already been proposed.  These include incorporating a normalization for finite systems~\cite{Masucci2013} and the development of a general framework that includes ideas from the radiation, gravity, and intervening-opportunities models~\cite{Simini2013}. 

We propose a novel null model for community detection based on the original formulation of the radiation model~\cite{Simini2012}. We use a similar formulation to Eq.~(\ref{eq:gravitynm}) to incorporate both the expected distance-dependent flux and the actual network structure. To avoid creating a directed network, we use a symmetrized predicted flux 
\begin{equation}
{\hat{T}_{ij} = (T_{ij} + T_{ji} )/2}
\label{eq:radiation-symmetric}
\end{equation}
 between nodes $i$ and $j$. 
\footnote{Although the directionality of fluxes is an important factor to study, we wish to keep our null models as simple as possible in order to focus on the effect of incorporating space into them. Additionally (and again for simplicity), we will construct our disease-correlation networks using Pearson correlations, so we will study the community structure of undirected networks. If one instead constructs a directed network --- e.g., by including a time delay when measuring the similarity of time series, considering ideas such as Granger causality, or otherwise measuring similarity in a way that produces a directed network (see, e.g., Ref.~\cite{Smith2011}), then it would also be desirable to construct a directed version of the radiation null model.  Clearly, this is an interesting future direction, but it is beyond the scope of our study.}
We thereby construct the \emph{radiation null model}
\begin{equation}
	P^{\mathrm{rad}}_{ij}= \hat{T}_{ij} \frac{ \sum_{\{k,l|d_{kl}=d_{ij}\}} {W_{kl}}} { \sum_{\{k,l|d_{kl}=d_{ij}\}}\hat{T}_{kl}}\,.
\label{eq:radiation-null}
\end{equation}

In Section \ref{Section:dengue}, we will study community structure in empirical data from several years of dengue fever occurrences in Peru. Because we do not possess detailed data on the commuting patterns in Peru (see the description of our data in Section~\ref{data}), we assume that commuters are distributed uniformly across space.  We can then simplify Eq.~(\ref{eq:radiation}) by substituting $T_i = T_f n_i$, where $T_f$ is the fraction of the population that commutes. Because the quantity $T_f$ is present in both the numerator and denominator of Eq.~(\ref{eq:radiation-null}), we can now cancel it out. However, if one possesses commuting data, it would be desirable to use it to improve the radiation null model.

We also extend the radiation null model to a multislice setting in an analogous manner to the gravity null model. The multislice radiation null model is
\begin{equation}
	\overbar{P}^{\mathrm{rad}}_{ijs}= \hat{T}_{ij} \frac{ \sum_{\{k,l|d_{kl}=d_{ij}\}} {\overbar{W}_{kls}}} { \sum_{\{k,l|d_{kl}=d_{ij}\}}\hat{T}_{kl}}\,.
\label{eq:radiation-nullm}
\end{equation}
Again, one can incorporate temporal data about population sizes and thereby replace $T_{ij}$ with $T_{ijs}$ to improve the null model.


\subsubsection{Spatial Null Models: Other Models}

The incorporation of spatial information into null models for community detection is an important problem, and several other ideas have been proposed recently. For example, Cerina et al.~\cite{Cerina2012} focused on disentangling the correlation between node attributes and space, so they used a simple exponential decay: $f(d_{ij}) = e^{-d_{ij}/\overbar{d}}$, where $\overbar{d}$ is the mean distance between nodes in a network. Shakarian et al.~\cite{Shakarian2013} focused on finding geographically-disperse communities, so they introduced a decay constant $\theta$ such that $f(d_{ij}) = e^{{-d_{ij}/\theta}^2}$. Another recently-proposed null model was used to attempt to find geographically-proximate communities \cite{Hannigan2013}. 

As the exact nature of the influence of spatial distance on interactions in the dengue fever data is unclear, we decided to focus only on null models that include a contribution from the data, rather than using null models with an arbitrarily chosen functional dependence. Thus, we do not test these null models in the present paper. 


\section{Synthetic Benchmark Networks}\label{Section:benchmarks}

To test the performance of the spatial null models, we develop novel synthetic benchmark networks that represent idealized spatially-embedded networks with planted community structure. 

In what we call the \emph{distance benchmark}, the probability of an edge between two nodes depends only on the geographical distance between nodes and on their community assignments. We assign $N$ nodes uniformly at random to positions on the lattice $\{1,2,\ldots,l\} \times \{1,2,\ldots,l\}$. 
We assign a population $n_i$ to each node $i$ (which is an idealized ``city''). We create two versions of the distance benchmark: the ``uniform population distance benchmark'' and the ``random population distance benchmark''. The uniform population version corresponds to the benchmark in Expert et al.~\cite{Expert2011}; we assign the same population ($n_i = 100$) to each node. In the random population benchmark, we assign an integer population uniformly at random from the set $\{1,\ldots,100\}$.

We also assign the nodes uniformly at random to one of two communities. In the distance benchmarks, the probability $p^{\mathrm{dist}}_{ij}$ that an edge exists between nodes $i$ and $j$ at distance $d_{ij}$ is inversely proportional to distance:
\begin{equation}
 	p^{\mathrm{dist}}_{ij} = \frac{\lambda(c_i,c_j)}{Z_1 d_{ij}}\,,
\end{equation}
where $c_i$ is the community that contains node $i$ and the function $\lambda (c_i, c_j) = 1$ if nodes $i$ and $j$ are in the same community and $\lambda (c_i, c_j) = \lambda_d$ otherwise. The ``inter-community connectivity'' $\lambda_d$ controls the degree of mixing between communities. When $\lambda_d = 0$, only nodes in the same community are adjacent to each other; when $\lambda_d = 1$, there are no distinct communities. The normalization constant $Z_1$ ensures that $\sum_{i>j} p^{\mathrm{dist}}_{ij} = 1$. We place $L = \mu N(N - 1)/2$ edges, where there is an edge between nodes $i$ and $j$ with probability $p^{\mathrm{dist}}_{ij}$ each, and the parameter $\mu \geq 0$ determines the network's edge density. We interpret multiple edges as weights. We normalize the weights in the network to $ \left[ 0,1 \right ]$ by dividing each entry by the maximum weight in the network. 
 
With our \emph{flux benchmark}, we aim to mimic the spread of disease on a network. We allocate its edge weights depending on the mean flux between pairs of nodes that is predicted by the radiation model. We place $N$ nodes uniformly at random on the lattice $\{1,2,\ldots,l\} \times \{1,2,\ldots,l\}$, and we assign populations and communities in the same manner as for the distance benchmark. Again as with the distance benchmark, we consider both uniform-population and random-population versions of the flux benchmark. Now, however, the edge probability $p^{\mathrm{flux}}_{ij}$ is directly proportional to the mean predicted radiation-model flux between nodes $i$ and $j$ ($\hat{T}_{ij}$, which is turn is inversely proportional to distance $d_{ij}$):
\begin{equation}
 	p^{\mathrm{flux}}_{ij} = \frac{\lambda(c_i,c_j) \hat{T}_{ij}}{Z_2}\,,
\end{equation}
where $Z_2$ is a normalization constant to ensure that $\sum_{i>j} p^{\mathrm{flux}}_{ij} = 1$. 

In Table \ref{table-benchmarks}, we summarize the four synthetic benchmark networks that we have just introduced.

\begin{table}[t]
\caption{Primary characteristics (i.e., population and edge probability) for the distance and flux benchmarks for static networks.
The quantity $\text{rand}(\{a,b\})$ signifies that select a number uniformly at random from the set $\{a,a+1,\ldots,b\}$.  Additionally, $\lambda (c_i, c_j) = 1$ if nodes $c_i$ and $c_j$ are in the same community and $\lambda (c_i, c_j) = \lambda_d$ otherwise, $d_{ij}$ is the distance between nodes $i$ and $j$ in space, and $Z_1$ and $Z_2$ are normalization constants.}
 \label{table-benchmarks}
\begin{tabular}{lll}
Benchmark & Population & $p_{ij}$ \\ 
\hline
Distance, uniform population & 100 & $p^{\mathrm{dist}}_{ij} = \frac{\lambda(c_i,c_j)}{Z_1 d_{ij}}$ \\
Distance, random population & $\text{rand}(\{1,100\})$ & $p^{\mathrm{dist}}_{ij} = \frac{\lambda(c_i,c_j)}{Z_1 d_{ij}}$ \\
\hline
Flux, uniform population& 100 & $p^{\mathrm{flux}}_{ij} = \frac{\lambda(c_i,c_j) \hat{T}_{ij}}{Z_2}$ \\
Flux, random population & $\text{rand}(\{1,100\})$ & $p^{\mathrm{flux}}_{ij} = \frac{\lambda(c_i,c_j) \hat{T}_{ij}}{Z_2}$ \\
\end{tabular}
\end{table}

We create both static (i.e., single-layer) and multilayer benchmarks networks. The static benchmarks enable us to study the performance of modularity maximization using a chosen null model in a simple setting without the additional complications of a multilayer network. However, the multilayer benchmarks are ultimately more appropriate for disease data because they can incorporate temporal evolution. 

We begin by placing nodes in space and assigning populations in the same manner as for the static benchmarks. We then assign nodes uniformly at random into one of two communities, and we extend this structure into a multilayer planted community structure with $m$ layers. For the ``temporally stable'' benchmarks, the planted community structure is the same for each layer. For the ``temporally evolving'' multilayer benchmarks, we change the community assignment of a fraction $p$ of the nodes. For each of these nodes, we select a new community assignment uniformly at random, and we change the community of the node in each layer; we start at a layer that we select uniformly at random, and we also change the assignment (to the same new community) in all remaining layers.

\begin{figure}
\centering
\includegraphics[width=0.5\linewidth]{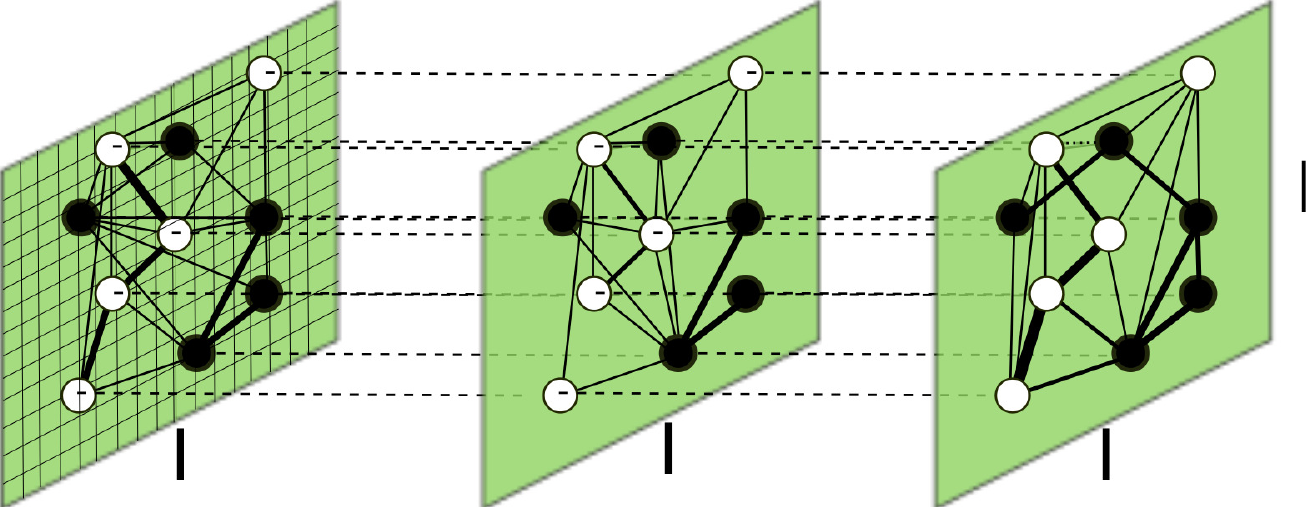}
\caption{Construction of temporally stable multilayer spatial benchmarks. We assign $N$ nodes uniformly at random to positions on a $l \times l$ lattice (which we show in layer 1) and divide them into two equal-sized communities (black and white) whose nodes we choose uniformly at random. Node $i$ has a population of $n_i$, and each slice has the same set of nodes. For each slice, we allocate edges uniformly at random according to a probability distribution that depends on the type of benchmark; for details, see the text and Table \ref{table-benchmarks}. We interpret multiple edges as weights, and we visualize these weights using edge thickness. We connect copies of nodes in adjacent layers with interlayer edges of weight $\omega$ (dashed lines).\label{Figure:benchmarks-multislice}}
\end{figure}

We then generate the edges for each layer independently, in the same manner as we generate a static benchmark and using identical parameter values $(N,l,\mu,\lambda_d)$ for each; see Fig.~\ref{Figure:benchmarks-multislice}. Independent generation of each layer based on the same starting conditions represents differences between observations due to noise and experimental variation.

For each of the above types of multilayer benchmarks, we set the value of the interlayer edges between corresponding nodes in consecutive layers to be $\omega \in[0,\inf]$. Thanks to normalizing the intralayer edge weights, we yield synthetic multilayer benchmark networks in which the relative magnitudes of interlayer edges and intralayer edges are comparable to those in the disease-correlation networks. 

Each of the reported community detection results for these benchmarks is an average over consensus community detection (over 50 repeats) for 50 independently drawn instances of a benchmark with the same values of the same parameter values $(N,l,\mu,\lambda_d)$, $(\gamma, \omega)$, and (when relevant) $p$.

We evaluate the performance of the NG, gravity, and radiation null models on our benchmarks by comparing algorithmic partitions with the planted community structure using normalized mutual information (NMI)~\cite{Strehl2002clusterensembles}. NMI is an information-theoretic similarity measure that is relatively sensitive to small differences in partitions, such as the move of a single node from one community to another, compared to pair-counting measures such as the Rand coefficient and z-Rand scores~\cite{Traud2008}. This sensitivity makes it suitable for assessing performance on benchmarks that are based on well-defined, ground-truth planted partitions. 

NMI is one of many normalized versions of mutual information (MI) \cite{Meila2007}. Both MI and NMI are based on the concept of \emph{information entropy}, which is a measure of uncertainty. MI measures the amount of information that one can predict about one random variable (which in the present paper is a partition of a network into communities) based on another one. For a partition $X = \{X_1, X_2, \ldots X_K\}$ with $K$ communities and a partition $Y = \{Y_1, Y_2, \ldots {Y_L}\} $ with $L$ communities, MI is defined as 
\begin{equation}
	I(X,Y) = \sum_{k=1}^{K} \sum_{l=1}^{L} P(k,l) \log_2 \left[\frac{P (k,l)}{P(k)P(l)}\right]\,,
\end{equation}
where $P(k)$ and $P(l)$ are the marginal probabilities of observing communities $k$ and $l$ in partitions $X$ and $Y$, respectively, and $P(k,l)$ is the joint probability of observing communities $k$ and $l$ simultaneously in partitions $X$ and $Y$. MI takes values between $0$ and $\min\{H(X),H(Y)\}$, where $H(X) = - \sum_{k=1}^{K} P(k) \log_2 P(k)$ is the entropy of $X$. 

Normalized mutual information (NMI)~\cite{Strehl2002clusterensembles} is defined as 
\begin{equation}	
	NMI (X,Y) = \frac{I(X,Y)}{\sqrt{(H(X) H(Y))}} \in [0,1]\,.
\end{equation}
The normalization to lie within the range $[0,1]$ facilitates interpretation and comparisons. We use NMI in the following sections, and we obtain the same qualitative conclusions using 
 variation of information~\cite{Kraskov2005}, which is a different normalized measure of similarity. See Appendix \ref{Appendix:VI} for our comparisons using VI.


\subsection{Results on Static Benchmarks}

\begin{figure}[tbp]
  \centering
\includegraphics[width=0.45\linewidth]{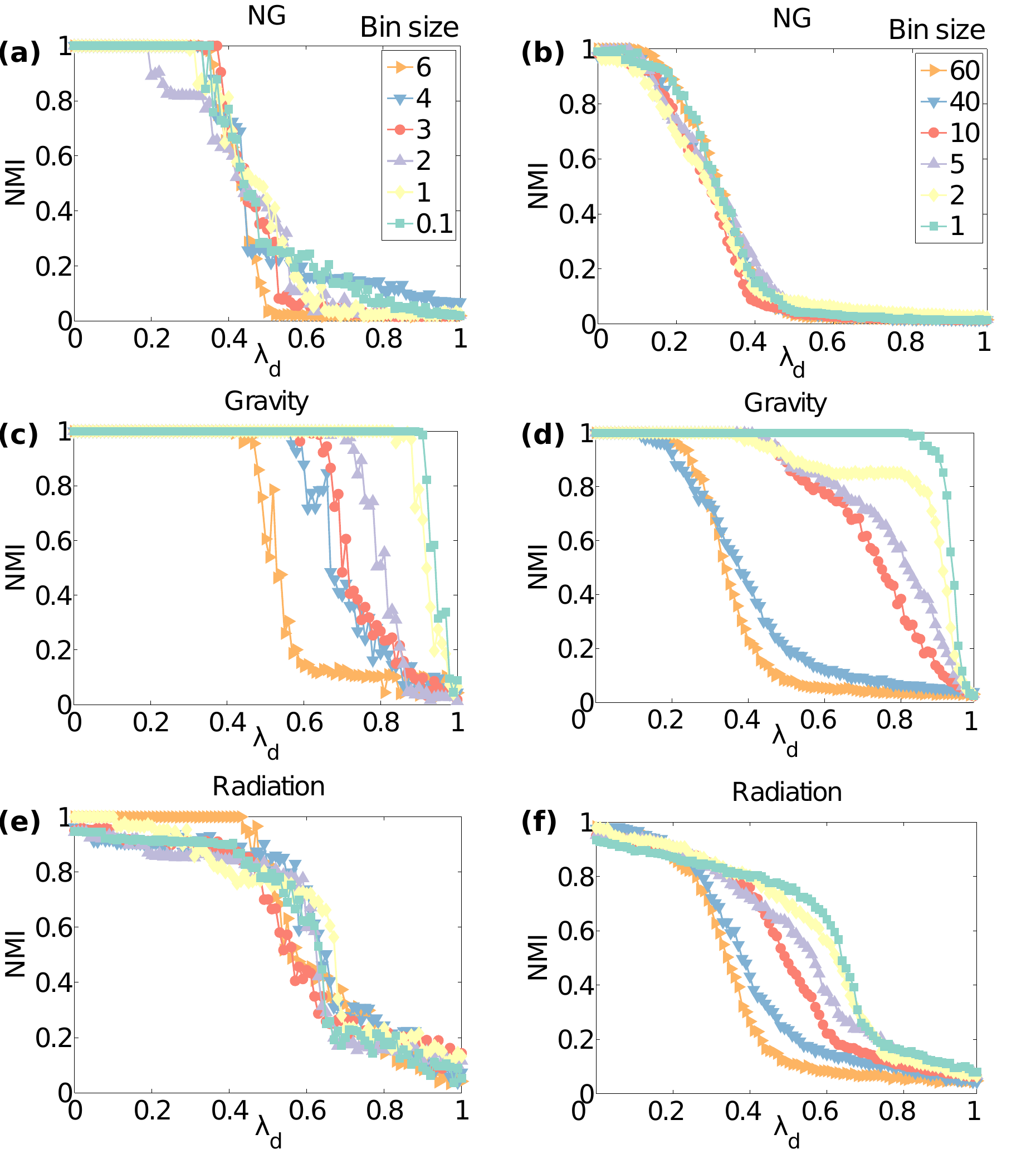}
\caption{Uniform population static benchmarks: Normalized mutual information (NMI) scores between algorithmically detected and planted community structures in static uniform population distance benchmarks for (left) $l = 10$, $N = 50$ and (right) $l = 100$, $N = 100$, edge density parameter $\mu = 100$ and uniform populations of $100$ for different bin sizes (colored curves). We detect communities by optimizing modularity using the (top) NG, (middle) gravity, and (bottom) radiation null models.
\label{Figure:Bench-static-even-d-NMI-spacesizevsnm}}
\end{figure}

To emphasize the difference between the gravity and radiation null models, we take $N = 50$ and $l = 10$ to obtain a relatively densely filled lattice. (See Appendix \ref{Appendix:cities} for the results for a synthetic network with parameter values $N = 10$ and $N = 90$.) We first compare this benchmark versus a situation with parameter values $N=100$ and $l=100$ (which are the parameter values that were used in Expert et al. \cite{Expert2011}). We test varying bin sizes in uniformly-spaced bins using the parameter values $b \in \{10^{-4},10^{-3},10^{-2},0.1\} \cup \{ 1,2,\ldots, 10\}$, $l=10$ and $b \in  \{1,2,\ldots, 100\}$, $l=100$. We find that bin width makes a large difference on both benchmarks: $b = 1$ produces the highest NMI scores (i.e., it has the ``best performance'') and increasing bin width leads to a decrease in performance of both spatial null models (see Fig.~\ref{Figure:Bench-static-even-d-NMI-spacesizevsnm}). This effect is especially pronounced for the gravity null model. 

In both cases, the best aggregate performance of the spatial null models at optimal bin sizes is similar for $l=10$ and $l=100$, so we henceforth use the $l=10$ benchmark with $b = 1$ to lower computational time and memory usage. However, one needs to keep the strong influence of bin size on algorithm results in mind for applications. 

\begin{figure}
	\centering
\begin{tabular}{m{0.03\linewidth}| m{0.16\linewidth} m{0.16\linewidth} m{0.16\linewidth} m{0.16\linewidth}}
Pop. & Uniform & Uniform & Random & Random \\
 & distance  & flux & distance & flux \\ 
 \hline
NG &
\includegraphics[width=\linewidth]{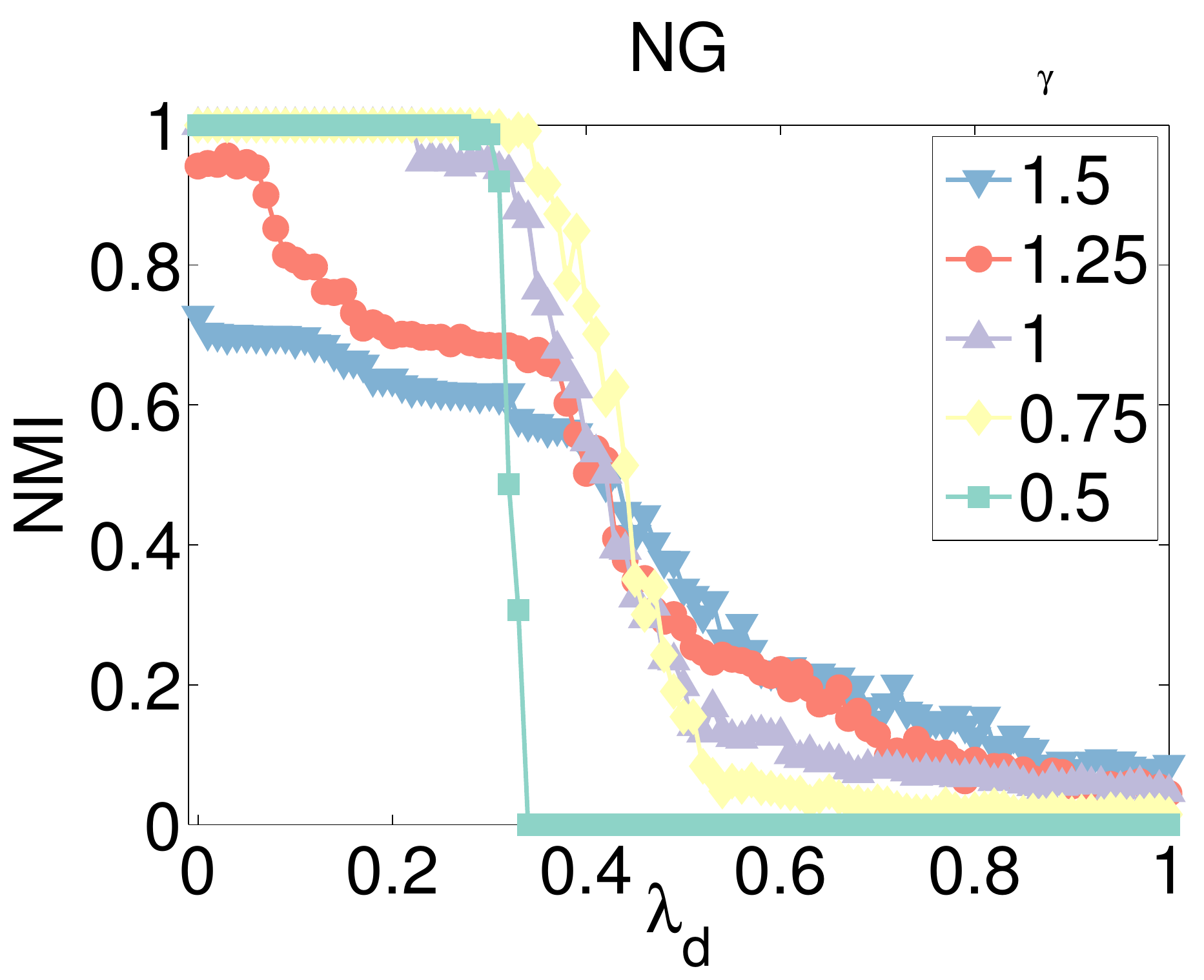} &
\includegraphics[width=\linewidth]{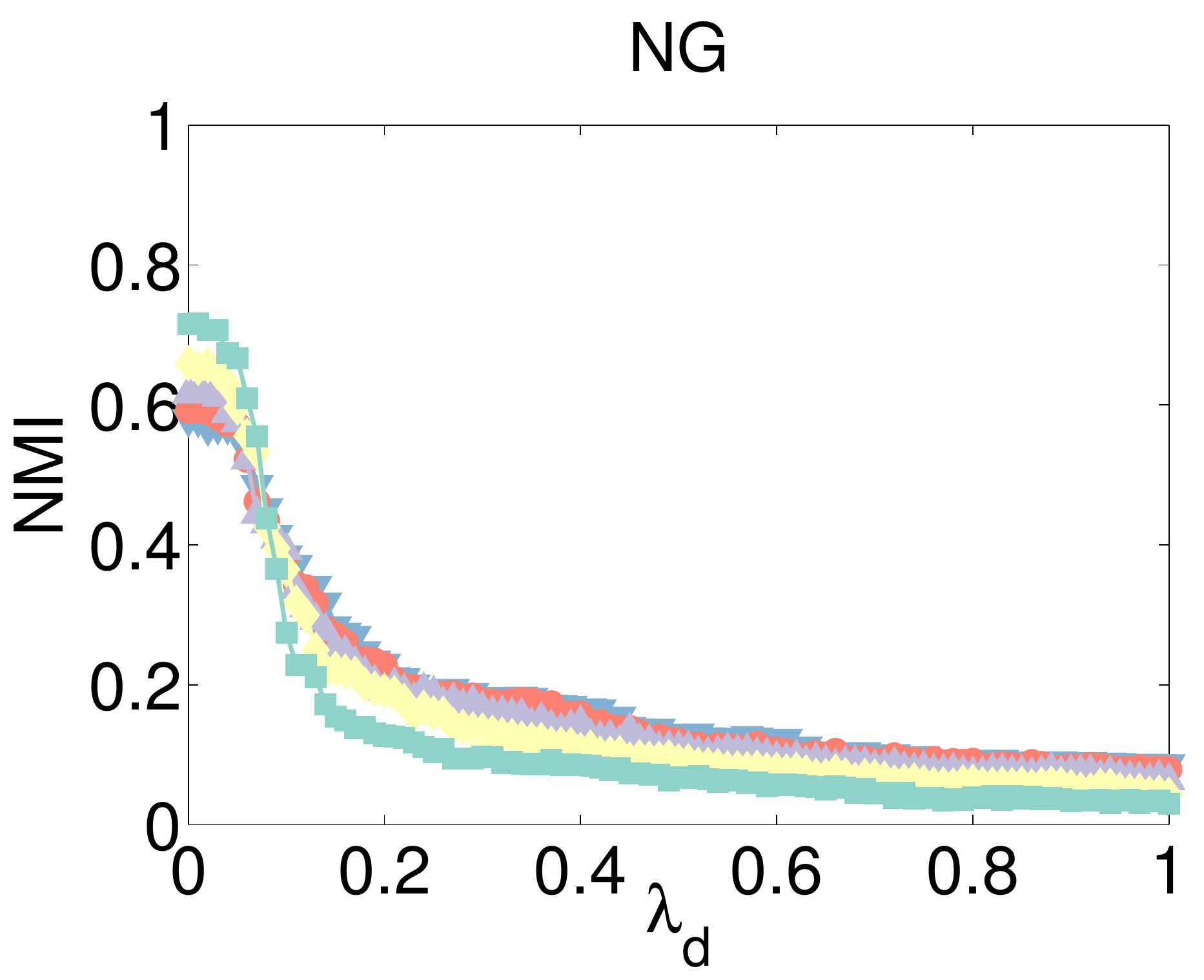} &
\includegraphics[width=\linewidth]{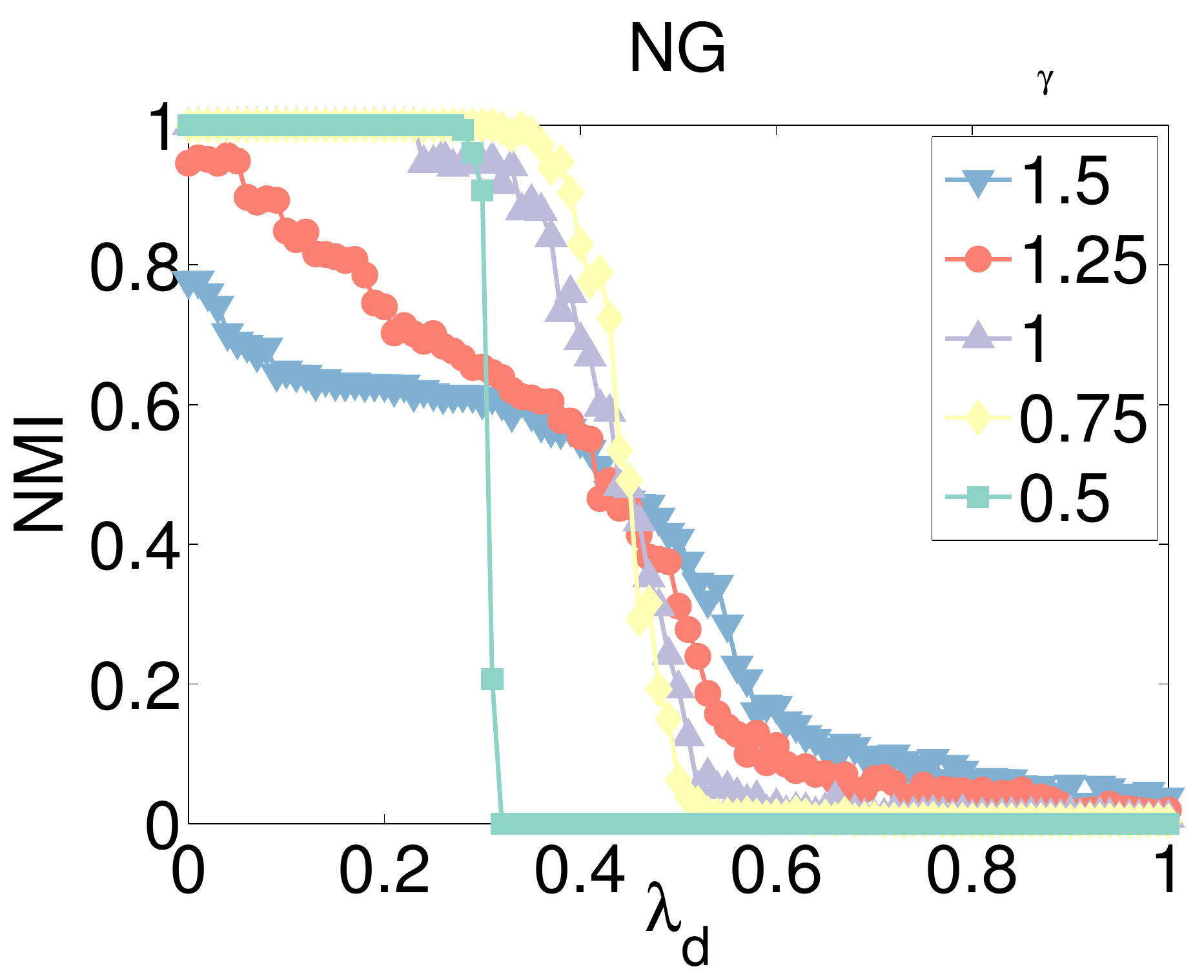} &
\includegraphics[width=\linewidth]{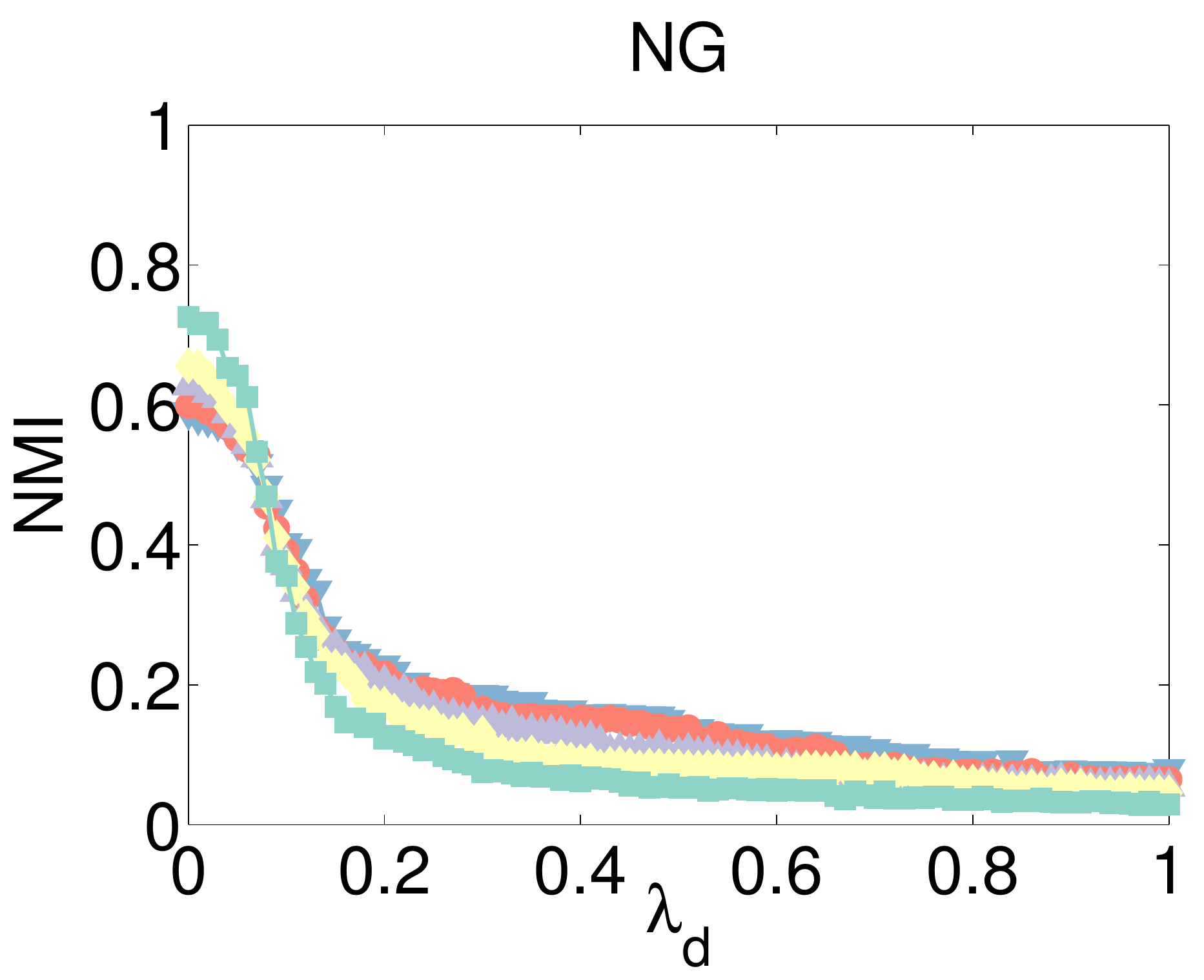}
\\
Grav.& 
\includegraphics[width=\linewidth]{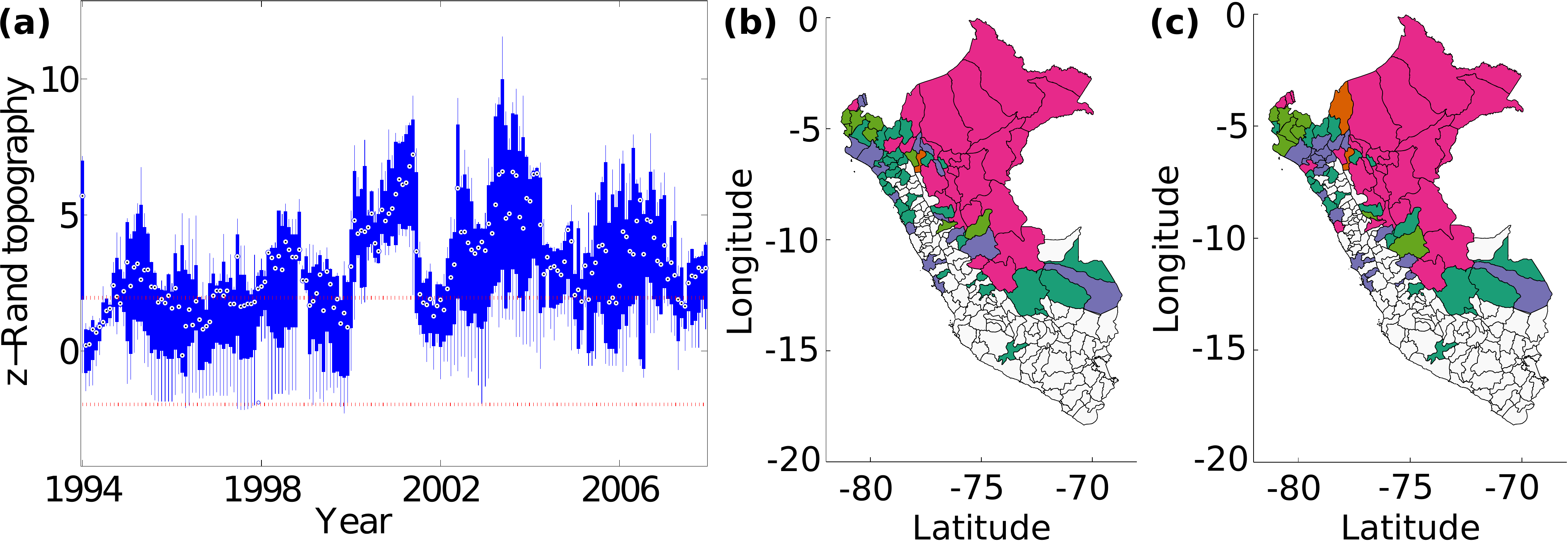} &
\includegraphics[width=\linewidth]{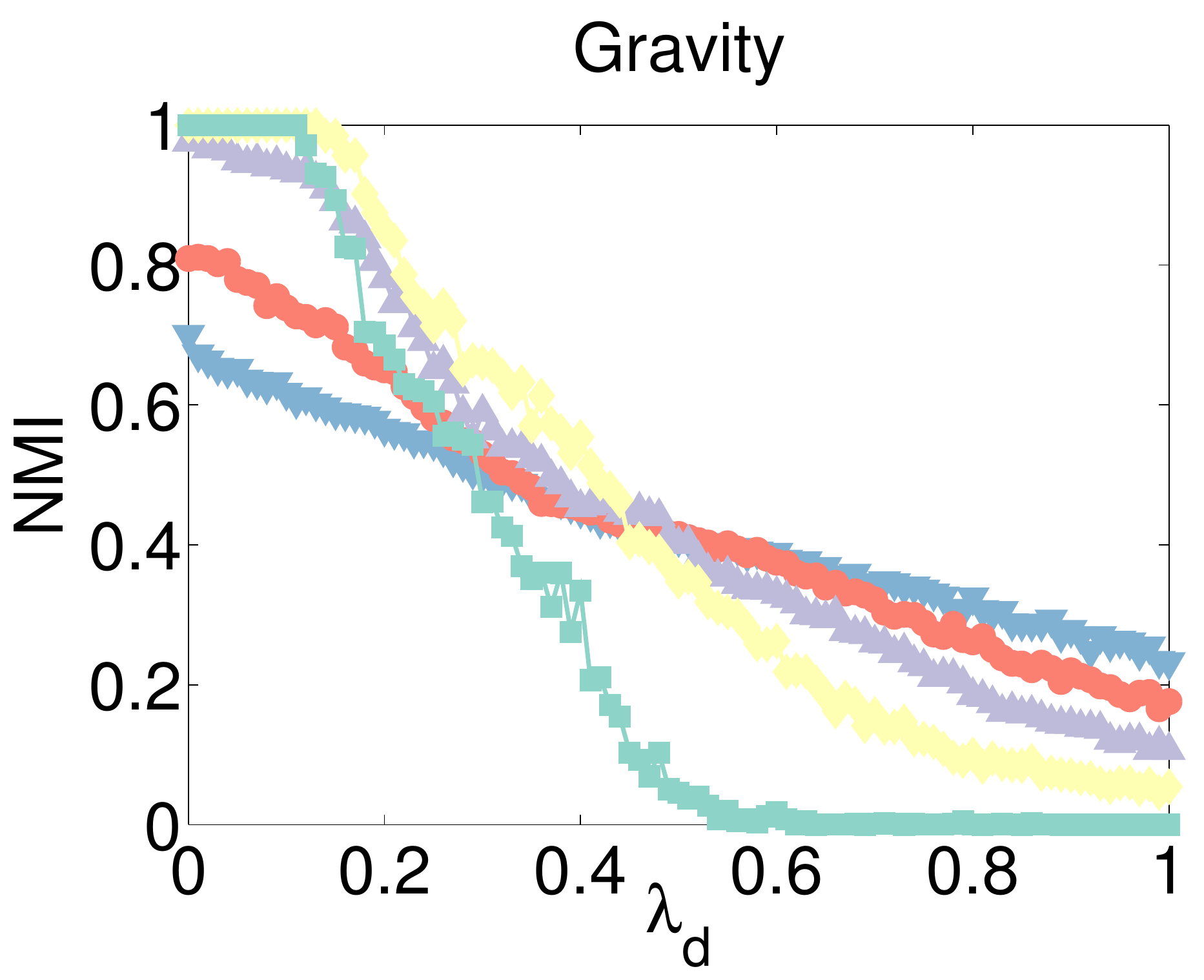} &
\includegraphics[width=\linewidth]{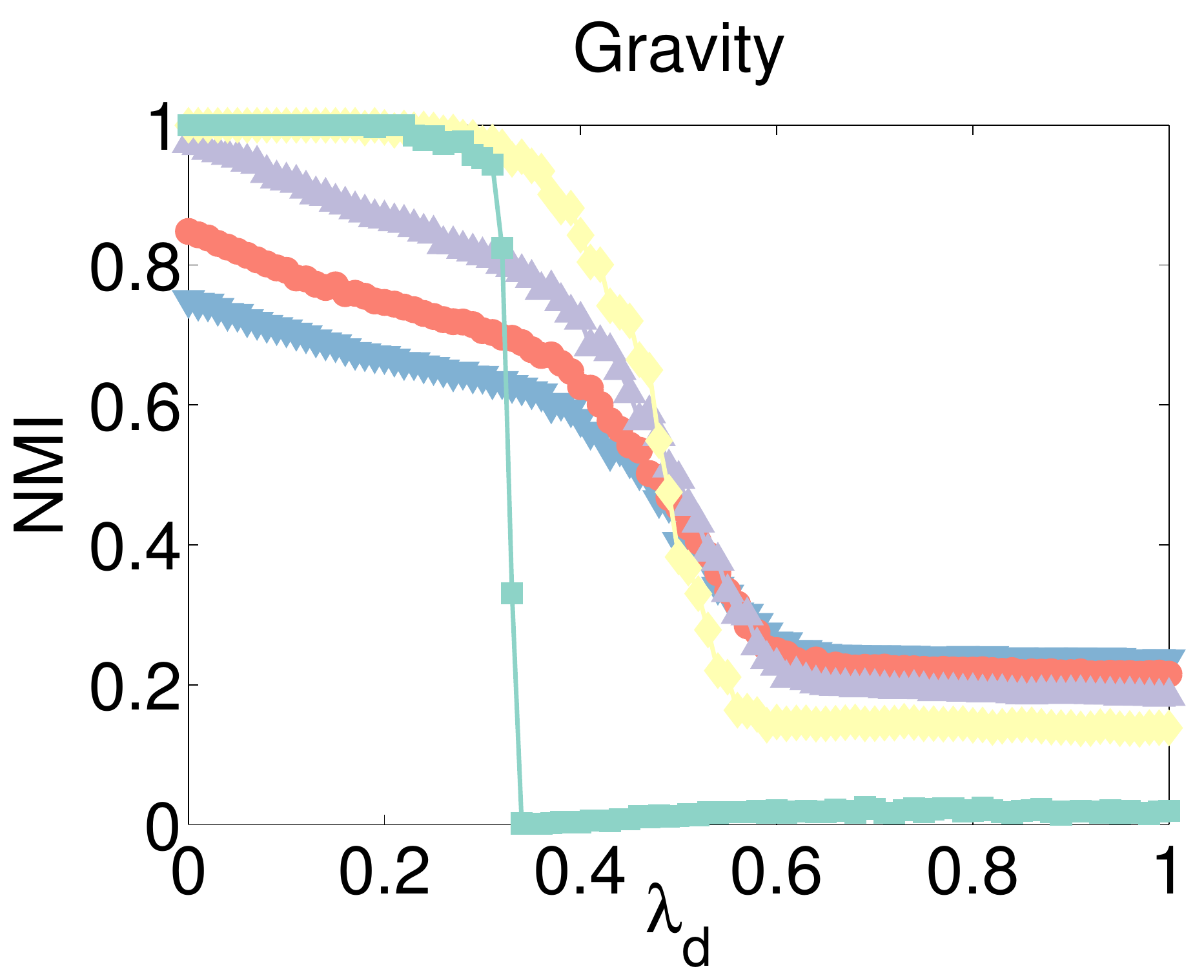} &
\includegraphics[width=\linewidth]{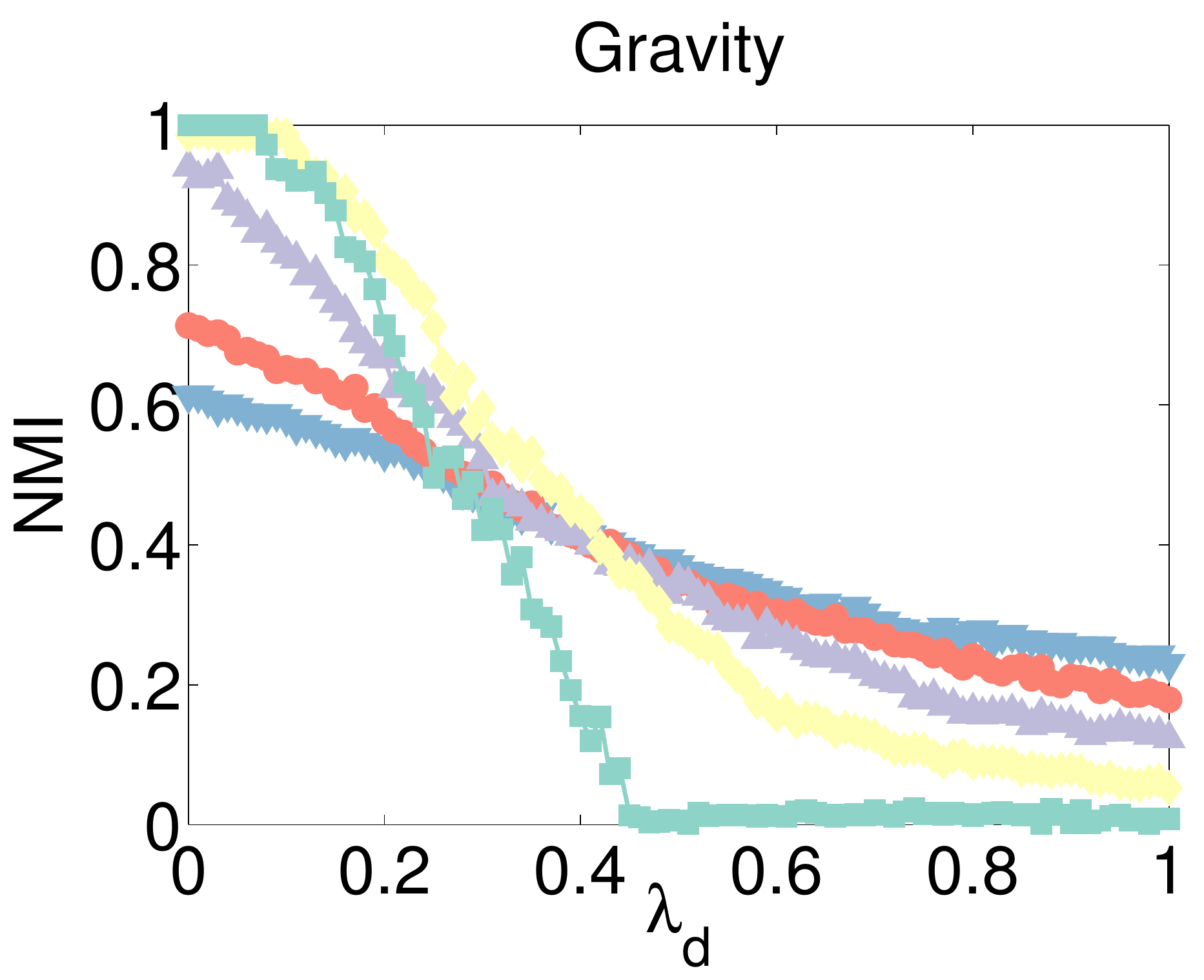}
\\
Rad.& 
\includegraphics[width=\linewidth]{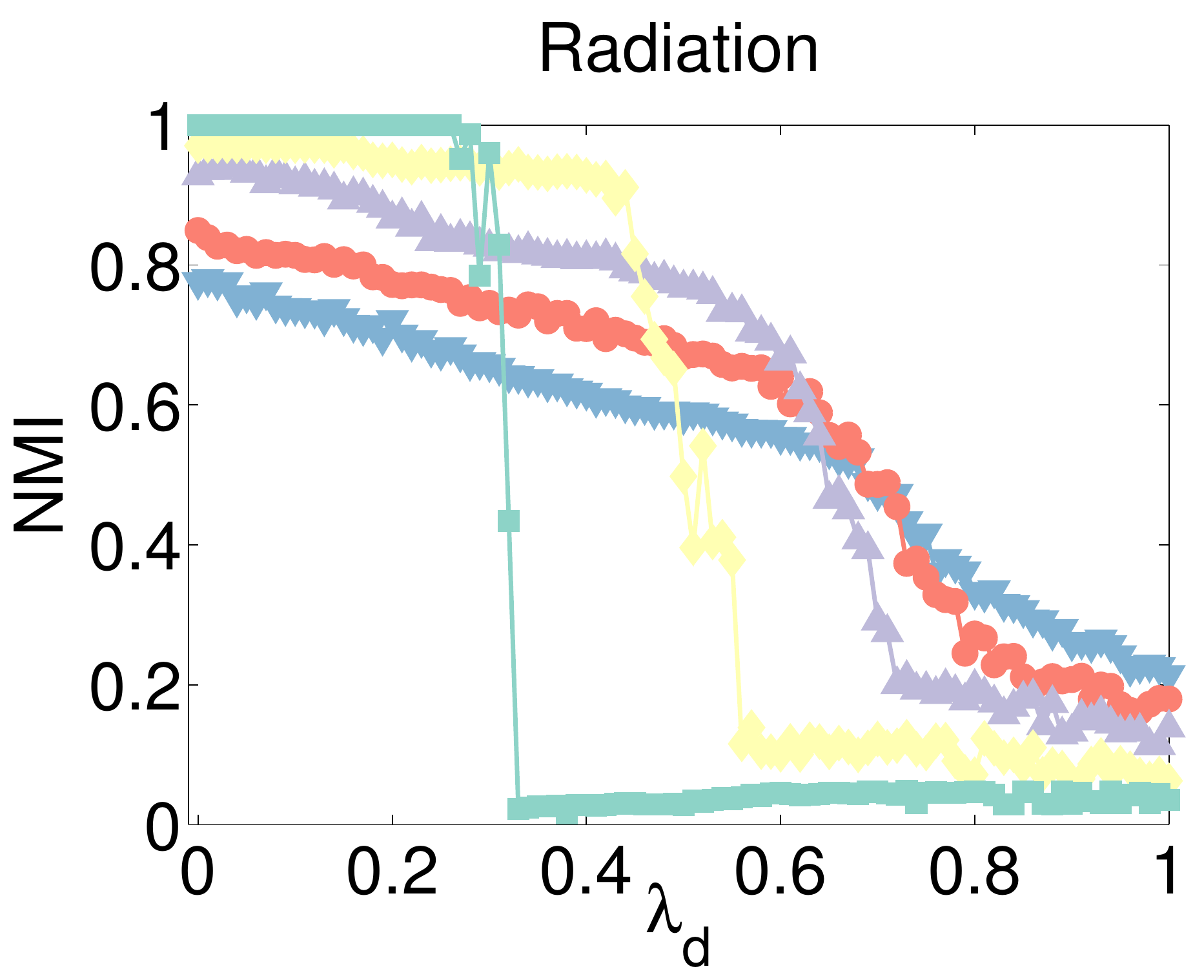} &
\includegraphics[width=\linewidth]{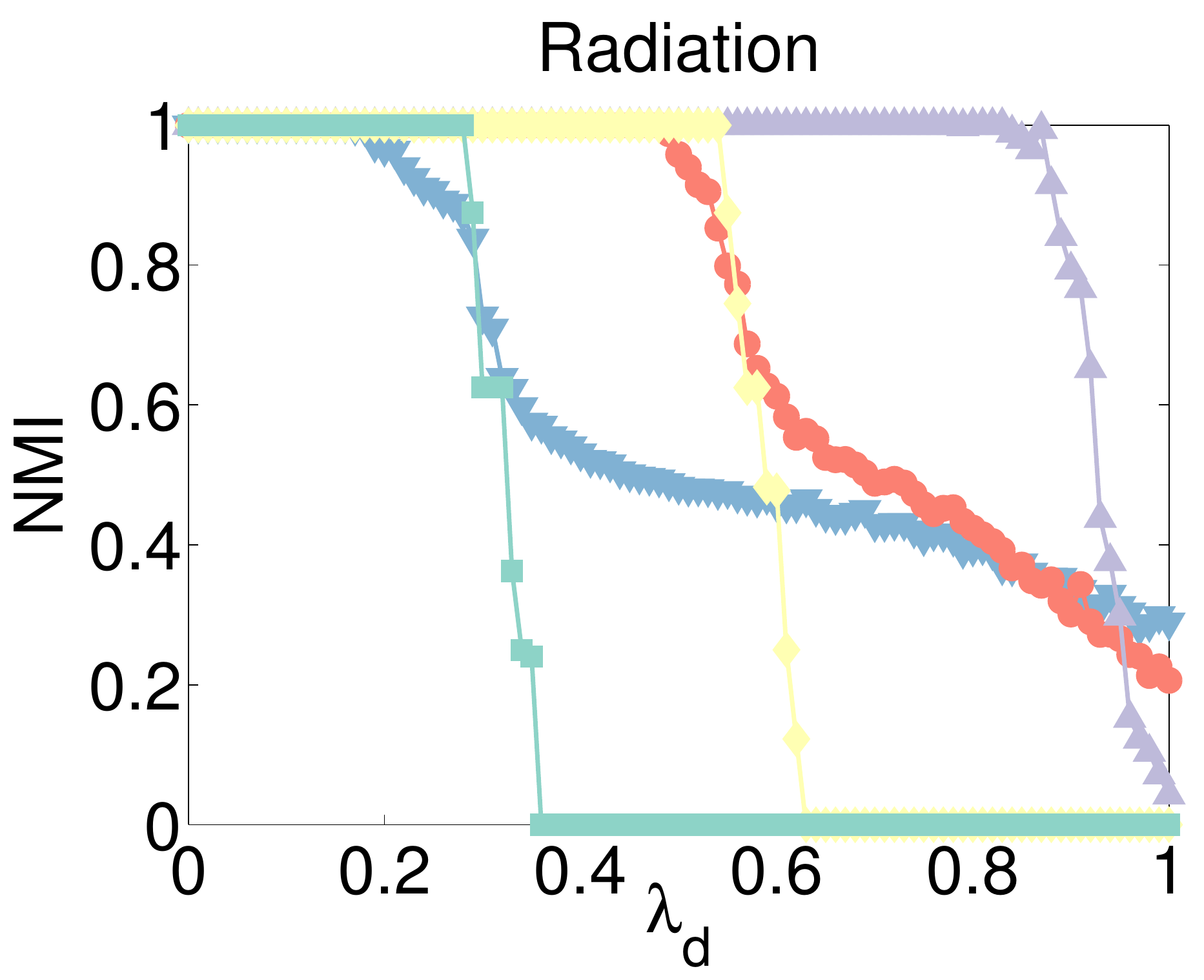} &
\includegraphics[width=\linewidth]{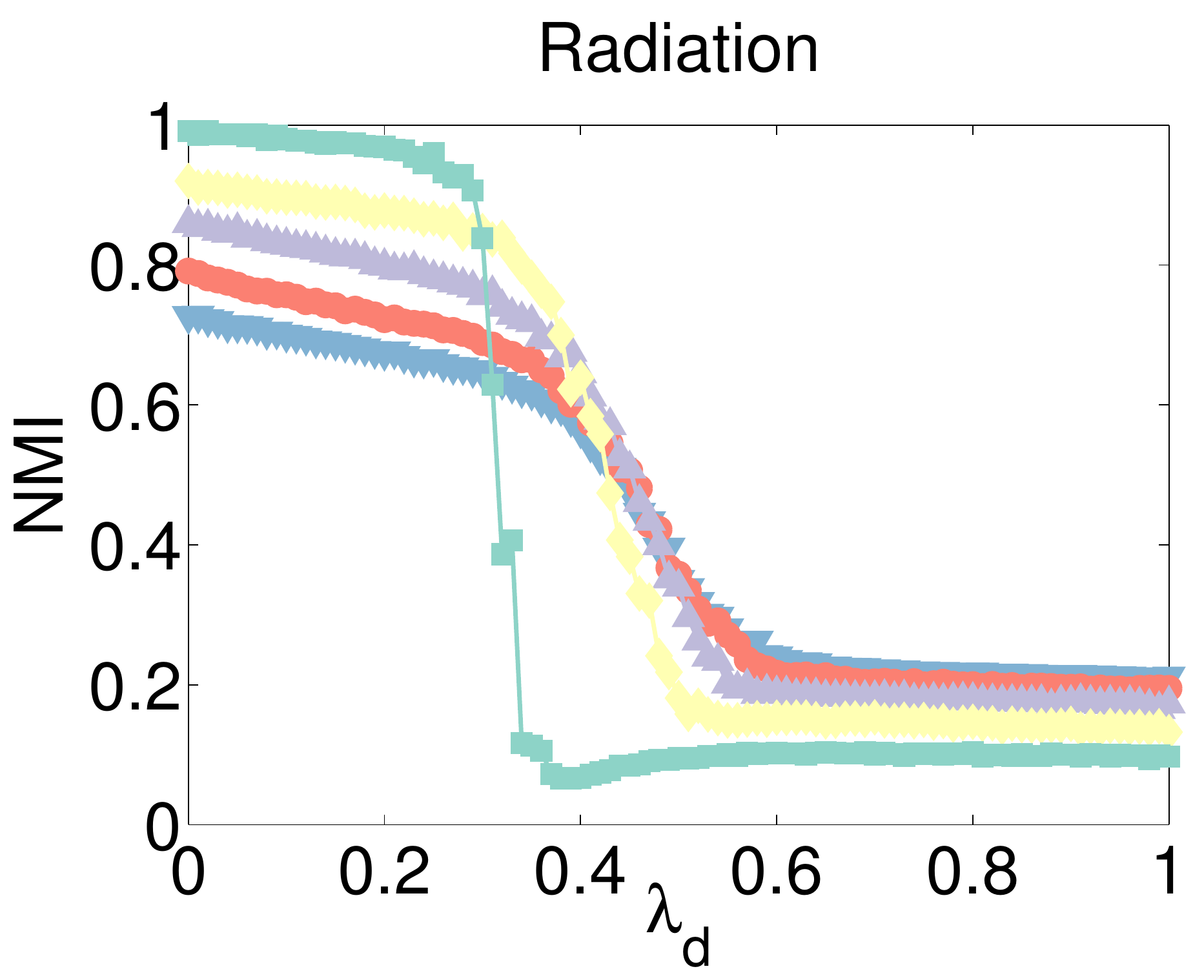} &
\includegraphics[width=\linewidth]{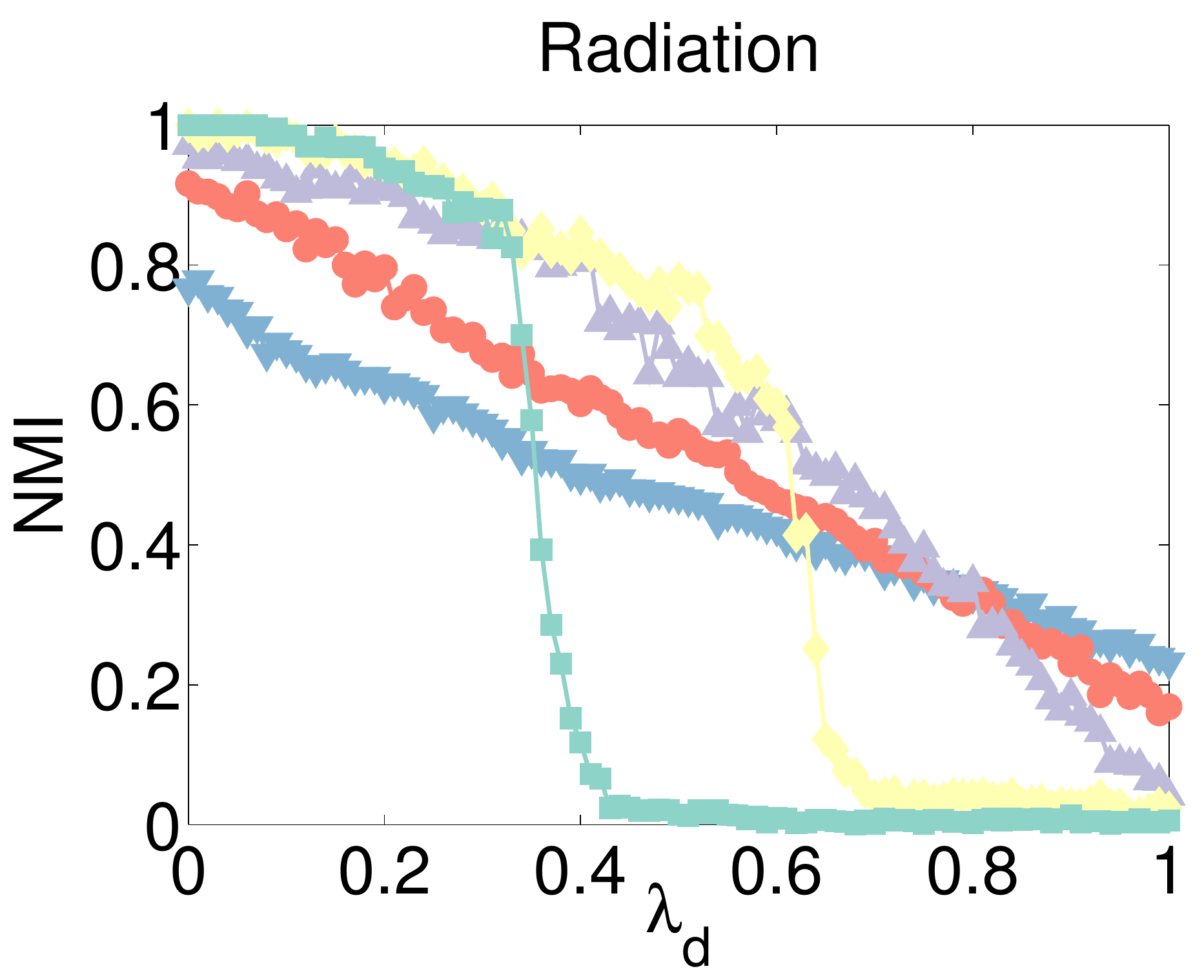}
\\ 
 \hline
\end{tabular}
\caption{Static benchmarks: NMI scores between algorithmically detected and planted community structures in static benchmarks with $l = 10$, $N=50$, $\mu = 100$ and (columns 1, 2) uniform populations of $n_i = 100$ or (columns 3, 4) populations $n_i$ determined uniformly at random from the set $\{0, \ldots, 100\}$ . We plot NMI for different values of the resolution parameter $\gamma$ (colored curves) as a function of inter-community connectivity $\lambda_d \in [0,1]$. We examine both distance benchmarks (in columns 1, 3) and flux benchmarks (in columns 2, 4). We detect communities by optimizing modularity using the (top) NG, (middle) gravity, and (bottom) radiation null models. \label{Figure:Benchmark:static-even-d-NMI-gammavsnm-tab}
}
\end{figure}

We then study the performance of the three null models using several values of the resolution parameter $\gamma \in \{0.5,0.75,1,1.25,1.5\}$ and the inter-community connectivity $\lambda_d \in \{0,0.01,\ldots,0.99,1\}$ on static benchmarks with $N = 50$ nodes and lattice size parameter $l = 10$. Smaller values of $\gamma$ tend to yield larger communities and vice versa. Considering larger $\lambda_d$ increases the level of mixing between the communities and makes community detection more difficult. For simplicity, we fix the density parameter $\mu = 100$. As we discuss in Appendix~\ref{Appendix:mu}, the value of $\mu$ has little effect on the results of community detection when it is above a certain minimum.

For the uniform population distance benchmark, the only factor that influences edge placement is the distance between nodes. On this benchmark, the gravity null model has the best performance, as it is able to find the correct partitions for $\lambda_d \lessapprox 0.82$ (see Fig.~\ref{Figure:Benchmark:static-even-d-NMI-gammavsnm-tab}). The radiation null model has the second best performance and is able to find partially meaningful partitions for $\lambda_d \lessapprox 0.74$, above which we observe a plateau of ``near-singleton'' partitions in which most nodes are placed into singleton communities.  (We use the term ``singleton partition'' to refer to a partition in which every node is assigned to its own community.)  The NG null model, which does not incorporate spatial information, does much worse than either of the spatial null models; it suffers a sharp decline in performance at $\lambda_d \approx 0.4$.  This demonstrates that, although incorporating spatial influence is beneficial for its own sake, we see that using a null model that incorporates population information to study community structure in networks whose structure does not depend on population decreases the performance of community detection. That is, incorporating spatial information is important, but it needs to be done intelligently.

On the uniform population flux benchmark --- in which we include the population density in the region between two nodes in the flux prediction (so the population density influences edge structure) --- the radiation null model outperforms the other null models. The gravity null model comes in second place, and the NG null model is a distant third. 

For the random population distance benchmark, we observe a fast deterioration in quality of the detected communities for $\lambda_d \gtrapprox 0.4$ for all null models, and all null models reach a ``near-singleton'' regime by $\lambda_d \approx 0.6$. The NG null model has the best performance among the three null models for $\lambda_d \lessapprox 0.43$. For $\lambda_d \gtrapprox 0.43$, the gravity null model has the best performance, although the partitions consist largely of singletons for $\lambda_d \gtrapprox 0.6$.

For the random population flux benchmark, the radiation null model has the best performance of the three null models. It has the slowest decrease in NMI scores with the increase in $\lambda_d$. The gravity null model has the second-best performance, and NG fails even when there is no mixing between the two communities (see Fig.~\ref{Figure:Benchmark:static-even-d-NMI-gammavsnm-tab}). However, even the best performance is much worse on random population benchmarks than it is on the uniform population benchmarks. Note additionally that including population information into the edge placement probability by taking $p^{\mathrm{distpop}}_{ij} = \frac{p_i p_j \lambda(c_i,c_j)}{Z_1 d_{ij}}$ (``distance and population benchmark'') brings back the advantage for the gravity null model (see Appendix \ref{app:benchmark-distpop}).

Among the parameter values that we consider ($\gamma \in \{0.5,0.75,1,1.25,1.5\}$), $\gamma = 1$ appears to give the best results (i.e., the largest NMI scores). In the near-singleton regime, $\gamma = 1.5$ outperforms it slightly (see Fig.~\ref{Figure:Benchmark:static-even-d-NMI-gammavsnm-tab}), however this partition is vastly different from the planted partition.


\subsection{Results on Multilayer Benchmarks}

\begin{figure}[tbp]
	\centering
	\includegraphics[width=0.45\linewidth]{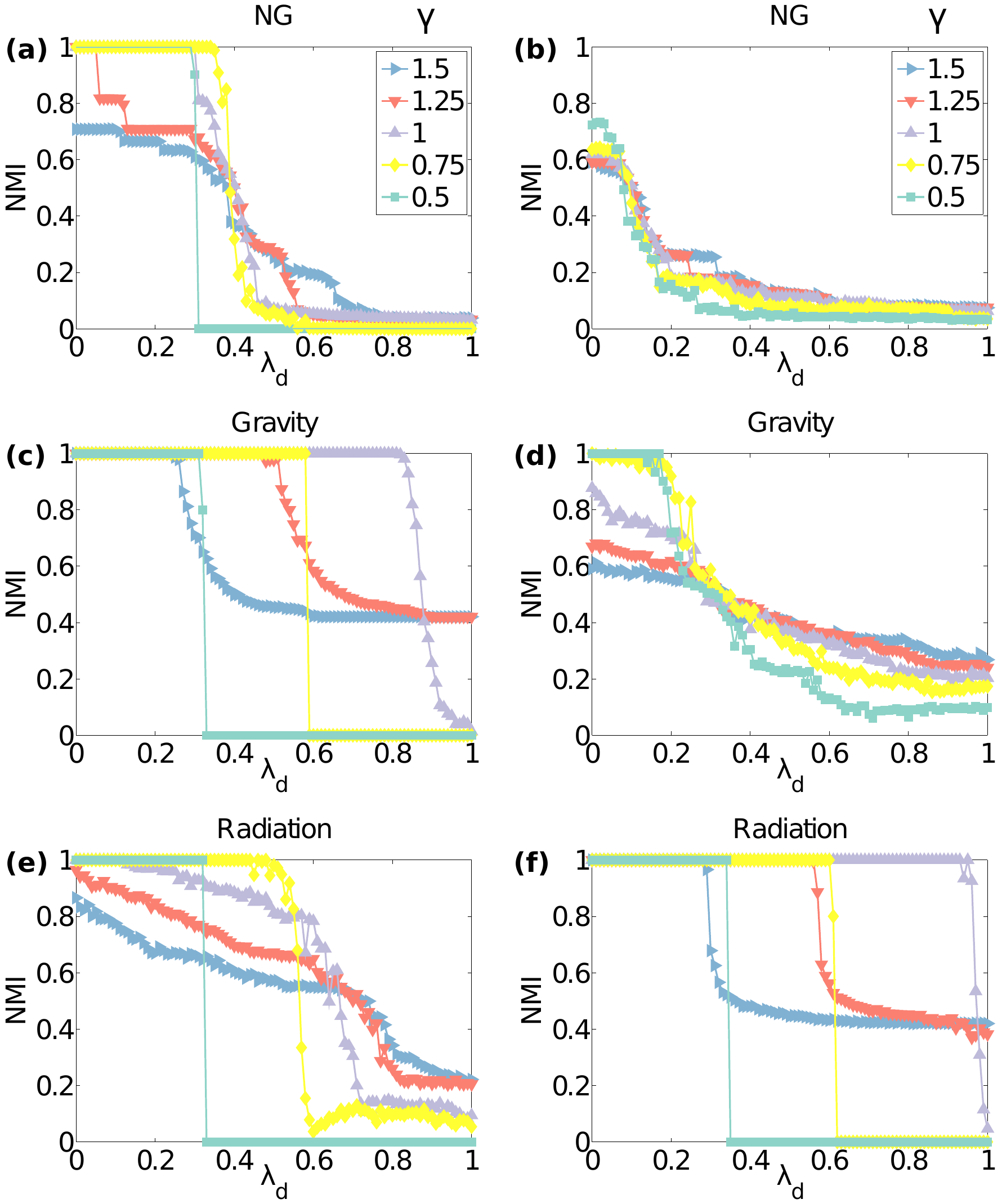}
\caption{NMI between algorithmically detected and planted community structures in uniform population ($n_i=100$ for all $i$) multilayer temporally stable spatial benchmarks with $N = 50$, $l = 10$, $m=10$, and $\mu = 100$ for $\omega=0.1$ and various values of $\gamma$ (colored curves) as a function of $\lambda_d$ for (left) the distance benchmark and (right) the flux benchmark. We detect communities by optimizing modularity using the (top) NG, (middle) gravity, and (bottom) radiation null models. 
 \label{Figure:Benchmark:multi-gammavsnm}}
\end{figure}

We now study the influence of the resolution parameters $\gamma$ and $\omega$ on the community quality of multilayer benchmarks. 
We first study the performance of the NG, gravity, and radiation null models on temporally stable uniform population benchmarks (see Fig.~\ref{Figure:Benchmark:multi-gammavsnm}) with parameter values $N = 50$, $l = 10$, and $m = 10$ layers using $\gamma \in \{0.5,0.75,1,1.25,1.5\}$ and $\omega \in \{10^{-3},0.1,0.25,0.5,0.75,1\}$. We expect that for larger $\omega$ values the weight of the interlayer edges outweighs the intralayer edges, leading to each node being assigned to the same community as its copies in other layers. However, for the temporally stable benchmarks we did not observe this effect; here, we only show figures for $\omega=0.1$, as different values of $\omega$ give very little difference in results (in some plots nearly unnoticeable). 

We also experimented with ``random population'' benchmarks (see Appendix \ref{Appendix:multi-random}) and smaller and larger values of $\omega$. Our results on multilayer benchmarks follow our findings from static benchmarks. Once again, we find that the choice of $\gamma$ has a large influence on the quality of the algorithmic partitions, and (as with our findings for static benchmarks) $\gamma = 1$ seems to yield the best performance (i.e., the highest NMI scores) in most cases, except the near-singleton regime, where $\gamma = 1.5$ outperforms it slightly.

\begin{figure}[tbp]
	\centering
	\includegraphics[width=0.45\linewidth]{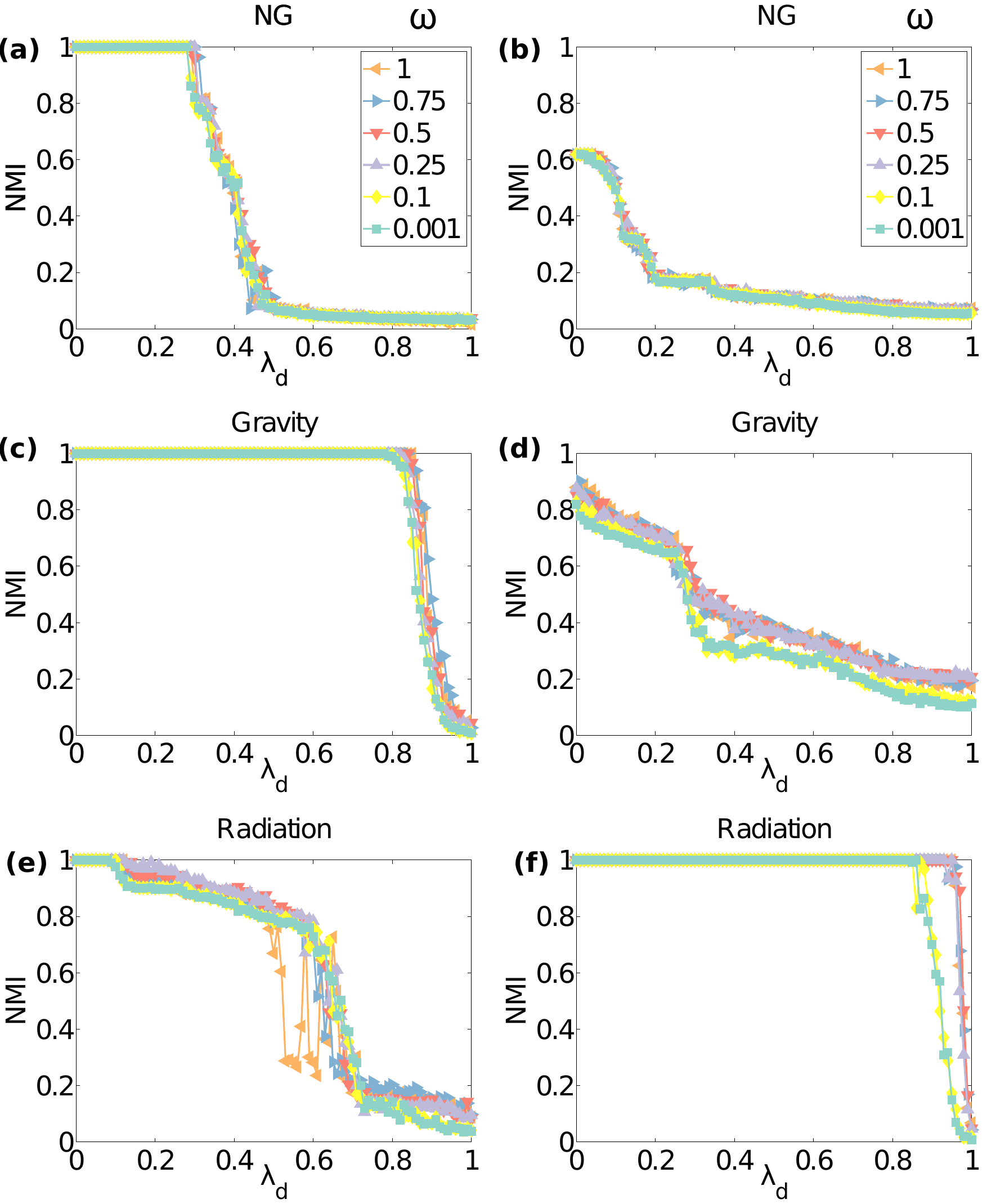}
\caption{NMI between algorithmically detected and planted community structures in uniform population ($n_i=100$ for all $i$) multilayer temporally stable spatial benchmarks with $N = 50$, $l = 10$, $m=10$, and $\mu = 100$ for $\gamma=1$ and different values of interlayer edge weights $\omega$ (colored curves) as a function of $\lambda_d$ for (left) the distance benchmark and (right) the flux benchmark. We detect communities by optimizing modularity using the (top) NG, (middle) gravity, and (bottom) radiation null models. 
\label{Figure:Benchmark:multi-omegavsnm}}
\end{figure}

We now examine the NMI between algorithmic versus planted partitions on temporally stable multilayer benchmarks while varying $\omega$ and $\lambda_d$ for fixed $\gamma = 1$. As we show in Fig.~\ref{Figure:Benchmark:multi-omegavsnm}, we find that the value of $\omega$ usually has little effect on our ability to detect the planted communities via modularity maximization on benchmarks with a temporally stable community structure.  This suggests that the small interlayer variation due to the independent creation of layers is not enough to observe the influence of $\omega$ on community detection.

\begin{figure}[tbp]
	\centering
	\includegraphics[width=0.45\linewidth]{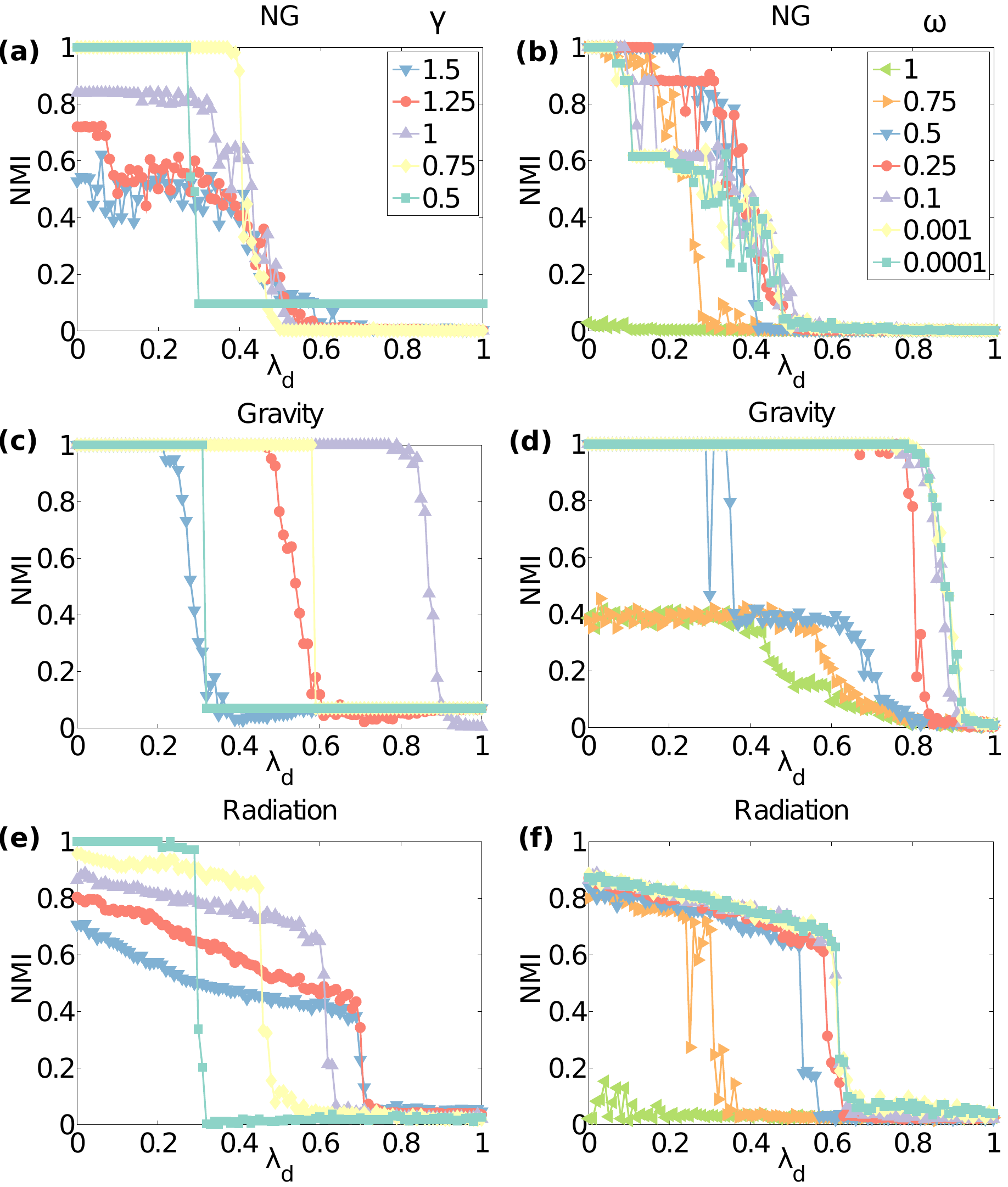}
\caption{NMI between algorithmically detected and planted community structures in uniform population ($n_i=100$ for all i) multilayer temporally evolving spatial distance benchmarks with $N = 50$, $l = 10$, $m=10$, and $\mu = 100$ for (left) $\omega=0.1$ and different values of the resolution parameter $\gamma$ (colored curves) and (right) $\gamma=1$ and different values of the interlayer weights $omega$ (colored curves) as a function of $\lambda_d$. We detect communities by optimizing modularity using the (top) NG, (middle) gravity, and (bottom) radiation null models. 
 \label{Figure:Benchmark:multi-t-gammavsnm}}
\end{figure}

We then study the performance of the three null models on temporally evolving uniform population benchmarks (see Fig.~\ref{Figure:Benchmark:multi-t-gammavsnm}) with parameter values of $N = 50$ nodes, a lattice parameter of $l = 10$, a fraction $p=0.4$ of nodes that change community over the whole timeline, and $m = 10$ layers. 
We show results for $\gamma \in \{0.5,0.75,1,1.25,1.5\}$ for $\omega=0.1$ and for $\omega \in \{10^{-3},0.1,0.25,0.5,0.75,1\}$ for $\gamma=1$.  Compare Fig.~\ref{Figure:Benchmark:multi-t-gammavsnm} to the left panels of Figs.~\ref{Figure:Benchmark:multi-gammavsnm} and~\ref{Figure:Benchmark:multi-omegavsnm}. On temporally evolving benchmarks varying $\omega$ makes a difference, where the structures for $\omega \lessapprox 0.1$  for the gravity null model and $\omega \lessapprox 0.5$ for the radiation null model are the most similar to the planted partitions. 
This is in accordance with our expectation that algorithmically detected community structure becomes overly biased towards connecting copies of nodes across layers above a critical $\omega$ value (which depends on network structure). 

We also perform a ``province-level'' community detection on the multilayer benchmarks in which we seek assignments of nodes (regardless of what layer they are in) to communities and compare the results to benchmark networks with planted community structure.  This is analogous to trying to detect community structure in disease data that persists over time --- e.g., to seek the influence of climate on disease patterns.  This is easiest to apply to temporally stable multilayer networks. We successfully detect the underlying communities, and we obtain similar performance results as with the multilayer communities that we discussed above (see the discussion in Appendix \ref{Appendix:bench-regions}). 

Our results on synthetic benchmark networks suggest that using a spatial null model on a spatial network does not necessarily assure a better result for community detection. The quality of results with different null models depends strongly on the data and the choice of parameter values. For example, incorporating population information into a null model in a situation in which the population is not influencing connectivity structure might cause community detection to yield spurious communities (as we discussed in the context of random population benchmarks). 

The level of influence of different node properties or events (such as disease flux on edge placement) and the extent of mixing between communities is often unknown for networks that are constructed from real data. For such networks, we recommend to try both spatial and non-spatial null models over a wide parameter range and to study the results carefully in light of any other known information about the network. In Section~\ref{Section:dengue}, we will present an example of using such a procedure to study the community structure of correlation networks that are created from time series of dengue fever cases.


\section{Application to Disease Data}\label{Section:dengue}

In this section, we assess the performance of the NG, gravity, radiation, and correlation\footnote{We discuss the \emph{correlation null model}, which was recently introduced in \cite{MacMahon2013arXiv}, in Section \ref{corr-null}.} null models on multilayer correlation networks that we construct from disease incidence data that describe the spatiotemporal spread of dengue fever (a tropical, mosquito-borne viral infection) in Peru from 1994 to 2008. 

Disease dynamics are strongly influenced by space, as the distance between regions affects the migration of both humans and mosquitos ~\cite{Jones2008}. Disease dynamics are also affected by climate due to the temperature dependence of the mosquito life cycle~\cite{Chowell2008}, and different regions of Peru have substantially different climates. Therefore it is important to examine and evaluate the performance of different spatial null models when examining communities in networks that are constructed from disease data.


\subsection{The Disease and the Data}\label{data}

Dengue is a human viral infection that is prevalent in most tropical countries and is carried primarily by the \textit{Aedes aegypti} mosquito~\cite{Guzman2010}. The dengue virus has four strains (DENV-1--DENV-4). Infection with one strain is usually mild or asymptomatic, and it gives immunity to that strain, but subsequent infection with another strain is usually associated with more severe disease~\cite{Guzman2010}.  

Although dengue was considered to be nearing extinction in the 1970s, increased human mobility and mosquito abundance have led to its resurgence in many countries --- often as recurrent epidemics with an increasing number of cases and severity of disease. Dengue is a rising threat in tropical and subtropical climates due to the introduction of new virus strains into many countries and to the rise in mosquito prevalence since the cancellation of mosquito eradication programs~\cite{Guzman2003}. It is currently the most prevalent vector-borne disease in the Americas~\cite{Guzman2003,Chowell2011}. 

Peru is located on the Pacific coast of South America. Its population of about 29 million people is distributed heterogeneously throughout the country. The majority live in the western coastal plain, and there are much smaller population densities in the Andes mountains in the center and the Amazon jungle in the east. The climate varies from dry along the coast to tropical in the Amazon and cold in the Andes. Such heterogeneities influence dengue transmission~\cite{johansson2009local}. For example, temperature ~\cite{jetten1997mosquitosdengue} and rain ~\cite{li1985rainfall} affect the life cycle of the main dengue vector \textit{Ae. aegypti}, and temperature affects its role in disease transmission~\cite{Depradine2004,Keating2001,Chowell2006}. The jungle forms a reservoir of endemic disease; from there, the disease occasionally spreads across the country in an epidemic~\cite{Chowell2008}. Additionally, as \textit{Ae. aegypti} typically only travels short distances~\cite{harrington2005dispersal}, human mobility can contribute significantly to the heterogeneous transmission patterns of dengue at all spatial scales~\cite{Stoddard2009}.

Our dengue data set consists of 15 years of weekly measurements of the number of disease cases across 79 provinces of Peru collected by the Peruvian Ministry of Health~\cite{PeruStats} between 1994 and 2008. These data have previously been analyzed by Chowell et al. to study the relationship between the basic reproductive number, disease attack rate, and climate and populations of provinces~\cite{Chowell2008}. 

Until 1995, the DENV-1 strain dominated Peru; it mostly caused rare and isolated outbreaks~\cite{Chowell2011}. The DENV-2 strain was first observed in 1995--1996, when it caused an isolated large epidemic~\cite{Kochel2002}. DENV-3 and 4 entered Peru in 1999 and led to a countrywide epidemic in 2000--2001~\cite{Montoya2003}, and there was subsequent sustained yearly transmission~\cite{Chowell2011}. The data contains a total of 86,631 dengue cases; most of them are in jungle and coastal provinces (47\% and 49\%, respectively), and only 4\% of the cases occur in the mountains. The disease is present in 79 of the 195 provinces across the data set, and never in all 79 provinces at once.

In this paper, we use the definition of ``epidemic'' from the US Agency for International Development (USAID): an \emph{epidemic} occurs when the disease count is two standard deviations above the baseline (i.e., mean) \cite{Lloyd2003}. When stating countrywide epidemics, we apply this definition when considering all nodes.  When stating local epidemics, we apply this definition to individual provinces (though one could also consider particular sets of provinces).


\subsection{Network Construction}\label{sec:netcreate}

\begin{figure}[tbp]
\centering
\includegraphics[width=0.45\linewidth]{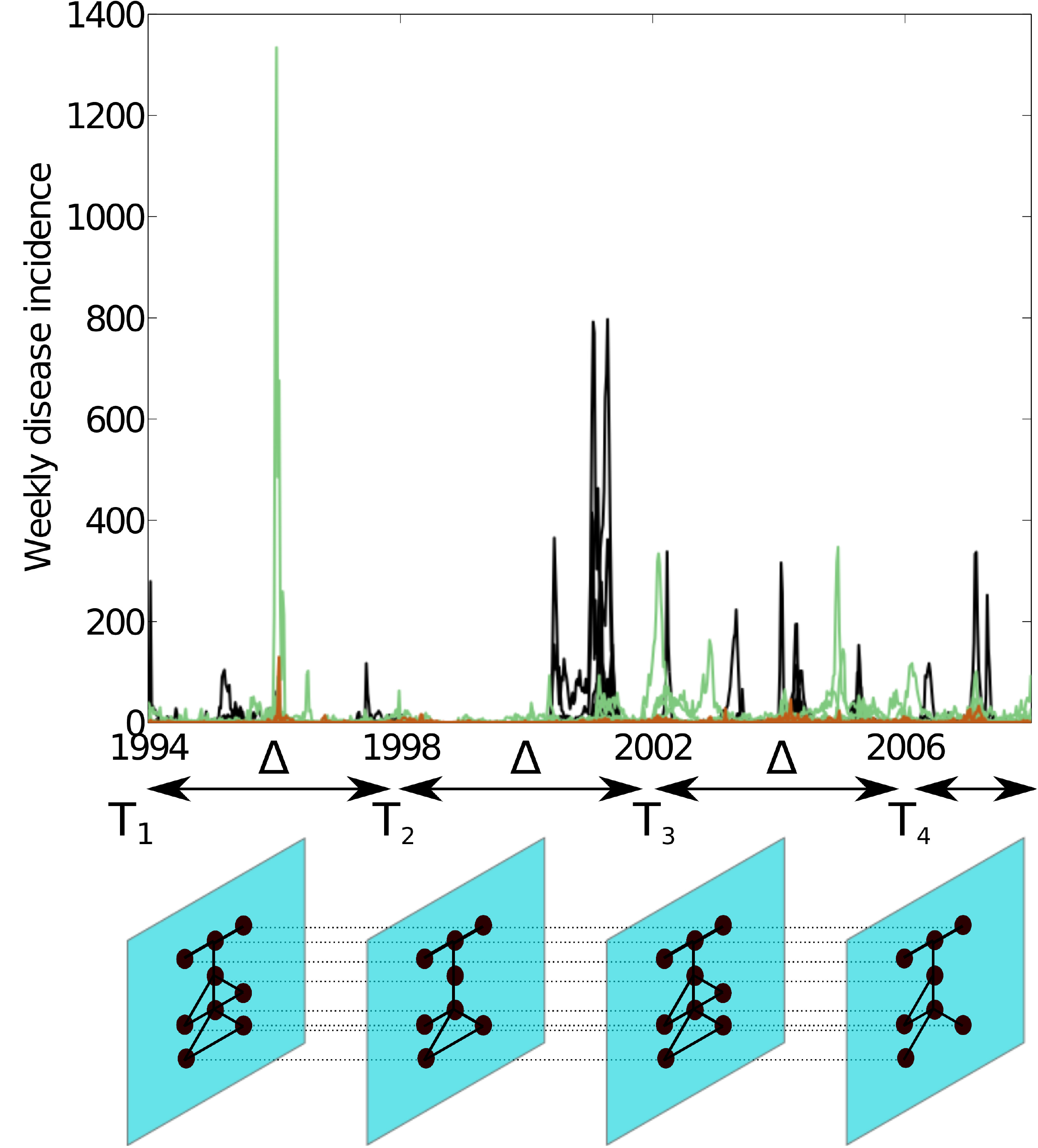}
\caption{Construction of multislice correlation networks from disease time-series data. The top panel shows the dengue fever time series for the 79 provinces of Peru.  We color the provinces by climate: coastal provinces are in black, mountainous provinces are in brown, and jungle provinces are in green. Observe the large epidemics in 1996 (focused in the jungle Utcubamba province) and 2000--2001 (countrywide, but primarily on the northern coast), and the recurrent post-2001 epidemics (which affect various jungle and coastal provinces). The bottom panel shows an example of the multislice network construction for 9 nodes with $\tau = \{1,209,417,625\}$ and $\Delta = 208$. (The time points correspond to 1/1/1994, 27/12/1997, 22/12/2001, and 17/12/2005). The nodes represent provinces and each intralayer edge weight is given by a  Pearson correlation between a pair of single-province time series in a given time window. One set of correlations gives one temporal layer, and we connect copies of each node in neighboring layers using interlayer edges of uniform weight $\omega \in [0,\infty)$ (dashed lines). The case $\omega = 0$ yields a set of static networks.  (All other aspects of our network construction are the same.)
\label{Figure:multislice}}
\end{figure}

Our data set $D$ consists of $N=79$ time series of weekly disease counts $\{D_1,D_2,\ldots,D_N\}$ over $T = 780$ weeks. The quantity $D_i(t)$ denotes the number of disease cases in province $i$ at time $t$. (See Fig.~\ref{Figure:multislice} for a plot of the number of cases versus time.) We create networks from this data by calculating the Pearson correlation coefficient between each pair of time series.~\footnote{Reference~\cite{Smith2011} compared several methods to calculate similarity networks from time-series data.  Our focus in the present paper is on generalizing and evaluating null models, so we use Pearson correlations for simplicity.}

We seek to study the temporal evolution of the correlations by constructing separate networks for different time windows ---  we  either construct a set of static networks or a multislice network. To create these networks, we divide each of the time series into $m$ time windows by explicitly defining a list of the starting points $\tau = \{\tau_1, \tau_2,\ldots,\tau_m\}$ for each time window and the time window width $\Delta = \tau_{t+1} - \tau_t$.  
In the present paper, we use $\tau_1=1$ unless we state otherwise. 

The starting point $\tau_t$ and window width $\Delta$ define a time window that we use to select a portion of the disease time series.  For example, for the time series of disease cases in province $i$, the time-series portion $E_i = \{D_i(\tau_t), D_i(\tau_t+1),\ldots,D_i(\tau_t+\Delta)\}$ represents the numbers of disease cases in province $i$ at times $\tau_t, \tau_t+1,\ldots,\tau_t+\Delta$.  By considering all provinces, one can use such time series either to construct a set of static networks or a multislice network.

For a static network, we define a set of $N$ nodes $\{1,2,\ldots,N\}$, where node $i$ corresponds to province $i$. The edge weight 
\begin{equation}
	W_{ij} = \frac{1}{2} (\rho_{ij} + 1) - \delta_{ij}\,,
\label{eqn:network}
\end{equation}
represents the similarity between the time series $E_i$ and $E_j$; the Kronecker delta $\delta_{ij}$ removes self-edges. The quantity $\rho_{ij}$ is the Pearson correlation coefficient between the disease time series for provinces $i$ and $j$.  That is, 
\begin{equation}
	\rho_{ij} = \frac{\langle E_i E_j \rangle - \langle E_i \rangle \langle E_j \rangle}{\sigma_i \sigma_j}\,,
	\label{equation:Pearson}
\end{equation}	
where $\langle \cdot \rangle$ indicates averaging over the time window under consideration, and $\sigma_i$ is the standard deviation of $E_i$. Our construction yields a fully connected (or almost fully connected) network $W$ with elements $W_{ij} \in [0,1]$. When studying static networks, we use $\tau = \{1,2,\ldots,T-\Delta\}$ to form a set of $T-\Delta$ overlapping static networks.

To construct a multislice network, we use the times $\tau = \left\{1,1+\Delta,1+2\Delta,\ldots,1+\Delta \times \left(\lfloor{\frac{T}{\Delta}\rfloor} -1 \right ) \right\}$ to create nonoverlapping time windows. 
The intralayer edge weights are
\begin{equation}
	W_{ijs} = \frac{1}{2} (\rho_{ijs} + 1) - \delta_{ij}
\label{eqn:networkm}
\end{equation}
for each layer $s$. We connect each node $i$ in the $r$th time window to copies of itself in an adjacent time window $s$ using interlayer edges of uniform weight $C_{isr} = \omega \in [0,\infty)$. This yields a weighted multislice correlation network. The case $\omega = 0$ in the multislice network corresponds to a sequence of static networks. See Fig.~\ref{Figure:multislice} for a schematic that shows the construction of a multislice network.

Similar constructions of (both static and multislice) networks from time series have been employed for systems such as functional brain networks~\cite{Bassett2011,Bassett2013}, currency exchange-rate networks~\cite{Fenn2009}, and political voting networks \cite{Mucha2010,Mucha2010a,Macon2012}.

Many features, such as the number of layers and the mean and variance of the Pearson correlation values, depend on the parameters that we use in constructing our networks. For example, it is important to consider the choice of the time window size $\Delta$. There is a trade-off between having many layers to obtain a good temporal resolution of events and ensuring that we construct each layer using enough time points to be confident of the statistical significance of the similarity values in the adjacency-tensor layer~\cite{Bassett2011}. Larger values of $\Delta$ yield smaller variations in mean correlation across the years and lessen the effects of small, regional epidemics on the number of cases and on the correlation between disease profiles in different provinces. Therefore, we want to use a sufficiently large value of $\Delta$ so that we can examine long-term, repetitive disease patterns. Studies based on random matrix theory (RMT) suggest an additional constraint of $\Delta/N > 1$, because correlation matrices generated from time series that are shorter than the number of time series being analyzed (i.e., than the number of nodes) are indistinguishable from the correlations that one calculates from short, uncorrelated sequences of noise~\cite{RMTconstraint}.
However, choosing a value of $\Delta$ that is too large risks over-smoothing data and losing important information.\footnote{See an analogous discussion of time-window choice in Ref.~\cite{Fenn2012} in the context of financial networks.} Unless we state otherwise, we use $\Delta = 80$ for the (overlapping) static networks and $\Delta = 60$ for multilayer networks (which never have more than 59 nodes in one slice) in order to obtain meaningful correlation matrices while preserving interesting disease patterns. 



\subsection{Community Structure in Disease-Correlation Networks}

It is well-known that geographical distance has an important influence on disease spread~\cite{Stoddard2009,Hay2006,Truscott2012}. Additionally, climate exerts a significant influence on dengue, and it is also necessary to consider Peru's particular topography (as its mountains form a barrier to disease spread)~\cite{Chowell2008, Chowell2011}. Therefore, we expect the community structure in the disease-spread networks to be strongly geographical. We also expect to observe large changes in community structure at certain time points --- such as when the introduction of the new disease strains around 1999 led to large epidemics and the onset of yearly countrywide epidemics ~\cite{Chowell2011}. In this section, we explore the similarity of algorithmically obtained community structures to spatial and temporal groupings of nodes across a range of parameter values. 

To compare the algorithmic partitions of the correlation networks versus manual partitions, we use the z-score of the Rand coefficient~\cite{Traud2008}. The Rand coefficient is 
\begin{equation}
	R  = (w_{11} + w_{00})/M\,, 
\end{equation}
where $w_{11}$ is the number of node pairs that are in the same community in both partitions, $w_{00}$ is the number of node pairs that are in different communities in both partitions, and $M$ is the total number of node pairs. 

We use z-Rand scores instead of NMI because the former measure is good at detecting similarities in coarse structure \cite{Traud2008,traud2012} but is less sensitive to minor changes such as one node changing community assignment. For the disease data, we do not possess ground-truth partitions as we did for our synthetic benchmark examples, so we seek to evaluate broad organizational similarities in the algorithmic and manual partitions rather than attempting to conduct a fine-grained evaluation of community structure versus a planted partition.  We thereby aim to inform our understanding of the general structural influences on the spatiotemporal patterns of disease spread.  One can also examine measures of spatial autocorrelation (e.g., Moran's $I$) \cite{Anselin1995}.

\begin{figure}
 \centering
  \hfill(a)
  \includegraphics[width=0.29\linewidth]{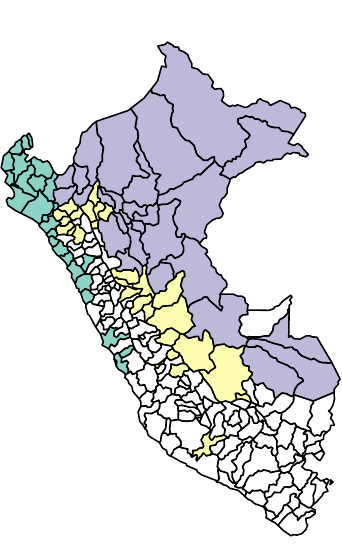}
  \hfill (b)
  \includegraphics[width=0.29\linewidth]{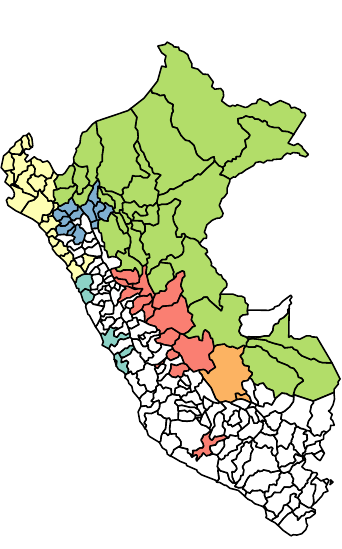}
  \hfill (c)
  \includegraphics[width=0.29\linewidth]{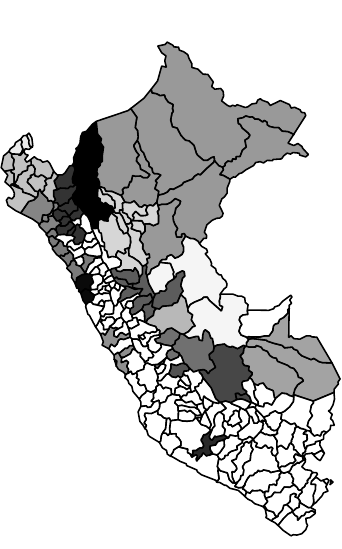}
 \caption{Visualization of the three different topographical partitions of Peru's provinces on a map. (Left) Broad climate partition into coast (yellow), mountains (brown), and jungle (green); (center) the further division of coast and mountains into northern coast, central coast, southern coast, northern mountains, central mountains, and southern mountains; and (right) the administrative partition of Peru.
 \label{Figure:Peru-regions}}
 \end{figure}

To examine the spatial community structures in the static and multilayer networks, we compare the results of the partitions that we obtain algorithmically to manual partitions using z-Rand scores. 
In the ``climate partitions'', we group nodes according to the topography of their associated provinces --- jungle, coastal, and mountainous provinces --- and then subsequently divide the coastal and mountainous communities into northern, central, and southern provinces [see Fig.~\ref{Figure:Peru-regions}(a,b)].  We use the detailed climate partition for the subsequent study. In the 19-community ``administrative partition'', we assign each node to its associated administrative region [see Fig.~\ref{Figure:Peru-regions}(c)]. We compare each of the 700 static networks against these two manual partitions to study the spatial element of the data. For the multilayer networks, we compare the algorithmic partition versus a manual partition by taking the same manual partition of nodes for all layers.

We use the term ``spatial partitions'' to describe partitions that yield high z-Rand scores in comparison to the climate or administrative manual partitions. For multilayer networks, we also compare the algorithmic partitions to partitions that contain a planted temporal change in community structure. For these comparisons, we group the multilayer nodes into ones that occur before or after a ``critical'' time $t_c$, and we use the term ``temporal partitions'' to describe partitions that yield high z-Rand scores in this comparison. Such temporal partitions contain two communities. We test all of the times $\tau = \left\{1,1+\Delta,1+2\Delta,\ldots,1+\Delta \times \left(\lfloor{\frac{T}{\Delta}\rfloor} -1 \right ) \right\}$ that we use to create the multilayer network, and we report the time with the highest z-Rand score as the critical time $t_c$. We also test for pairs of critical times (yielding a partition into three communities) by examining all possible pairs of critical times in the same manner.

	  
\subsubsection{Community Structure Using the NG Null Model}
\label{Section:dengue-NG}

Before looking at multilayer networks, we first study the community structures of the $700$ overlapping static networks formed by taking $\tau = \{1,2,\ldots,700\}$ and using $\Delta = 80$. We select the networks for which the algorithmic partitions score the highest against manual spatial partitions of the network for further study.

\begin{figure*}[tbp]
	\centering
	\hfill (a) \includegraphics[height=4.7cm]{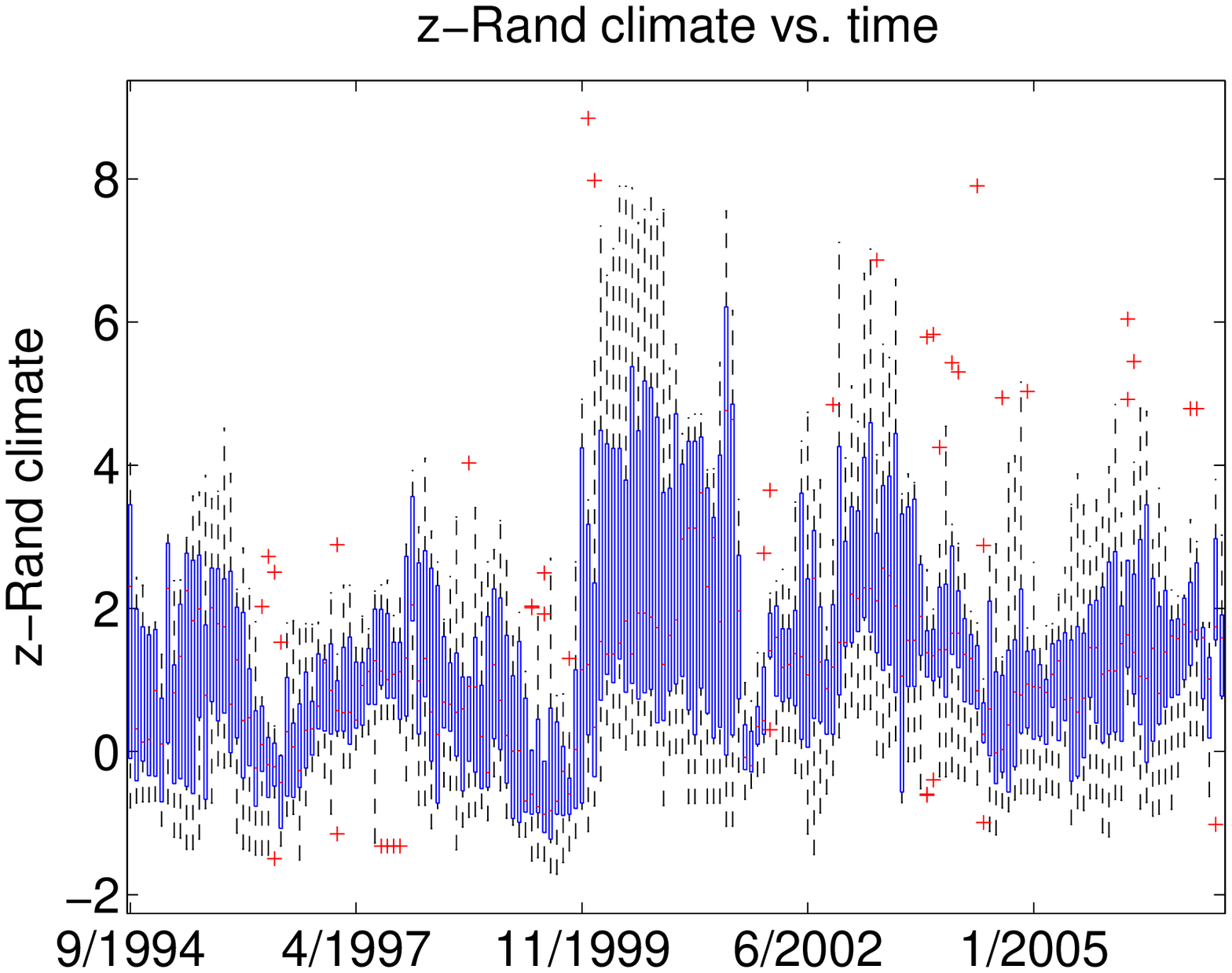}
	\hfill (b) \includegraphics[height=4.7cm]{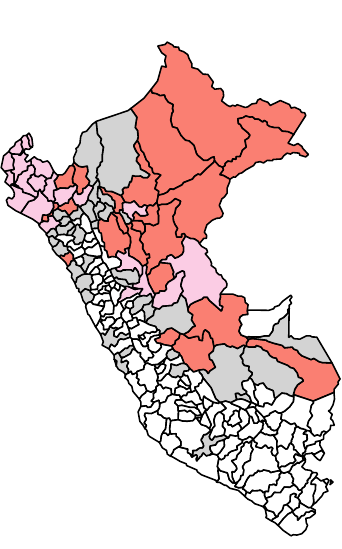}
	\hfill (c) \includegraphics[height=4.7cm]{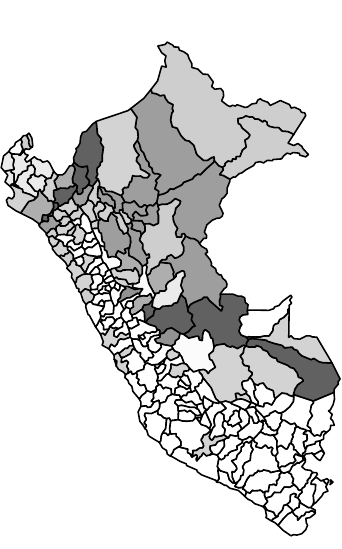}
	\caption{Properties of algorithmic community structure, which we obtained by maximizing modularity using the NG null model for the dengue fever static correlation networks with window size $\Delta=80$. (a) A box plot of the z-Rand scores versus the detailed climate partition at different $\gamma$ values ($\gamma \in {0.1,0.2,\ldots,3}$), for the 700 static networks covering the whole time period (horizontal axis), 
	(b) community structure with the highest z-Rand score when compared to the climate partition.
The resolution-parameter value is $\gamma = 1$, the layer is 293 (which occurs in December 1999), the z-Rand score is 8.85,
(c) community structure with the highest z-Rand score when compared to the administrative partition.
The resolution-parameter value is $\gamma = 1.2$, the layer is 453 (which occurs in October 2002), the z-Rand score is 8.76, and we show the largest community in brown.
Our visualization in panels (b,c) uses a map of Peru in which we color provinces according to their community assignment. White provinces are ones in which our data does not include any reported cases of dengue fever in the indicated time window. 
	\label{Figure:NG-static-boxes}}
\end{figure*}

The community structures that we obtain from maximizing modularity have a strong spatial element, as suggested by the high z-Rand scores when compared to topographical partitions.  As one can see in Fig.~\ref{Figure:NG-static-boxes}(a), which shows a compact box plot of the z-Rand scores versus climate partitions for resolution parameter values of $\gamma \in {0.1,0.2,\ldots,3}$ (each box) across the 700 networks covering the data set (the horizontal axis), the spatial element is especially evident after the year 2000. 

As one can see from a plot of number of epidemic cases over time (see Fig.~\ref{Figure:multislice}), this transition seems to occur around the time of the largest countrywide epidemic in the data, and the subsequent period includes recurring yearly epidemics that have been linked to climatic patterns in prior studies~\cite{Chowell2011}. There are two periods of significantly spatial partitions: one corresponds to the 2000--2001 epidemic and it contains the spatial partition with the highest z-Rand score against climate [see Fig.~\ref{Figure:NG-static-boxes}(b)], and the second occurs in 2002--2004 and it contains the spatial partition with the highest z-Rand score against administrative partition [see Fig.~\ref{Figure:NG-static-boxes}(c)]. Note that the topographical z-Rand scores decrease after 2004 despite the continuing yearly dengue epidemics.

In Figs.~\ref{Figure:NG-static-boxes}(b,c), we plot the partitions that have the highest z-scores with respect to the manual climate and administrative partitions on a map of Peru. We observe that the high-scoring climate partition consists of one community that is dominated by the jungle (red community) and another community that is dominated by the coast (pink community), whereas the high-scoring administrative partition is composed of 7 smaller communities. 
The jungle nodes form the largest communities in both of these spatial partitions, and it is these communities that contribute the most to the high spatial scores. There was a dengue epidemic in most of the provinces in these large communities during the time periods covered by the relevant networks. It is possible that their proximity drove the large amount of synchrony in the epidemic spread in these provinces.  

\begin{figure*}[h!tbp]
  \centering
 \hfill (a) \includegraphics[height=3cm]{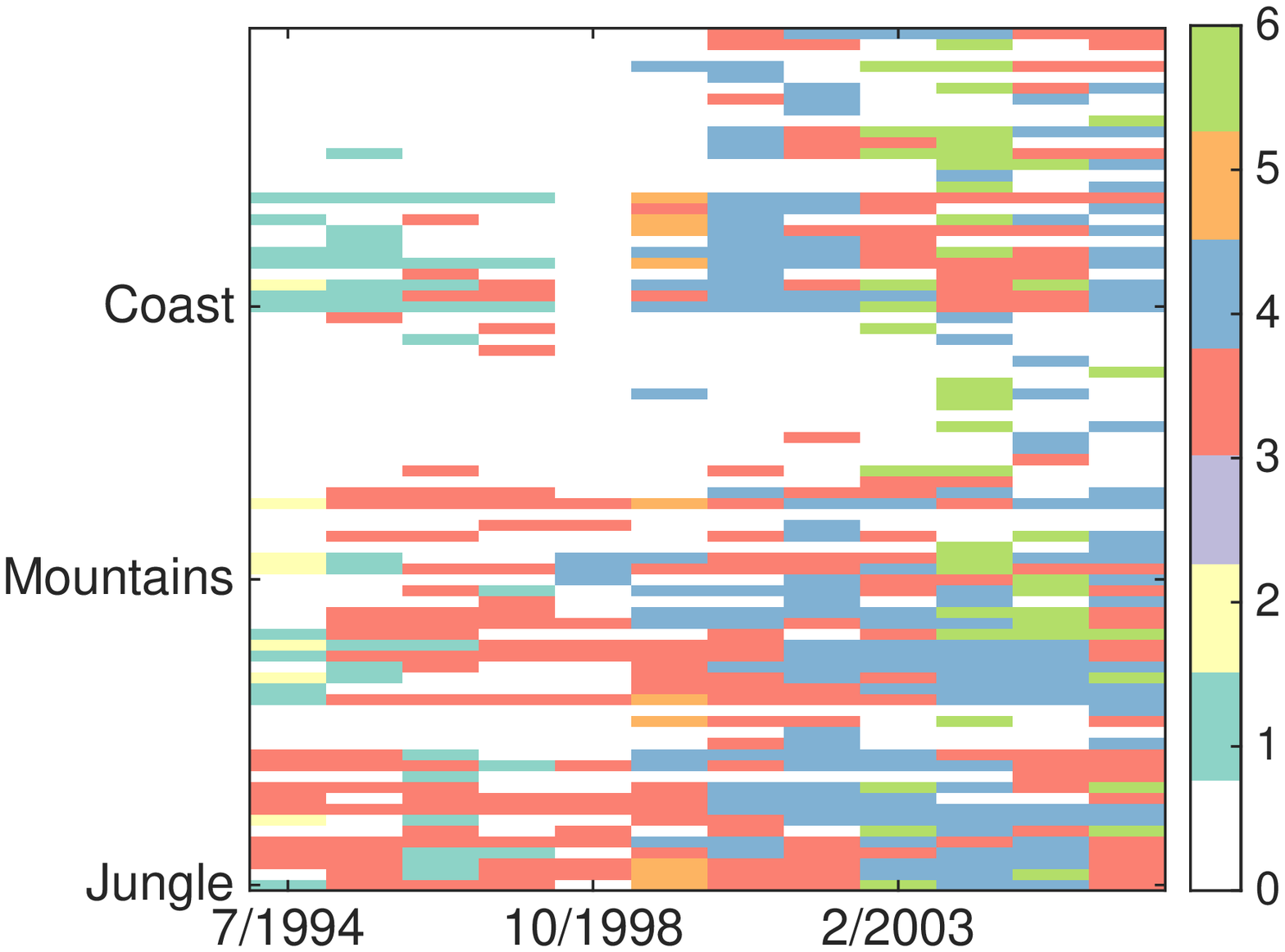}
\hfill (b) \includegraphics[height=3cm]{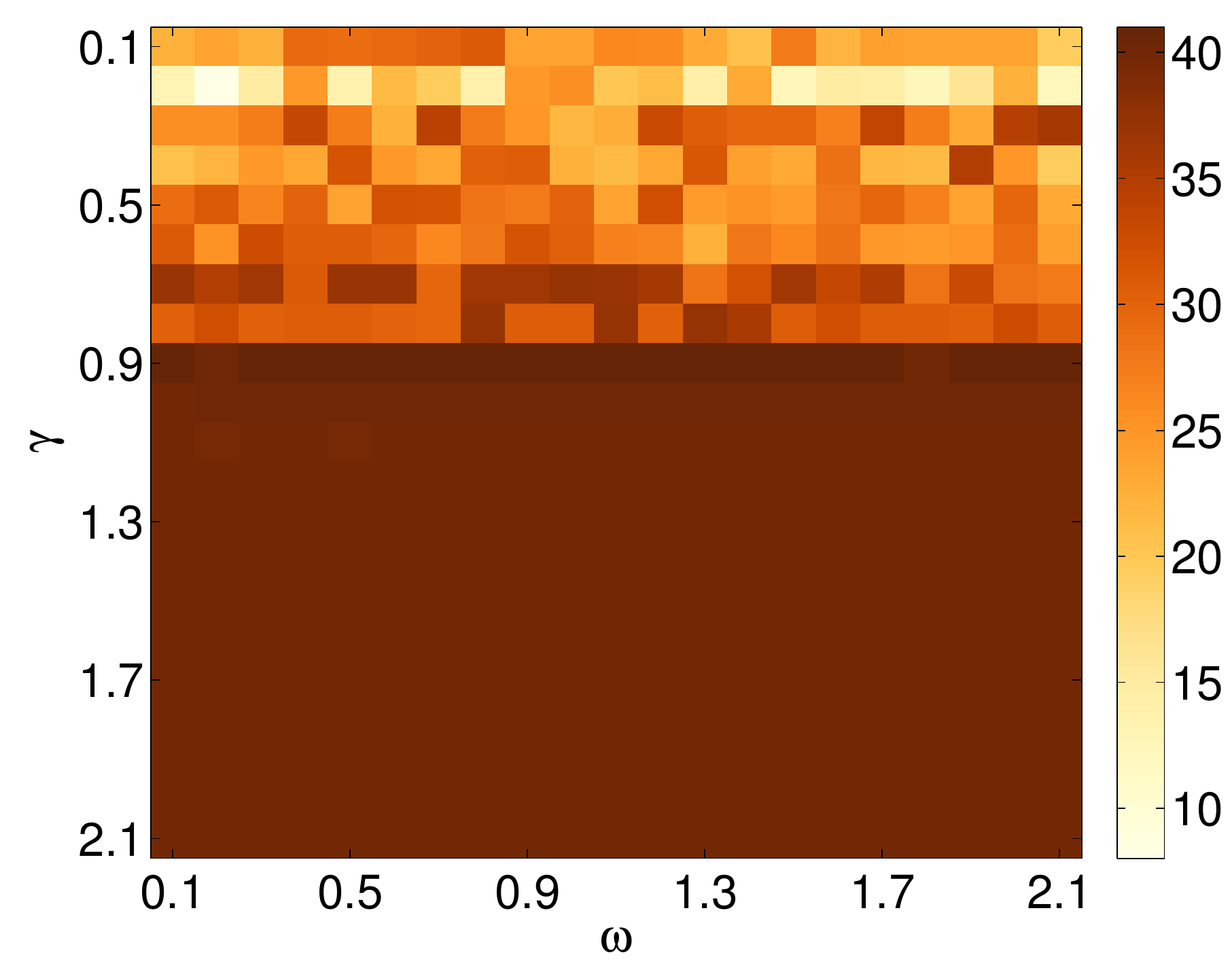}
\hfill(c) \includegraphics[height=3cm]{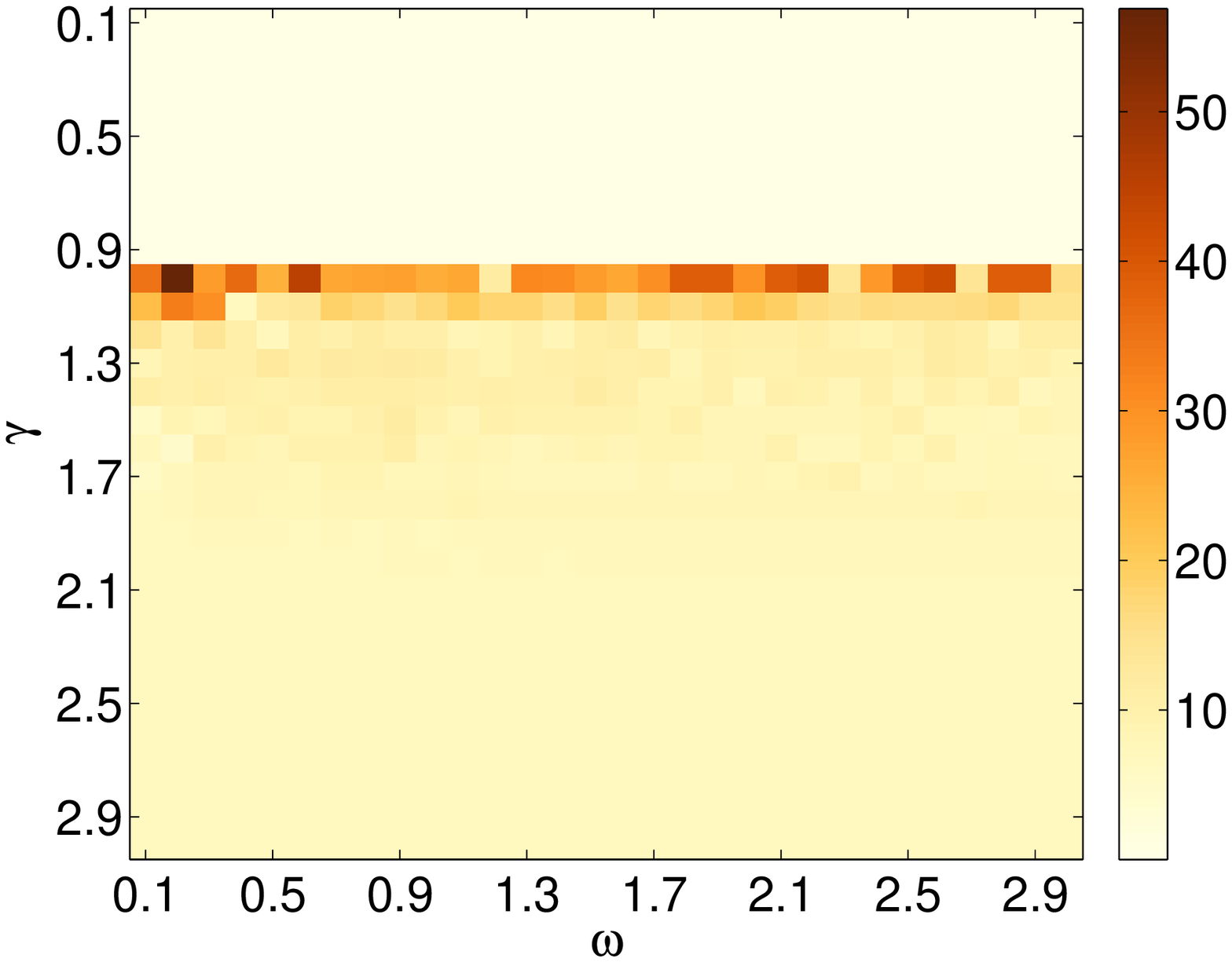}
\caption{Algorithmic partitions, which we obtain by maximizing modularity using the NG null model, of the dengue fever multilayer disease-correlation network that we construct using $\Delta=60$. (a) An example of a consensus community structure that we obtain for a resolution-parameter value $\gamma=1$ and $\omega=0.1$ across 100 repeats of community detection. Mid-layer timepoints are plotted on the horizontal axis, and nodes are on the vertical axis. Node community membership is indicated by color.
We observe several times when communities die and new ones are born. (b,c) Results of varying the parameters $\gamma$ and $\omega$.  We show the z-Rand scores for similarity to (b) ``spatial'' partitions by climate and (c) temporal partitions before and after a critical time $t_c$. (For this figure and for each set of parameter values, we select the highest scoring $t_c$; in the majority of cases, $t_c$ corresponds to January 2002.) 
\label{Figure:NGog}}
\end{figure*}

We now consider community structure in the multilayer disease network with nonoverlapping layers that we construct using the time points $\tau = \{1,61,\ldots,721\}$ and the time window width $\Delta = 60$.  
To find interesting parameter regimes, we compare the algorithmically computed community structure of the dengue fever multilayer disease-correlation network to manual partitions across a range of $\omega$ and $\gamma$ parameter values between $0$ and $3$ (see Fig.~\ref{Figure:NGog}). For $\gamma \lessapprox 1$, all nodes are in one community. For $\gamma \in [1, 1.2] $ and $\omega \lessapprox 1$, the algorithmically detected partitions have a relatively high z-Rand score when compared to the temporal partition [see Fig.~\ref{Figure:NGog}(c)]. 
 When looked at in detail, the partitions exhibit a mixture of spatial and temporal features. [See Fig.~\ref{Figure:NGog}(a) for an example.] 


When studying the qualitative features of the partitions for $\gamma \in [1,1.2]$ (where the endpoints of this interval are approximate) and $\omega \lessapprox 1$, we observe that community detection repeatedly finds 2001 as the single critical time, and 2001 and 2005 as the most common pair of critical times $t_c$ (i.e., the strongest change points in temporal community structure), which agrees with the visual observations.  
This finding suggests that a strong shift in the patterns of disease correlations occurred around these times. Indeed, Peru experienced a large countrywide dengue epidemic in 2000--2001, and this period also marks the onset of new yearly epidemic dynamics~\cite{Chowell2011}. Thus, our method recovers the most important biological event in this data set in addition to providing additional information about spatial influences on disease spread. We also observe several other time when new communities are born, but we do not know the biological significance of these dates. 
Notably, in this parameter regime, our community structure does not identify the large epidemic in the jungle Utcubamba province in 1996 (see Fig.~\ref{Figure:multislice}), which is the other large event in this data set.  

The community structure that we detect depends heavily on parameter values. In many parameter regimes --- especially when $\gamma \gtrapprox 1$ and $\omega \gtrapprox 0.5$ --- communities appear to be predominantly spatial, and we find high z-Rand scores when compared to the climate and administrative partitions [see Fig.~\ref{Figure:NGog}(b)]. 
The high influence of spatial proximity on the community structure is unsurprising, as spatial distance is an important influence on disease spread~\cite{Stoddard2009,Truscott2012}. Previous studies have also noted that the community structure of spatial networks obtained by maximizing modularity using the NG null model tends to be strongly influenced by geographical factors~\cite{Thiemann2010, Ratti2010, Expert2011}. If there are other interactions that shape the dengue fever correlation network, they might be obscured by the strong influence of spatial proximity. However, such interactions might be revealed by using a spatial null model that incorporates the expected effect of space on interactions.  We pursue this idea in Section \ref{space}.


\subsubsection{Community Structure Using Spatial Null Models}\label{space}

We apply spatial null models to the dengue fever correlation networks. We obtained province locations from the \url{Geonames.org} website~\cite{geonames}, and we obtained their populations from the Peruvian Instituto Nacional de Estad\'{i}stica e Inform\'{a}tica (INEI)~\cite{PeruStats}. We were only able to obtain the 1994 and 2007 populations; due to the limited range of data and the several changes in Peruvian administrative structures between the two times, we only use the 2007 populations.

The maximum inter-province distance is about 1300 km. We report numerical experiments using a bin size of 400 km after testing the spatial deterrence for several other sizes (ranging between 50 and 500 km) in the same manner as in Ref.~\cite{Expert2011}: that is, we study the shape of the deterrence function [see Eq.~\ref{eq:gravitymodel} and the nearby discussion] with changing distance across bin sizes, and we then examine the community structures that we obtain using different bin sizes. We find that bin sizes have an effect on the shape of the deterrent function (with lower sizes giving smoother results), but all of the bin sizes that we tested produced very similar partitions for both the gravity and radiation spatial null models. We selected the smallest bin size that guaranteed more than 5 nodes in each bin.

Recall from Section \ref{data} that only 79 of the 195 provinces had reported cases of dengue fever in our data, so we use the location and population data only for those provinces.

\begin{figure*}[h!tbp]
\centering
\hfill (a) \includegraphics[height=3.6cm]{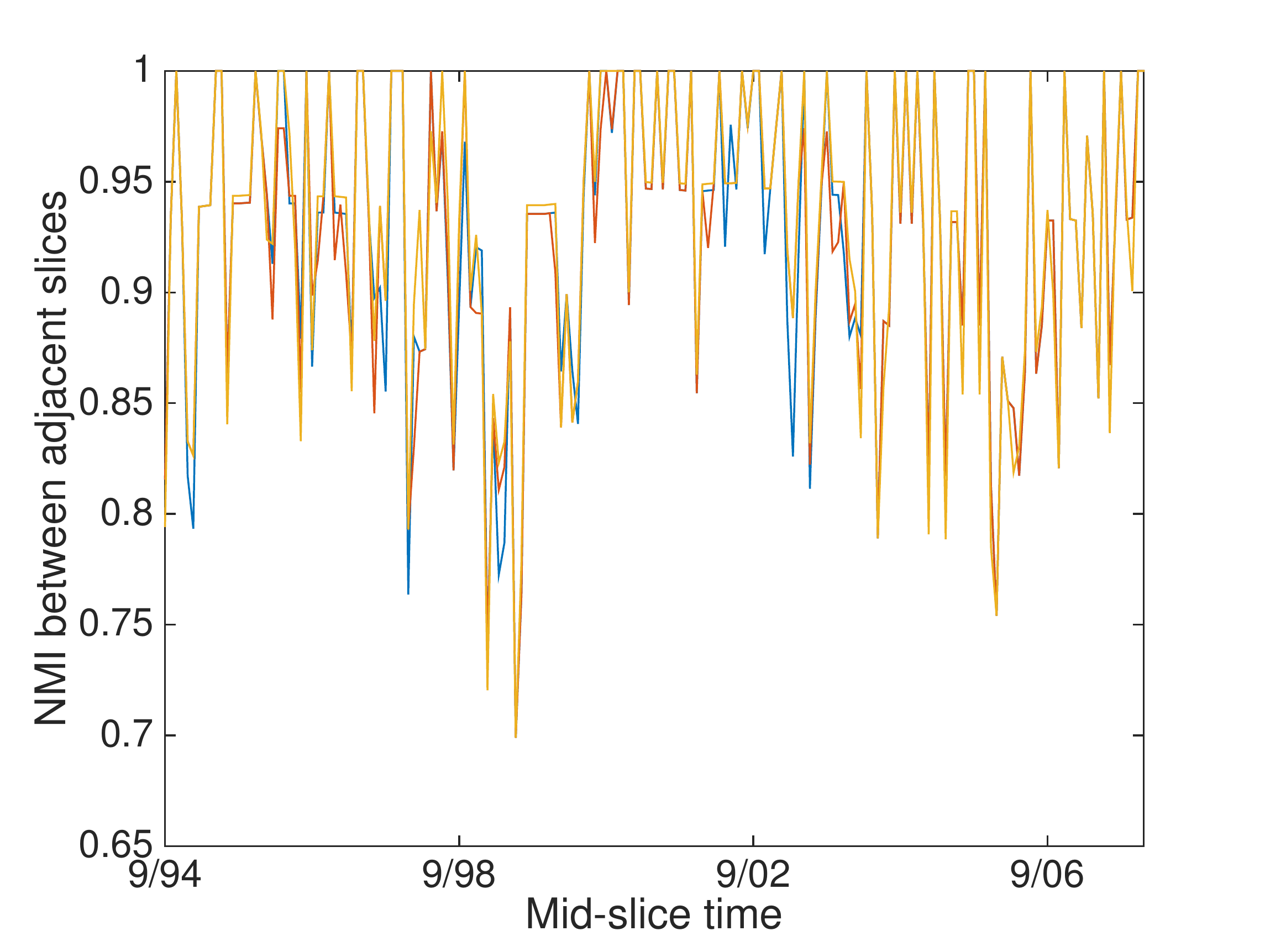}
\hfill (b) \includegraphics[height=3.6cm]{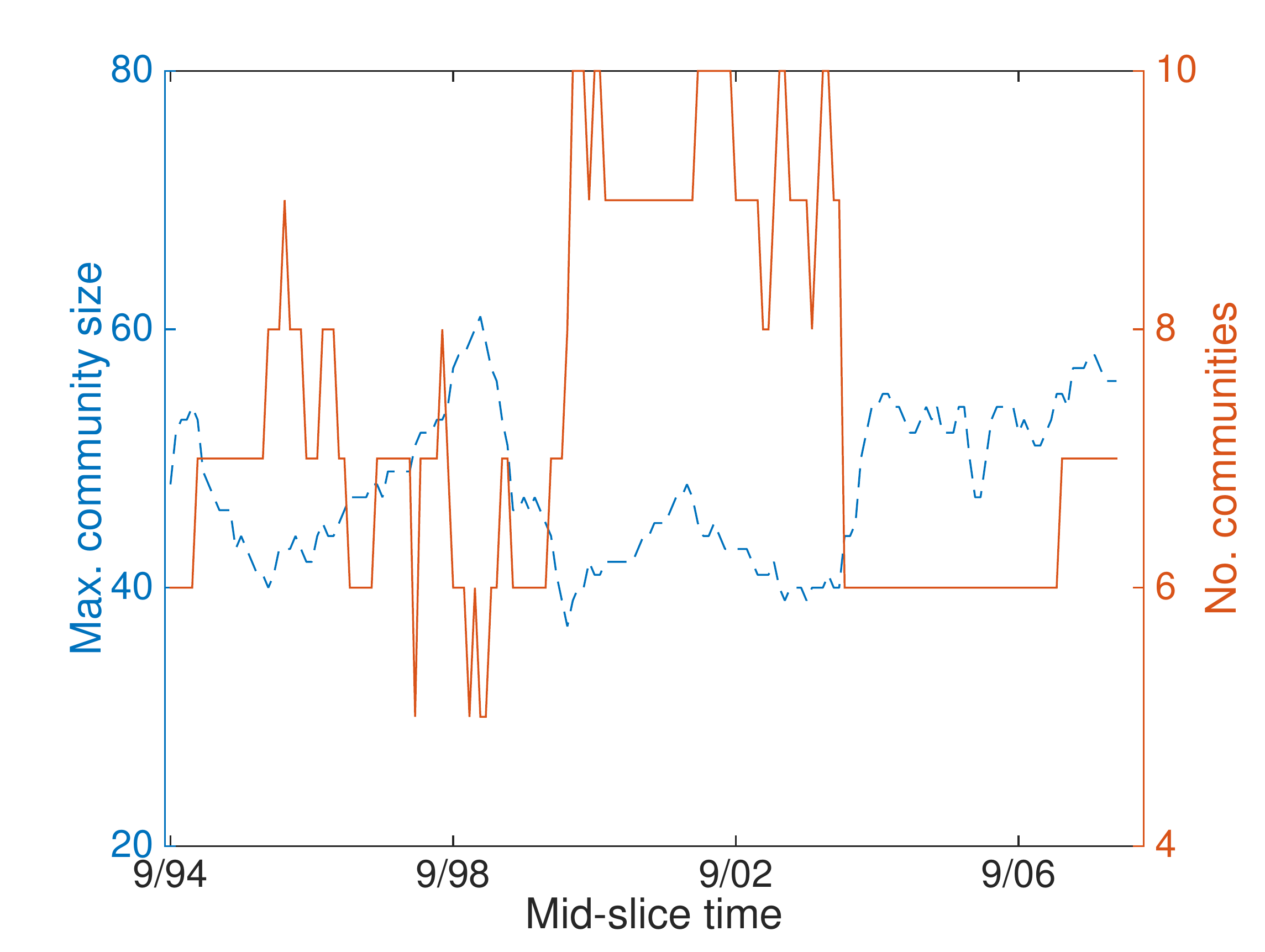}
\hfill (c)\includegraphics[height=3.6cm]{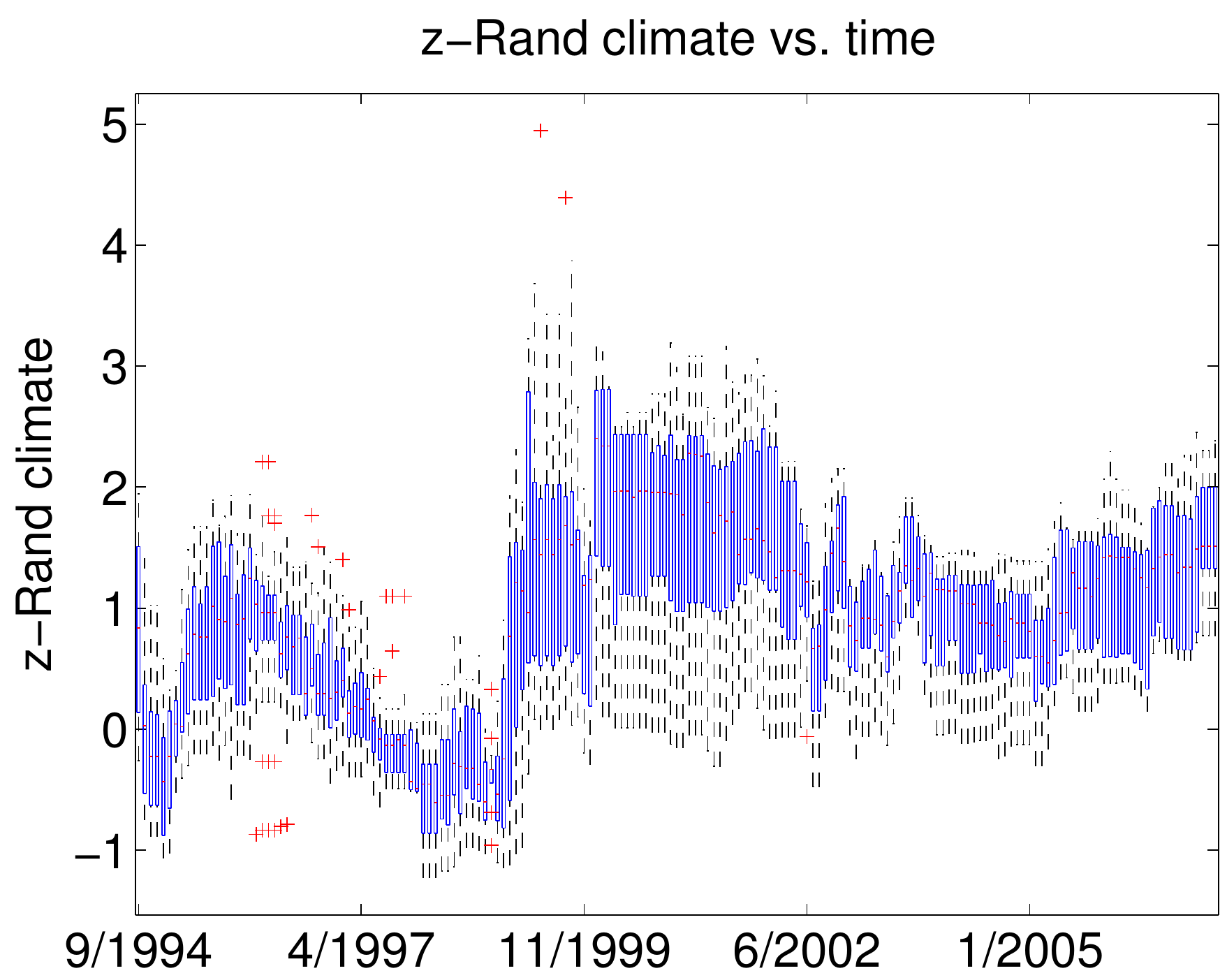}

\caption{Properties of the algorithmic community structure, which we detected by maximizing modularity using the gravity null model, of the dengue fever static correlation networks that we construct using a time window of $\Delta=80$. (a) NMI between adjacent layers for $\gamma \in \{0.9,1,1.1\}$. (b) Maximum community size (blue dashed curve) and number of communities (green solid curve) for $\gamma=1$. (c) Community structure scoring the highest z-Rand score versus climate among the dengue fever static correlation networks that we construct using $\Delta=80$. (The resolution-parameter value is $\gamma = 2.9$, the layer is 66, and the z-Rand score is 4.94.) 
We show the structure on a map of Peru, and we color provinces according to their community assignment. 
White provinces are ones in which our data does not include any reported cases of dengue fever in the indicated time window. 
 \label{Figure:spatial-static-boxes}}
\end{figure*}

We first study the community structure on static disease-correlation networks using the gravity and radiation null models. Both null models seem to remove most of the spatial element of the community structures (including temporal variation in the spatial correlations), as indicated by low values and low variation of spatial z-Rand scores (not shown). For both the gravity and radiation null models, we observe high similarity between layers for a variety of values of the resolution parameter $\gamma$ [see Fig.~\ref{Figure:spatial-static-boxes}(a)]. These structures contain one dominant community with the majority of nodes present at any given time, and several singleton communities [see Fig.~\ref{Figure:spatial-static-boxes}(b)]. By examining the partitions directly using a map of Peru, we see that the
singleton communities tend to consist of the highest-populated nodes. 

\begin{figure}[h!tbp]
  \centering
\hfill(a) \includegraphics[height=4cm]{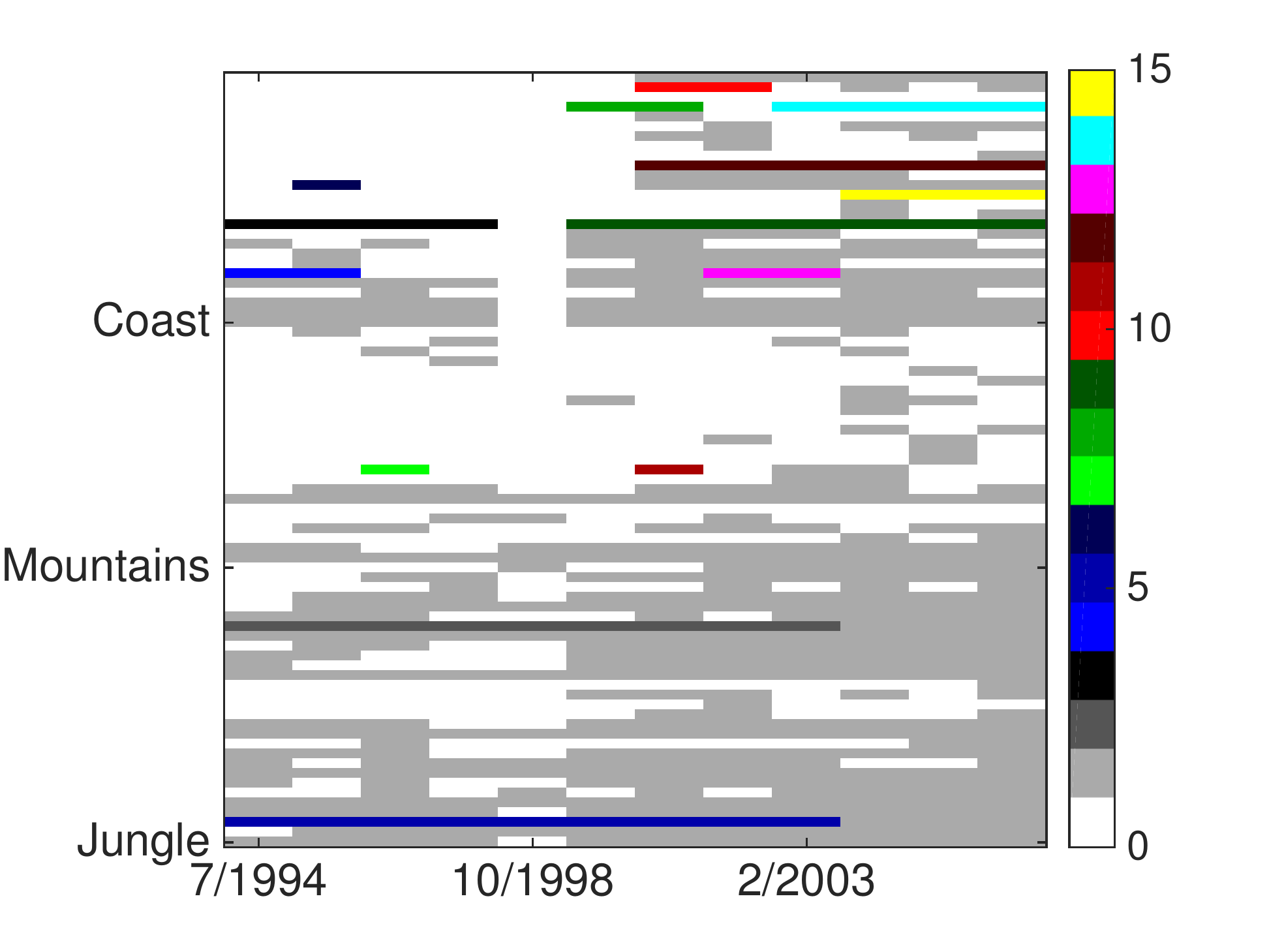}
\hfill(b) \includegraphics[height=4cm]{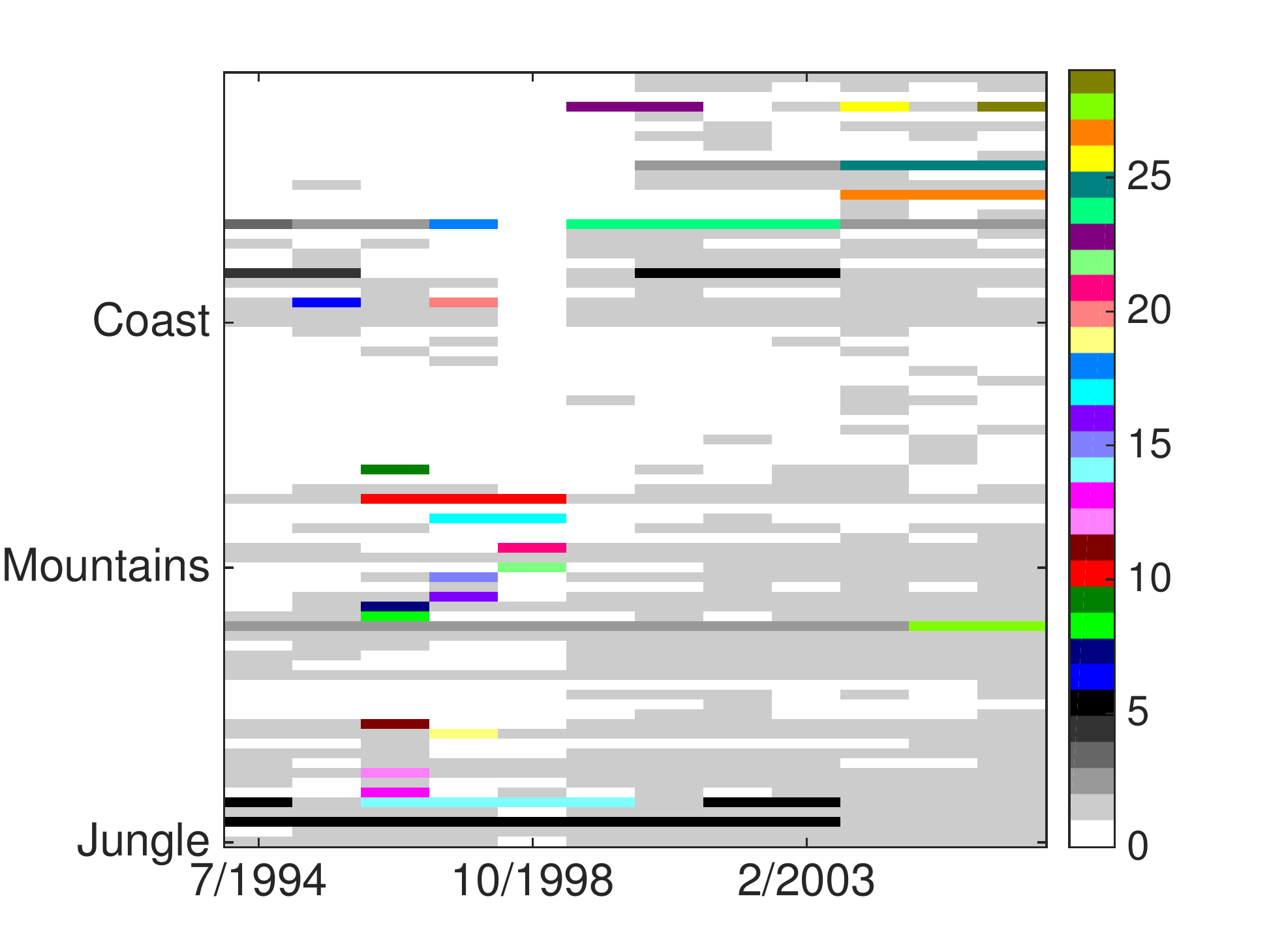}

\caption{Consensus community structure, which we obtain by maximizing modularity using (a) the gravity null model and (b) the radiation null model, of the dengue fever multilayer disease-correlation network that we constructed using a time window of $\Delta=60$ across 100 repeats of community detection. We use a resolution-parameter value of $\gamma=1$ and consider $\omega =0.1$. Mid-layer timepoints are plotted on the horizontal axis, and nodes are on the vertical axis. Node community membership is indicated by color. \label{Figure:GRog}}
\end{figure}

We also examine the spatial null models for multilayer correlation networks. The community structures again exhibit one large community containing the majority of multilayer nodes [see Fig.~\ref{Figure:GRog}(a,b)], and several multilayer nodes corresponding to provinces with highest populations form singleton communities across time. This situation occurs for all of the tested parameter values. Additionally, we do not observe any clear pattern in the z-Rand scores as we change $\gamma$ and $\omega$.

Our findings suggest that the addition of space into a null model for modularity optimization might remove the majority of the variation in the correlation structure of the dengue fever correlation networks, such that the influence of population size could be the only major factor that remains. This could relate to the concept that a minimum population size is required for sustained disease transmission; it has been estimated that this size is between 10,000 and 500,000 for dengue~\cite{Kuno1997,Chowell2008}. There are only 5 provinces with populations over 500,000, and these provinces are often assigned to singleton communities when we use a spatial null model.  This suggests that they have different disease patterns from the other provinces.


\subsection{Community Detection Using a Correlation Null Model}\label{corr-null}

Recently, MacMahon et al.~\cite{MacMahon2013arXiv} proposed a new null model that they designed specifically for modularity maximization for networks that are constructed based on the pairwise Pearson correlations between time series. They used ideas from RMT \cite{mehta} to generate a null model that represents the ``random'' component of a correlation matrix and can take into account the single most strongly influencing factor on the correlation structure. In the context of financial systems, which was the focal example of Ref.~\cite{MacMahon2013arXiv}, this factor is often called a ``market mode''.  Given that we often found a single large community when we used spatial null models, it is interesting to see what results we obtain using such a correlation null model.

To use a correlation null model, we need to construct our network directly from pairwise correlations without subsequently shifting them to $[0,1]$ and removing self-edges. We construct networks by selecting time windows and calculating Pearson correlations in the same manner as in Section~\ref{sec:netcreate}, but the here edge weights are left as raw correlations: $C_{ij} = \rho_{ij}$ (Eq.~\ref{equation:Pearson}).

Because of the special structure of correlation matrices, modularity using the standard NG null model assigns importance to pairs of nodes $i$ and $j$ whose Pearson correlation is larger than the product of the correlations of each node with the time series of the 
total number of disease cases in the country over the chosen time window: $E_{\textrm{tot}}$, where $E_{\textrm{tot}}(t) = \sum_{i=1}^{N} E_i (t)$. 

By contrast, the correlation null model that we adopt from Ref.~\cite{MacMahon2013arXiv} uses ideas from RMT to detect communities of nodes that are more connected than expected under the null hypothesis that all time series are independent of each other. 

For a given correlation matrix constructed from $N$ time series that each have length $T$ (with $T/N > 1$), one posits based on RMT that any eigenvalues that are smaller than the eigenvalue $\xi_+ = (1+ \sqrt{N/T} )^2$ are due to noise. Here, $\xi_+$ is the maximum eigenvalue predicted for a correlation matrix that is constructed from the same number of entirely random time series.

Additionally, for many empirical correlation matrices, the largest eigenvalue $\xi_m$ is much larger than the others, and its corresponding eigenvector has all positive signs~\cite{MacMahon2013arXiv}. In this situation, there is a common factor, which is called the ``market mode'' in financial applications, that influences all of the time series~\cite{Sinha2011}. 

We can thus decompose our correlation matrix $C$ as follows: $C = C^{(r)} + C^{(g)} + C^{(m)}$, where $C ^{(r)}$ is the ``random'' component of the matrix, $C^{(m)}$ is the ``market mode'', and the ``group mode''  $C^{(g)}$ is embodies the meaningful correlations between time series. We write $C^{(m)} = \xi_m v_m \otimes v_m $ and $C^{(r)} = \sum_{\{i | \xi_i \leq \xi_+\}}\xi_i v_i \otimes v_i $, where $\xi_i$ and $v_i$ are an eigenvalue and its corresponding eigenvector, $v_m \otimes v_m$ is the outer product of the two vectors (a special case of the Kronecker product for matrices), and $\xi_m$ is the maximum observed eigenvalue in the correlation matrix $C$. We can construct a correlation null model either by removing both the ``random'' component of the matrix and the influence of the ``market mode'' (i.e., by using the null model $P^{\textrm{corr}} = C ^{(r)} + C ^{(m)}$) or by only removing the random component (i.e., by using the null model $P^{\textrm{corr}} = C^{(r)}$).

To satisfy the $T/N > 1$ requirement to applying the RMT approach of Ref.~\cite{MacMahon2013arXiv}, we require $\Delta \geq N$. 
For subsequent calculations, we use $\Delta = 80$ for static networks, and $\Delta = 60$  for multilayer networks (which have a maximum of 59 nodes per slice, as not all provinces experience disease at the same time), unless stated otherwise. 

Although our maximum eigenvalue is larger than the other eigenvalues and every component of the associated eigenvector is positive, the eigenvector does not appear to affect all nodes to the same extent. The above construction thus yields a non-uniform null model for our data in practice, so we are unable to identify the analog of a market mode.  
We thus do not incorporate such a mode into the null models that we employ for community detection. We use the correlation null model
\begin{equation}
	P^{\textrm{corr}} = C ^{(r)} = \gamma \sum_{\{i| \xi_i \leq \xi_+\}}\xi_i v_i \otimes v_i \,,
	\end{equation}
where $\gamma$ is the resolution parameter. For the multilayer setting, we write
\begin{equation}
		\overbar{P}^{\textrm{corr}_s}= C_s ^{(r)} = \gamma \sum_{\{i| \xi^s_i \leq \xi^s_+\}}\xi^s_i v^s_i \otimes v^s_i \,,
\end{equation}
where $\xi^s_i$ and $v^s_i$ are an eigenvalue and its corresponding eigenvector for layer $s$.

\begin{figure*}[tbp]
\centering
	\hfill (a) \includegraphics[height=4.7cm]{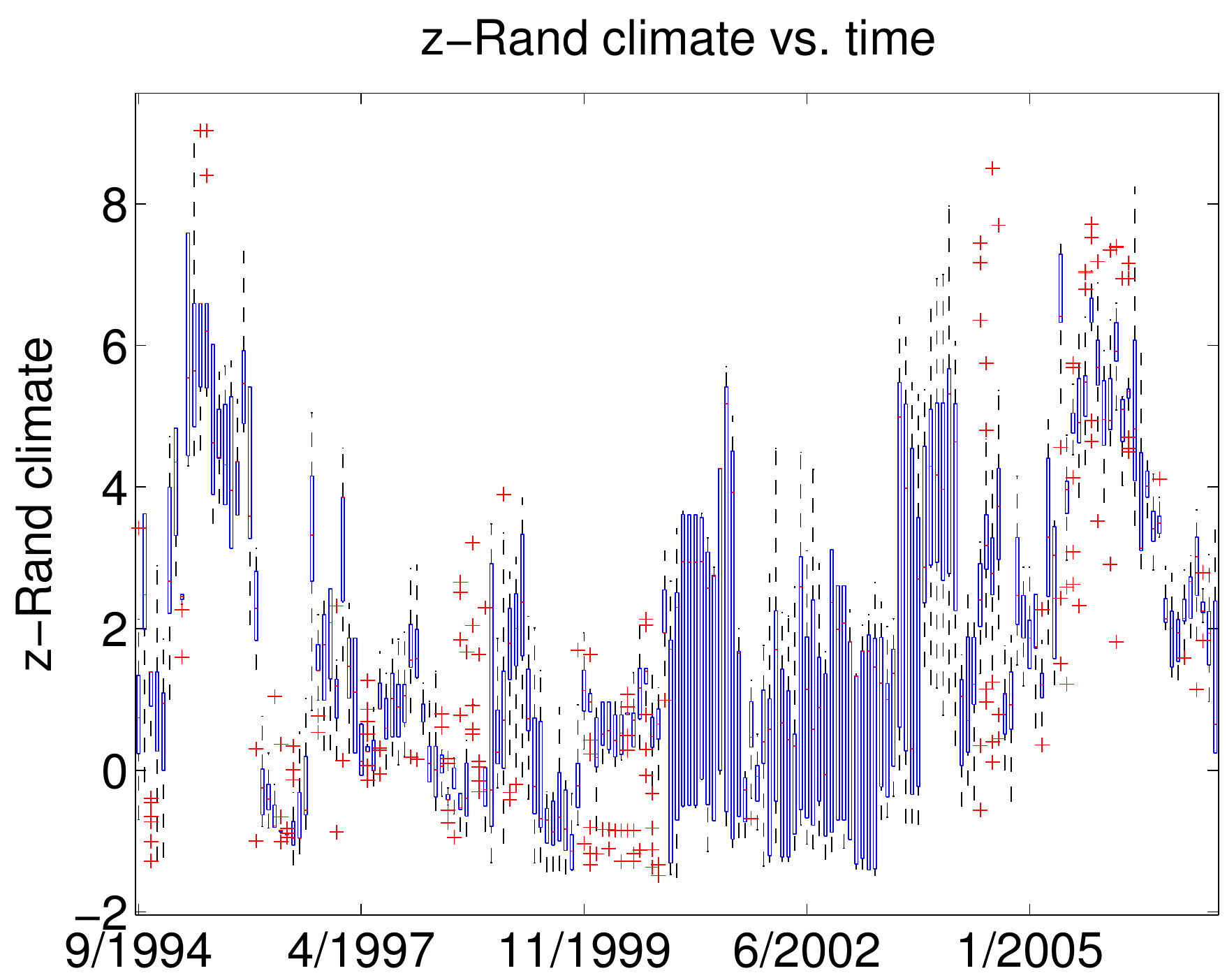}
	\hfill (b) \includegraphics[height=4.7cm]{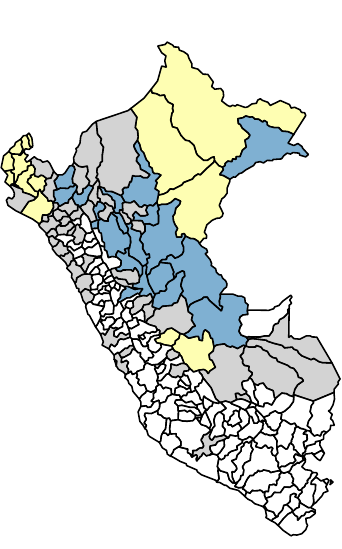}
	\hfill (c) \includegraphics[height=4.7cm]{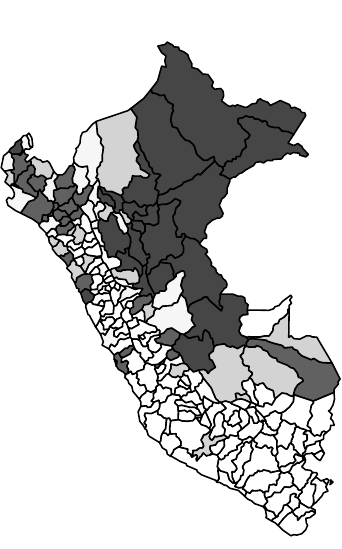}
\caption{Algorithmic community structure, which we obtain by maximizing modularity using a correlation null model, for the static dengue fever correlation networks that we construct using $\Delta=80$. 
(a) A box plot of the z-Rand scores versus the detailed climate partition at different $\gamma$ values ($\gamma \in {0.1,0.2,\ldots,3}$), for the static networks covering the whole time period (horizontal axis).  
In panels (b,c), we show partitions of the network with the highest z-Rand score based on (b) climate ($\gamma = 2.4$; layer 37, which corresponds to October 1995; and a z-Rand score of 9.04) and (c) administrative divisions ($\gamma = 2.3$; layer 132, which corresponds to April 1996; and a z-Rand score of 8.9). We color provinces according to their community membership on a map of Peru. White provinces are ones in which our data does not include any reported cases of dengue fever in the indicated time window. 
\label{Figure:temporal-static-boxes}}
\end{figure*}

We test the performance of this correlation null model on correlation networks that we construct from dengue fever time series with $\Delta = 80$. In most of the static networks, the community structures appear to be affected by spatial proximity --- especially for post-2000 networks, as illustrated by the high z-Rand scores versus the climate partition (particularly in 1995--1996, 2000--2001, 2003--2004, 2005--2006).
See Fig.~\ref{Figure:temporal-static-boxes}(a). These high z-Rand scores result from (1) the classification of the majority of jungle provinces into one community and (2) the existence of a community that contains many of the northern coastal provinces [see Figs.~\ref{Figure:temporal-static-boxes}(b,c)]. 

\begin{figure*}[tbp]
  \centering
\hfill (a)\includegraphics[height=3.1cm]{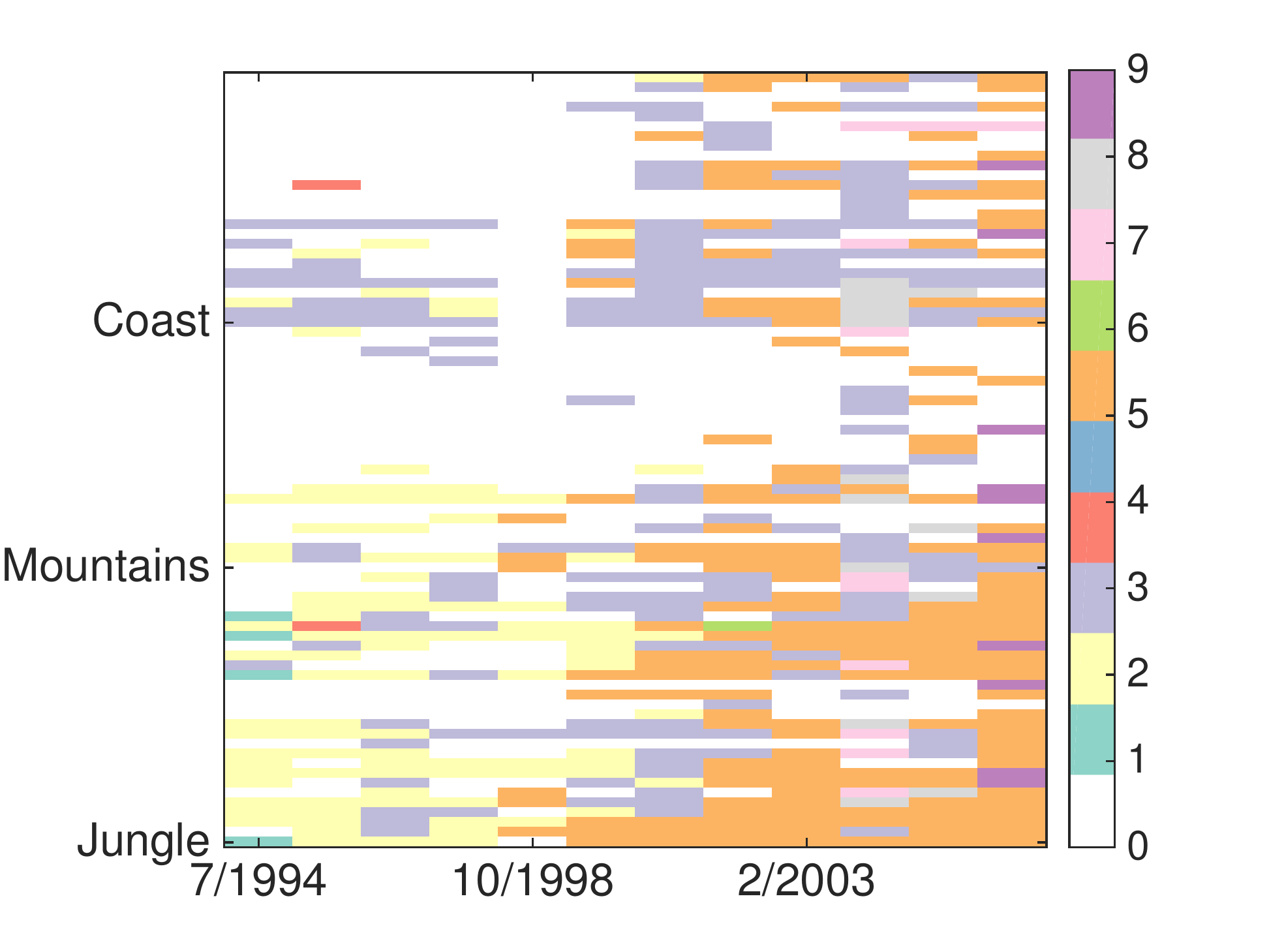}
\hfill (b)\includegraphics[height=3cm]{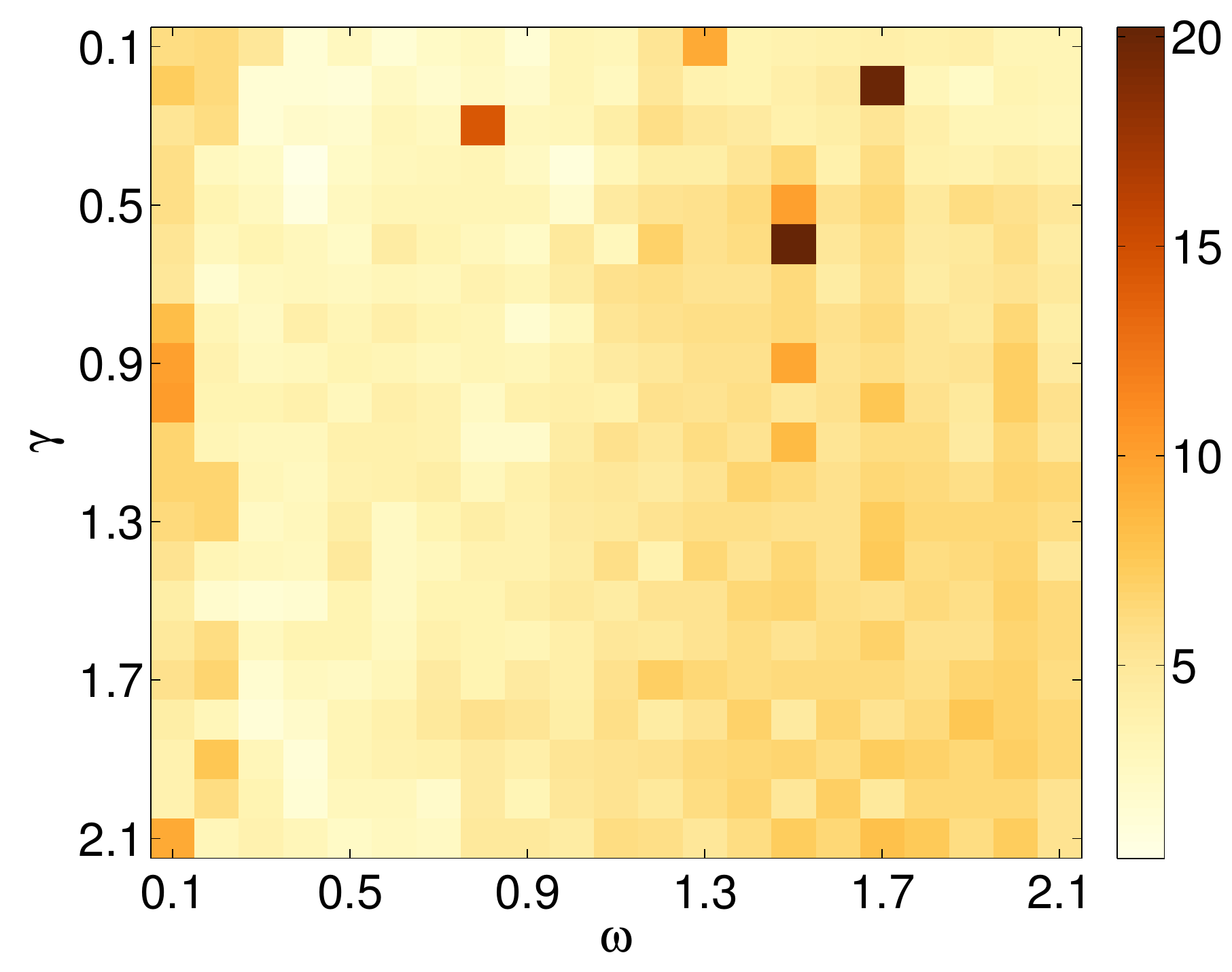}
\hfill (c)\includegraphics[height=3cm]{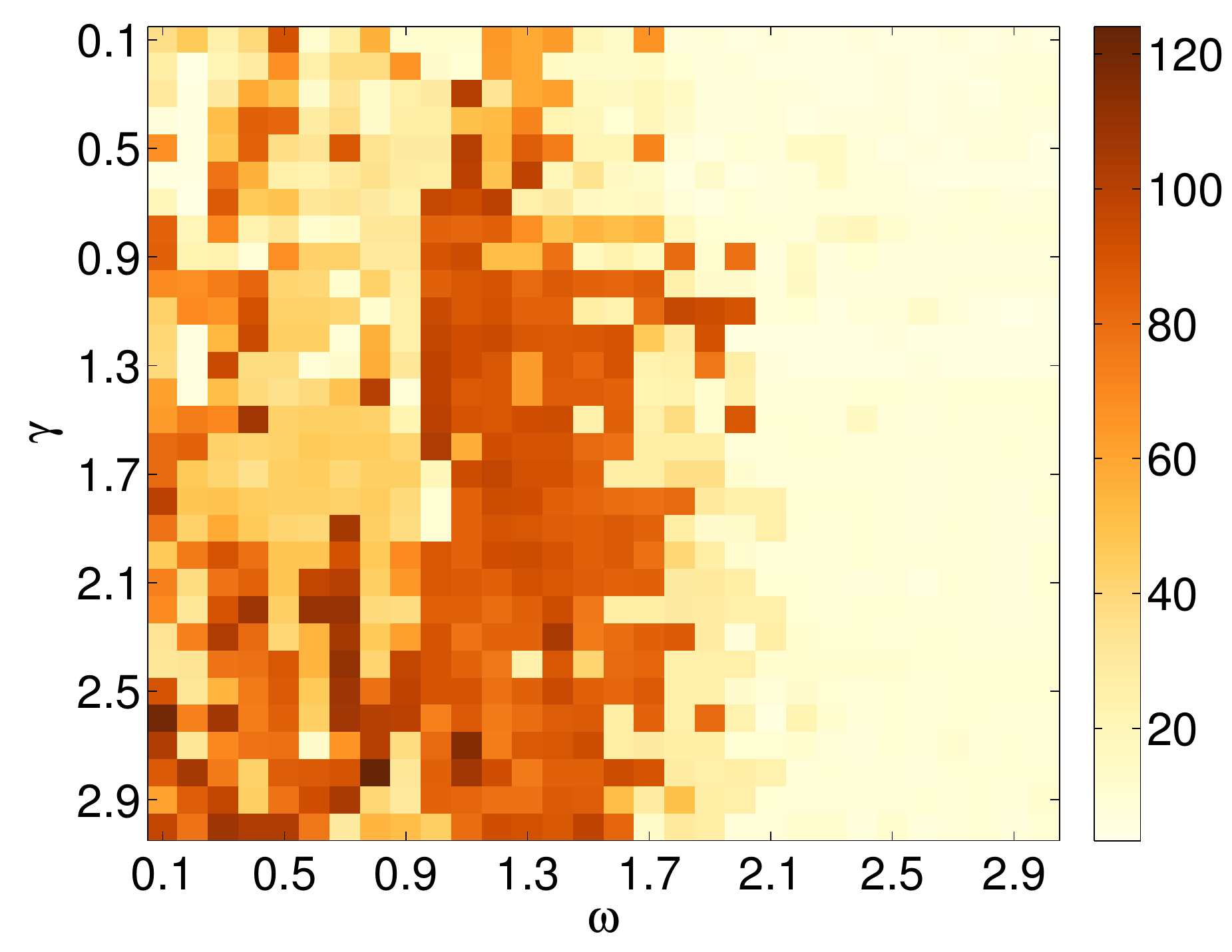}
\\
\centering
\hfill (d)\includegraphics[height=3cm]{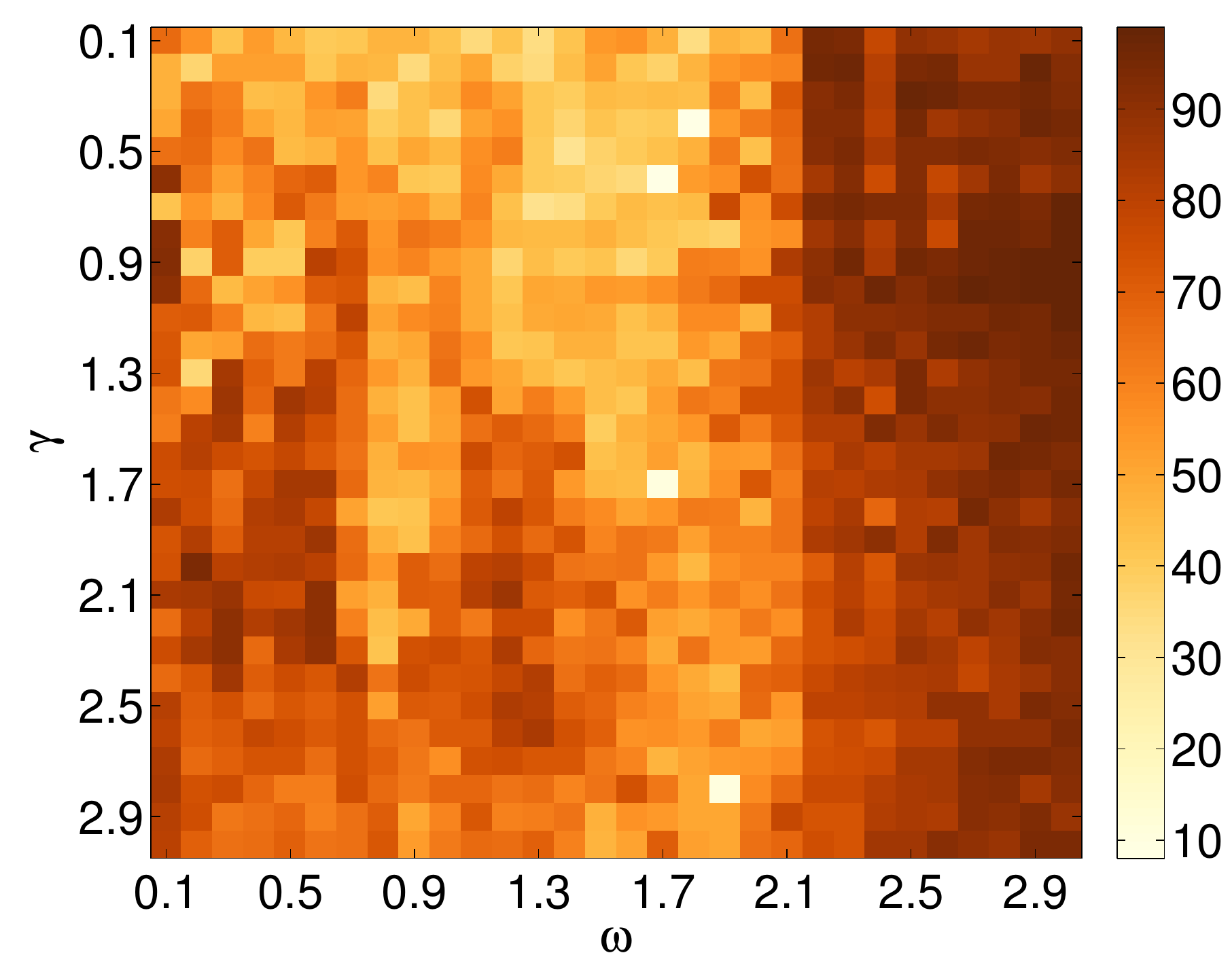}
\hfill (e)\includegraphics[height=3cm]{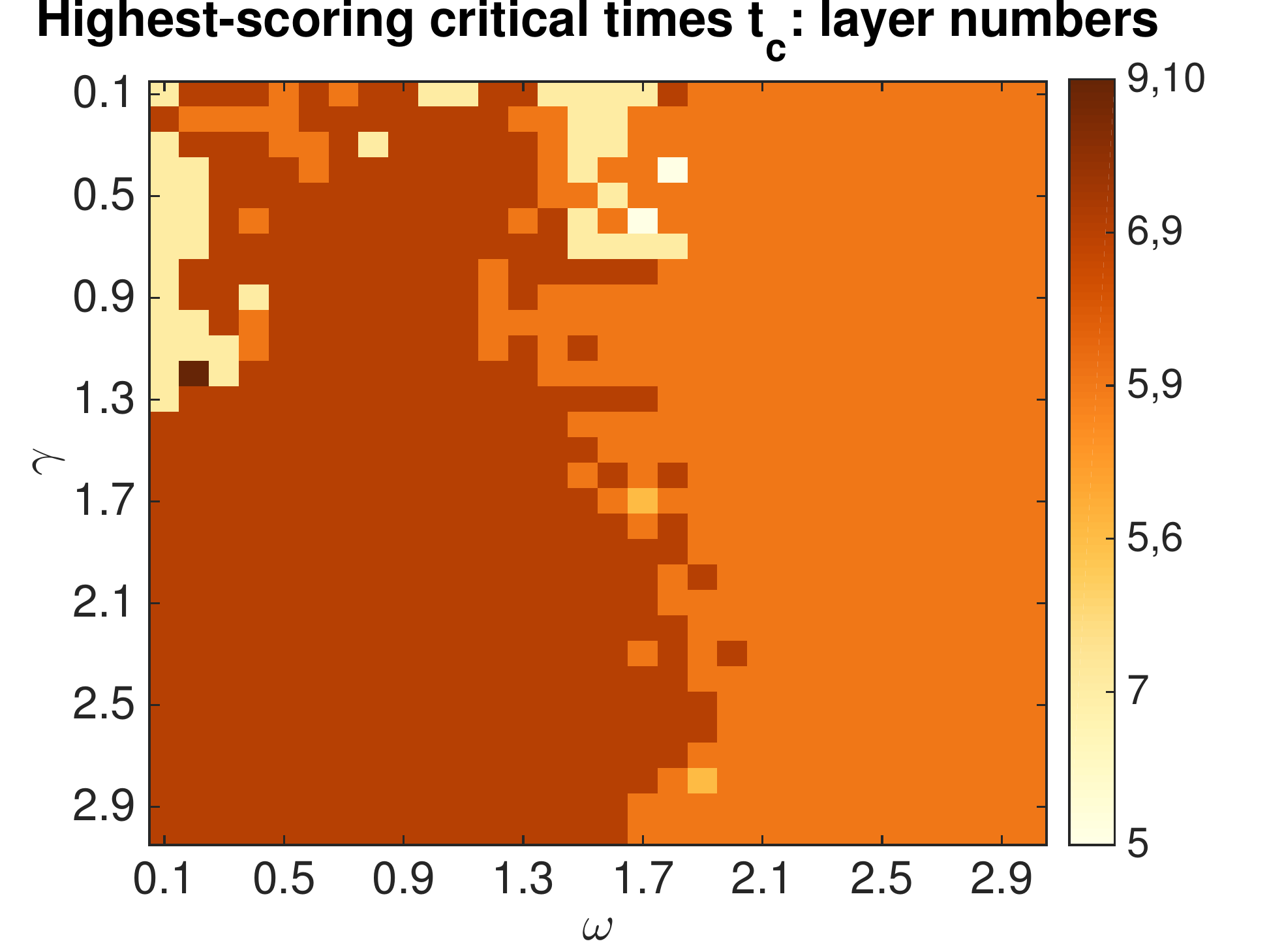}
\caption{Algorithmic community structure, which we obtain by maximizing modularity using a correlation null model, of the dengue fever multilayer disease-correlation network that we construct using $\Delta=60$. (a) Consensus community structure for $\gamma=1$ and $\omega =0.1$ calculated across 100 repeats. (b,c) Results of varying $\gamma$ and $\omega$.  We show the z-Rand scores for similarity to (b) ``spatial'' partitions by administrative region, (c) temporal partitions before and after a single critical time $t_c$. (For each parameter set, we select the single highest scoring $t_c$). In (d) we show the z-Rand scores for similarity to temporal partitions with 3 communities determined using a pair of critical times. (e) Pairs of critical times $t_c$ with the largest z-Rand scores.
 \label{Figure:Tempog}}
\end{figure*}

We also perform community detection on multilayer networks using the correlation null model $\overbar{P}^{\textrm{corr}_s}$ for $(\gamma,\omega) \in [0.1,3] \times [0.1,3]$. We obtain partitions with a mixture of temporal and spatial features. We calculate a consensus community structure across 100 realizations for each value of $\gamma$ and $\omega$. In Fig.~\ref{Figure:Tempog}(a), we show the best-looking partition,
 which we obtain for $\gamma=1$ and $\omega =0.1$. This partition includes 9 communities. Although several communities coexist in each layer, the primary divisions appear to be largely temporal. For example, community 2 shrinks after layer 6 (January 2001). However, the highest z-Rand score versus a temporal partition (with either one or two critical times) for this partition is one that has a single critical time $t_c$ at the end of layer 7 (i.e., in July 2002). That is, this manually-constructed temporal partition has one community that contains layers 1--7 and a second community that contains layers 8--12.

We obtain a temporal partition with one critical time for $\gamma \lessapprox 1$ and $\omega \lessapprox 0.3$. The other parameter regimes have different critical times. For $\omega \lessapprox 1.5$ and the allowance of a pair of critical times, we obtain the highest temporal z-Rand score when the critical times occur immediately after layer 6 (i.e., January 2001) and immediately after layer 9 (i.e., March 2004). See Fig.~\ref{Figure:Tempog}(d). For $\omega \gtrapprox 1.5$, we obtain the highest temporal z-Rand scores with a pair of critical times when those times occur immediately after layer 5 (i.e., June 1999) and immediately after layer 9. (Note that these z-Rand scores tend to be lower than those for $\omega \lessapprox 1.5$.) See Fig.~\ref{Figure:Tempog}(e).
In our sweep over the different values of $\gamma$ and $\omega$, we obtain lower climate z-Rand scores compared to what we obtained using the NG null model (compare Fig.~\ref{Figure:Tempog}(b) with Fig.~\ref{Figure:NGog}(b)). 
We do not observe any clear patterns in the spatial z-Rand scores as we vary $\gamma$ and $\omega$.

\subsubsection{Province-level Multilayer Communities}
\label{dengue-region-NG}

We now examine the province-level information that we can glean from the data. The simplest approach is to construct a single static network from the entire length-$T$ time series, but our multilayer approach allows us to aggregate data less severely.  This, in turn, allows us to lose less information. 

\begin{figure*}[tbp]
\centering
\hfill (a)\includegraphics[width=.21\linewidth]{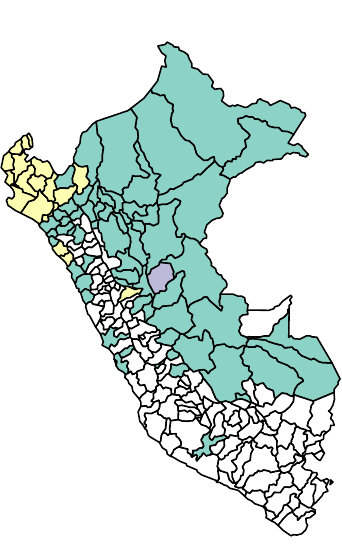}
\hfill (b)\includegraphics[width=.21\linewidth]{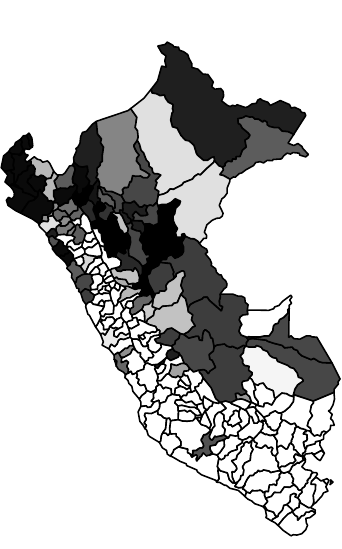}
\hfill (c)\includegraphics[width=.21\linewidth]{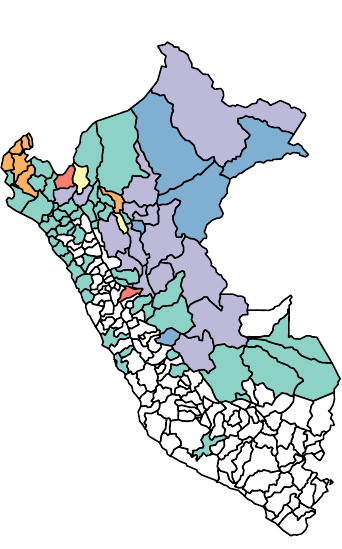}
\hfill (d)\includegraphics[width=.21\linewidth]{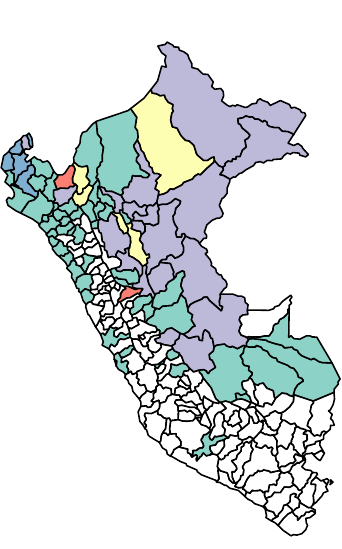}

\caption{Province-level algorithmic community structure, which we obtain by maximizing modularity, for the static and multilayer dengue fever correlation networks. We color the provinces according to their community assignments. White provinces are ones in which our data does not include any reported cases of dengue fever in the indicated time window. (a) NG null model that is fully aggregated (i.e., $\tau = 1$ and  $\Delta=779$) with a resolution-parameter value of $\gamma=1$. (b) NG null model that is fully aggregated with $\gamma=1.1$. (c) NG null model in a multilayer network with province-level communities that we obtain from the multilayer network with a time window of width $\Delta=60$. (d) Correlation null model in a multilayer network with province-level communities that we obtain with a time window of width for $\Delta=60$.
\label{Figure:Dengue-regionlevel-map}}
\end{figure*}

When we aggregate all time series to construct a single similarity network (i.e., we choose $\tau=1$ and $\Delta=779$), we find that the community structures that we obtain via modularity maximization with the spatial and correlation null models all consist of a single large homogenous community with up to three outlier nodes (see Fig.~\ref{Figure:Dengue-fullyaggregated} in Appendix \ref{Appendix:regions}). Only the NG null model is able to detect meaningful-looking communities, especially for $\gamma = 1$ and $\gamma = 1.1$ [see Fig.~\ref{Figure:Dengue-regionlevel-map}(a)]. For $\gamma =1$, the we find three communities; one is a singleton, and the middle one consists almost exclusively (15 of 17 nodes) of northern coastal provinces. This partition has z-Rand score versus climate of 7.3. For $\gamma = 1.1$, using the NG null model yields 28 communities, and many of them are small. 

Nodes grouped in the community of northern coastal provinces are the provinces of Peru that were most strongly involved in the 2000--2001 dengue epidemic; 15 nodes in this community experienced this epidemic, whereas only two other nodes experienced it. 

The data aggregation over the whole time series results in the 2000--2001 epidemic dominating all other events in the time series. If we use the community structure of the temporally evolving multilayer network to create the province-level structure, we might be able to shed some more light on other interactions between provinces. 

We then study the structure of province-level communities that we obtain from community detection using the uniform null model on an association matrix $A^{\mathrm{province}}$. As we discussed in Section \ref{Section:networks}, we create this matrix by counting the number of multilayer nodes that are classified together in a consensus community detection on a multislice network.  We consider the parameter values $\gamma = 1$ and $\omega = 0.1$ over 100 repeats.

\begin{figure*}[h!tbp]
	\centering
	\hfill (a) \includegraphics[width=.45\linewidth]{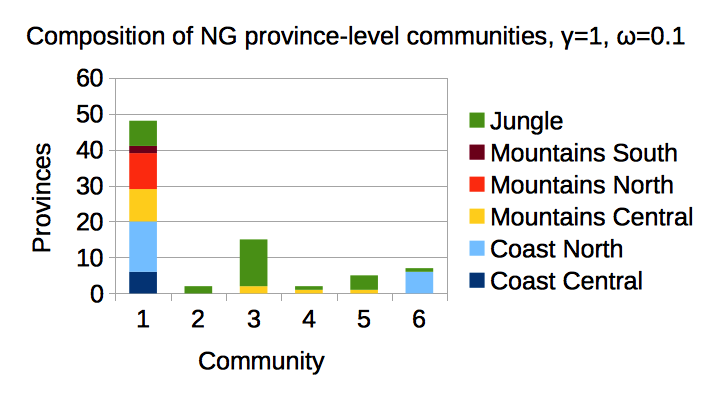}
	\hfill (b) \includegraphics[width=.45\linewidth]{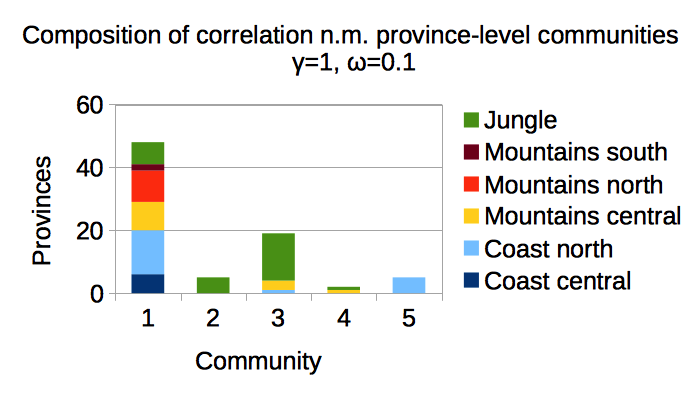}
%
\caption{Membership of the consensus province-level communities, which we computing by maximizing modularity, in multilayer dengue fever networks for $\gamma=1$. In panels (a) and (b), we compare the climate composition of the communities using (a) the NG null model and (b) a correlation null model.  We order communities according to their size, and the horizontal axis gives the community number. 
}
\label{Figure:Dengue-regionlevel-climates}
\end{figure*}

\begin{figure*}[h!tbp]
	\centering
	\hfill (a)
	\includegraphics[width=.45\linewidth]{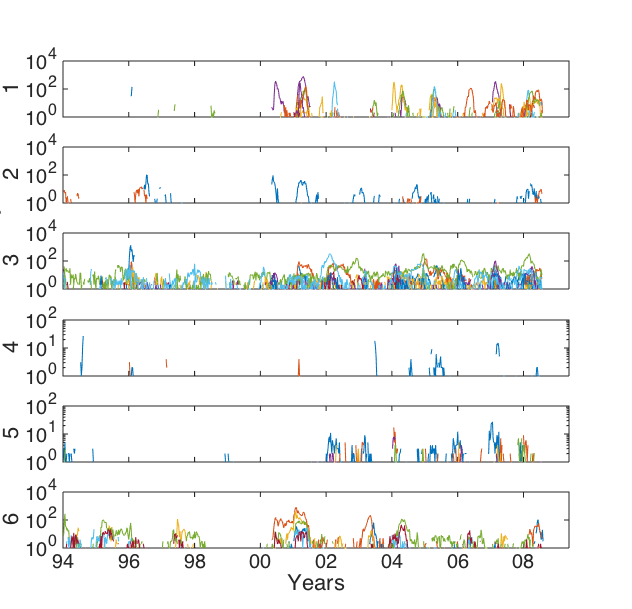}
	\hfill (b) \includegraphics[width=.45\linewidth]{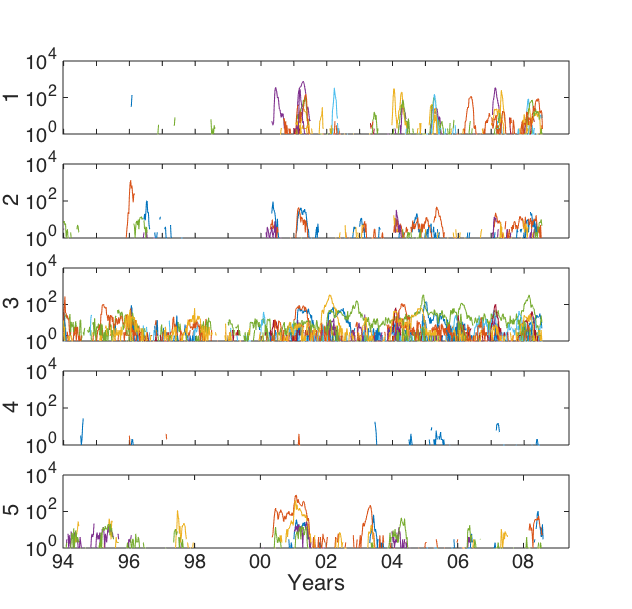}

\caption{Time series for disease occurrences in the provinces that belong to consensus province-level communities, which we computing by maximizing modularity, in multilayer dengue fever networks for $\gamma=1$ using (a) the NG null model and (b) a correlation null model.}
\label{Figure:Dengue-regionlevel-timeseries}
\end{figure*}

Comparing the province-level communities that we obtain using the NG and correlation null models versus the broad topographical categories of coast, mountain, and jungle reveals large-scale climatic influence on disease patterns. The two null models yield similar results: more than 40 nodes are grouped into one large community that includes central coast, northwestern and southern jungle, and eastern jungle; and coastal north nodes form smaller, strongly spatial communities [see Fig.~\ref{Figure:Dengue-regionlevel-map}(c,d) and Fig.~\ref{Figure:Dengue-regionlevel-climates}]. When we study the disease time series of the provinces grouped into the province-level communities, we observe distinct types of disease incidence patterns. The NG null model finds one more pattern type (nodes with late onset of disease, as we illustrate in Fig.~\ref{Figure:Dengue-regionlevel-timeseries}) than the correlation null model.


\section{Conclusions}\label{Section:conclusions}

In conclusion, we examined time-dependent community structure and the effect of different null models --- including ones that incorporate spatial information --- on the results of modularity maximization.  We conducted our computational experiments using novel synthetic benchmark spatial networks and correlation networks constructed from spatiotemporal dengue fever incidence data in provinces of Peru (a system that is strongly influenced by spatial effects).  We compared our results for the standard Newman-Girvan null model versus two null models that incorporate spatial information: a gravity null model~\cite{Expert2011} and a novel radiation null model.  We also compared the NG null model on disease-correlation networks with a recently-developed correlation null model (and a multilayer generalization of it) that is designed specifically for studying correlation networks that are derived from time series~\cite{MacMahon2013arXiv}. 

Our results indicate that it is very important to incorporate problem-specific information such as spatial information into the null models for community detection. Our results also illustrate that there are many nuances to consider. That is, it is not simply a matter of incorporating spatial information in an arbitrary way but rather developing spatial null models that are motivated by application-appropriate generative models. For example, the NG null model performs better than the spatial null models (which both use population data) on the random population distance benchmark where populations vary but edge weight does not depend on them. However, when we remove the variation in population or modify the benchmark to include population in edge placement probabilities, we find that the gravity null model performs best (as expected). 

Parameter choices can also be extremely important, as demonstrated by the large influence of bin size (when binning distances for the spatial null models) on community detection results, the failure to find meaningful communities with any of the null models at low edge densities, and the strong influence of resolution parameter $\gamma$ on the results.

To summarize, one needs to consider seriously what variables that influence the connections in a system of interest one wants to include in a null model, be careful about including spurious variables, and test how the results change for many parameter values. 

Finally, not incorporating space at all can be more appropriate than incorporating it in a manner that is overly naive. (See, for example, our results on the random population benchmarks.)

In our consideration of dengue fever data, we observed for static networks that the NG and correlation null models find structures that are strongly spatial --- especially after the onset of yearly epidemics in 2000. In our study, we observed that spatial partitions are often dominated by large communities of neighboring jungle nodes that experience local epidemics during the time window. 

On a multilayer network, maximizing NG modularity can result in either spatial or temporal partitions (depending on the parameter regime). Temporal partitions successfully find the most important time point in the history of the disease --- namely, the introduction of a new disease strain that caused a large epidemic in 2000--2001 and a subsequent shift in disease patterns --- and several other potentially interesting time points and periods of high spatial correlation. 

When studying province-level connectivity, we illustrated that consensus province-level communities from an association matrix that we constructed from the multilayer network across time is a far preferable approach to complete data aggregation. For the aggregation into a static network, maximizing modularity using any of the test null models except the NG null model failed to detect any meaningful communities; the NG community structure corresponds to the large 2000--2001 epidemic. Aggregating networks results in loss of information that is desirable to study for meaningful patterns \cite{mikko-review,holme12}. 

When we constructed multilayer networks and computed consensus communities, the computed ``spatial'' multilayer partitions and province-level partitions highlight the importance of climate to the disease patterns of dengue, as the jungle provinces are placed into distinct communities from most mountainous and coastal provinces. This is sensible, as the yearly epidemic patterns tend (on average) to exhibit an earlier epidemic onset in the jungle~\cite{Chowell2008,Chowell2011} and the jungle climate is rather distinct from the climate in coastal and mountainous provinces. The main climatic difference between jungle provinces and other provinces is temperature, and the influence of temperature on dengue transmission~\cite{Keating2001,Depradine2004,johansson2009local} and attack rate and persistence has been documented~\cite{Chowell2006,Chowell2008}. 

The province-level communities that we detect using both the NG and the correlation null models yield distinct temporal disease incidence patterns. The NG null model also finds one additional jungle community (with a late disease onset) than the correlation null model. The assignment of different jungle nodes into separate communities hints that the variables that influence jungle epidemics may be different than those in other climates. Moreover, the variability in disease patterns between jungle provinces is high, perhaps due to the year-round disease presence (in contrast to the existence of a summer disease season on the coast). Chowell et al. \cite{Chowell2008} reported that the coastal and mountainous provinces exhibit more spatial heterogeneity of disease incidence than the jungle provinces, and population size appears to play a larger role in disease persistence in the jungle. Additionally, the jungle climate is more homogenous (especially in the north-south direction) than the other two climates. 

When we attempt to remove the influence of space by using the gravity and radiation null models, we obtain one large community that contains all but the highest-population provinces (which are assigned to singleton communities). In contrast to the linguistic example in Ref.~\cite{Expert2011}, this suggests for our disease networks that the incorporation of space into the null model accounts for the majority of the structure present in the network. The spatial structure that we removed likely includes the structure that corresponds to the climate variation that causes different epidemic patterns in the jungle, coastal, and mountainous provinces. The only variable that we were able to identify as influencing community structure when using spatial null models is province population: the highly populated (and typically coastal) provinces forming singleton communities. These highly populated provinces are likely to be economic centers, with many people traveling there from the other provinces and thereby transmitting the disease~\cite{osorio2004travel,Martens2000,harrington2005dispersal,Stoddard2009}. 

These provinces could then be the seeds of epidemics for the other coastal and mountainous provinces, and two studies have in fact reported (so-called) ``hierarchical'' transmission of dengue from populous regions to those with low populations in both Peru and Thailand~\cite{Chowell2008,Cummings2004}. 
This situation could lead to high correlations across atypically long distances compared with the majority of the data, which could in turn cause populous provinces to be assigned to singleton communities. Additionally, it is known that population size influences dengue transmission: the basic reproductive number $R_0$ and disease persistence (i.e., the fraction of weeks with disease cases) are positively correlated with population size, and the attack rates are negatively correlated with it~\cite{Chowell2008,Chowell2011}.

The incorporation of spatial information into null models for community detection is both interesting and desirable.  As we have illustrated in the present paper, however, there are many nuances that it is important to consider. We have also demonstrated that it is important to develop null models that incorporate generative mechanisms for human mobility and flux.  We similarly expect that domain-dependent, mechanistic null models will also be crucial in many other applications. 


\section*{Acknowledgements}

MS was supported by the EPSRC Systems Biology Doctoral Training Centre. GC acknowledges support from grant number 1R01GM100471-01 from the National Institute of General Medical Sciences (NIGMS) at the National Institutes of Health and the Multinational Influenza Seasonal Mortality Study (MISMS) led by the Fogarty International Center, National Institutes of Health~\cite{NIH}. MAP was supported by the European Commission FET-Proactive project PLEXMATH (Grant No. 317614) and a grant (EP/J001759/1) from the EPSRC. 

We thank Marya Bazzi, Paul Brodersen, Andrew Elliott, Paul Expert, Lucas Jeub, Heather Harrington, Felix Reed-Tsochas, and Jes\'us San Mart\'in for helpful discussions. We thank Vittoria Colizza and Chiara Poletto for several extensive discussions and comments on the manuscript, and we thank Andrew Elliott for reading and commenting on this manuscript and for his help with creating the figures. Additionally, we thank Yulian Ng and Tom Prescott for their contributions at early stages of this project.
The dengue fever data was collected by the Peruvian Ministry of Health~\cite{PeruHealthMinistry}.  We obtained province locations from the \url{Geonames.org} website~\cite{geonames}, and we obtained their populations from the Peruvian Instituto Nacional de Estad\'{i}stica e Inform\'{a}tica (INEI)~\cite{PeruStats}.






\newpage


\appendix

\section{Spatial benchmarks: Variation of Information} \label{Appendix:VI}

\begin{figure}[h!tbp]
	\centering
\includegraphics[width=0.5\linewidth]{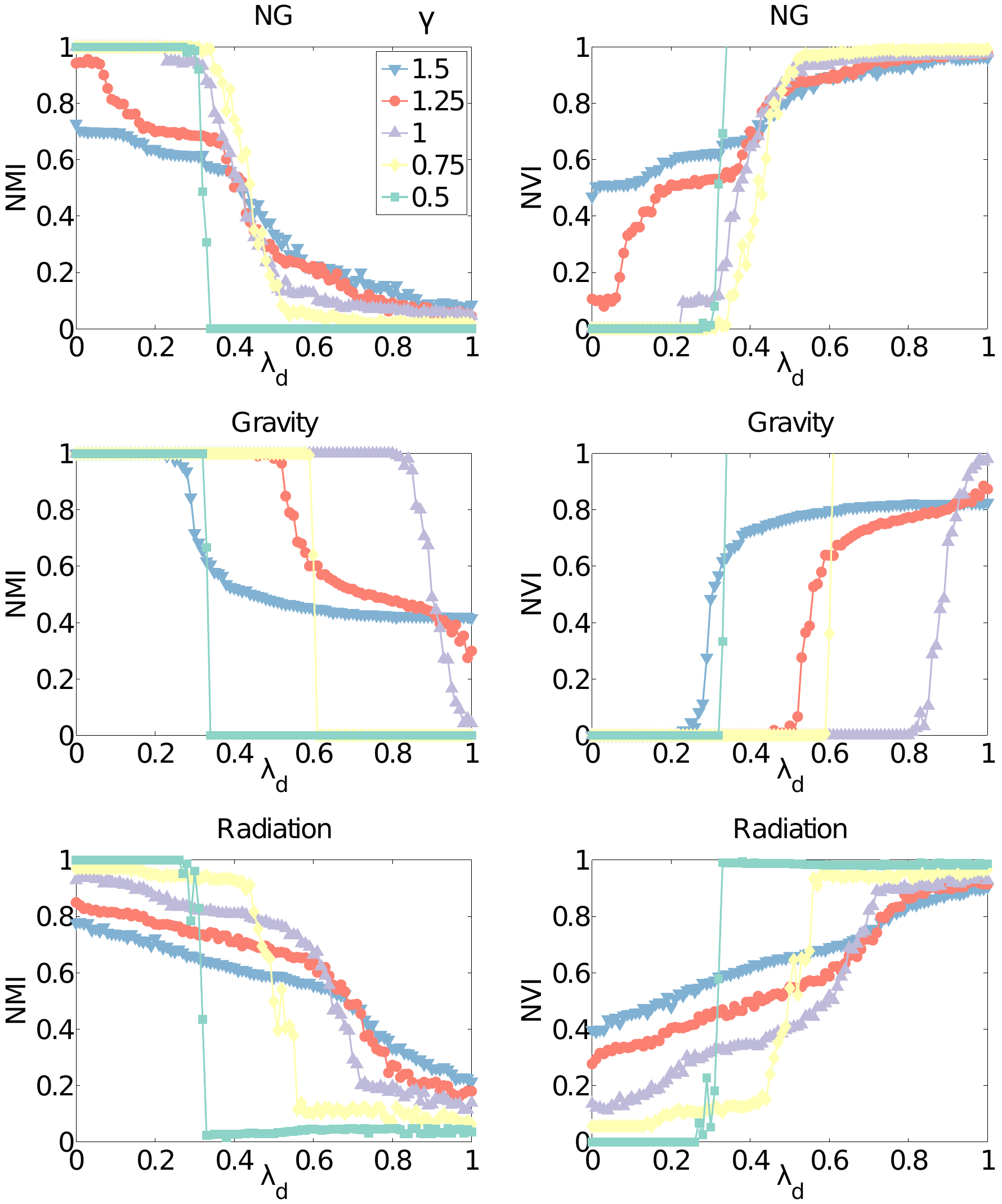}
\caption{(Left) Normalized mutual information (NMI) and (right) normalized variation of information (NVI) between algorithmically-detected partitions, which we obtain by maximizing modularity, and planted partitions in the uniform population distance static spatial benchmarks with $N = 50$ cities, a grid size of $l = 10$, and a density parameter of $\mu = 50$. We examine the partitions for different values of the resolution parameter $\gamma$ as a function of inter-community connectivity $\lambda_d$ using the (top) NG null model, (middle) gravity null model, and (bottom) radiation null model.
 \label{Figure:Benchmark:static-VI-gammavsnm}}
\end{figure}

Normalized variation of information (NVI)~\cite{Meila2007} is a viable alternative similarity measure to NMI for the spatial benchmark networks. In contrast to NMI, variation of information (VI) and NVI are metrics in the mathematical sense. Both measures are related to mutual information. For a partition $X = \{X_1, X_2, \ldots X_K\}$ with $K$ communities, VI is defined as 
\begin{equation}
	 VI(X,Y) = H(X) + X(Y)  - 2I (X,Y)\,,
\end{equation}
where $H(X) = - \sum_{k=1}^{K} P_k \log_2 P_k$ is the entropy of the random variable associated to partition $X$, the quantity $I(X,Y) = \sum_{k=1}^{K} \sum_{l=1}^{L} P(k,l) \log_2 \left[\frac{P (k,l)}{P(k)P(l)}\right]$ is the mutual information, $P(k)$ and $P(l)$ are the respective marginal probabilities of observing communities $k$ and $l$ in partitions $X$ and $Y$, and $P(k,l)$ is the joint probability of observing communities $k$ and $l$ simultaneously in partitions $X$ and $Y$). VI is equal to $0$ if partitions $X$ and $Y$ are identical, and $VI(X,Y) < \log_2{N}$, where $N$ is the number of nodes in the whole network. 
Normalizing VI yields NVI, which is given by~\cite{Kraskov2005}
 \begin{equation}
	 NVI(X,Y) =  \frac{1-VI(X,Y)}{H(X,Y)} \in [0,1]\,.
 \end{equation}
See Refs.~\cite{Meila2007,Kraskov2005} for additional discussions. As one can see in Fig.~\ref{Figure:Benchmark:static-VI-gammavsnm}, both NMI and NVI perform similarly and neither gives visibly better precision. 

\section{Spatial benchmarks: Variation of the number of cities $N$} \label{Appendix:cities}

\begin{figure}[h!tbp]
  \centering
  \includegraphics[width=0.5\linewidth]{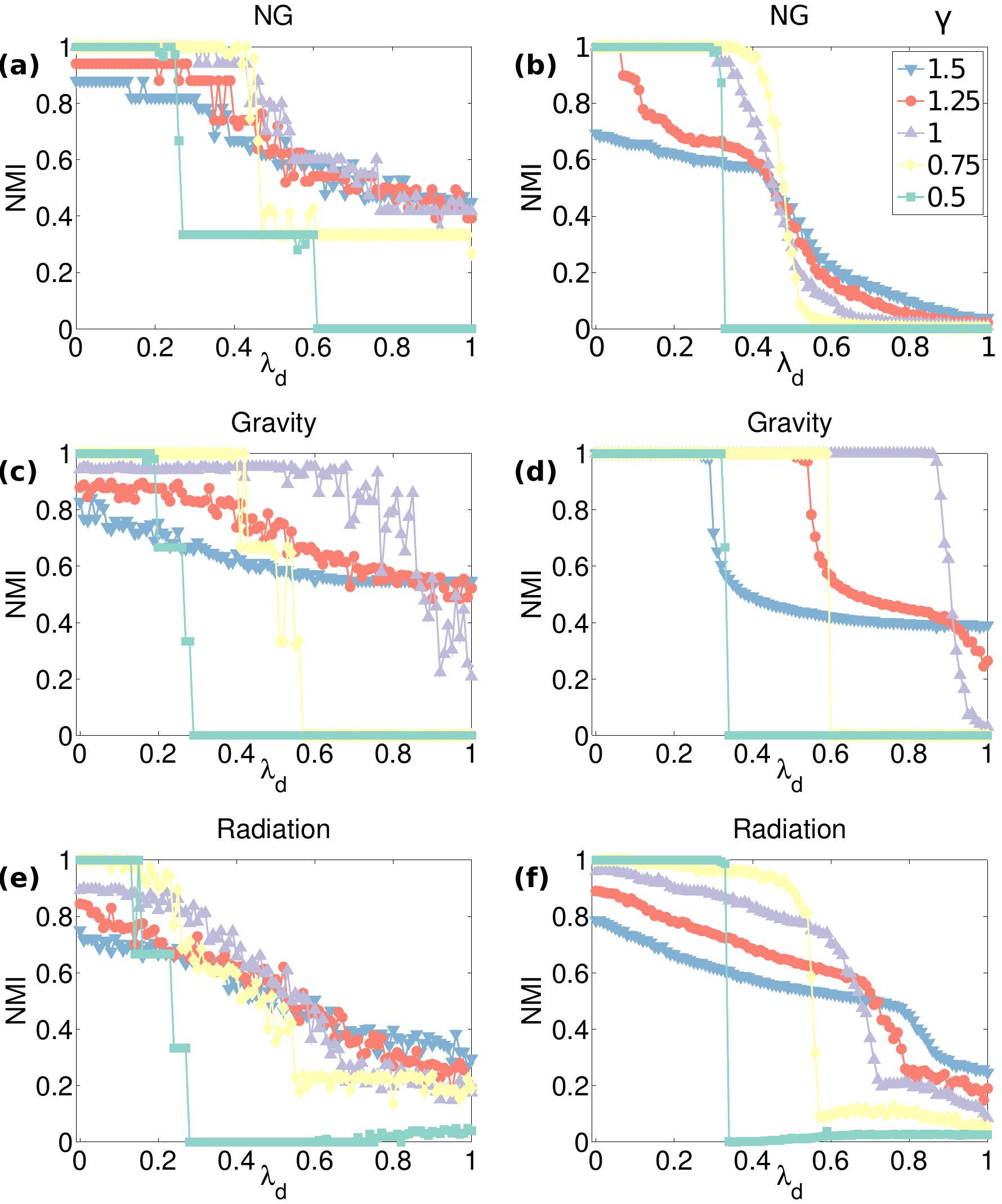}
\caption{Uniform population static benchmarks: NMI scores between algorithmically detected partitions, which we obtain by maximizing modularity, and planted partitions in static uniform population distance benchmarks with $l = 10$, a density parameter of $\mu = 100$, and uniform populations of $100$ for different numbers of cities in an underlying space of the same size. The number of cities is (left) $N = 10$,  and (right) $N = 90$. We use the NG (top), gravity (middle), and (bottom) radiation null models.  See Fig.~\ref{Figure:Benchmark:static-even-d-NMI-gammavsnm-tab} in the main text for plots with $N = 50$.
\label{Figure:Benchmark:static-even-d-NMI-binsizevsnm}}
\end{figure}

We now vary the number $N$ of cities in benchmarks with a fixed size of $l=10$, density parameter of $\mu = 100$, and a uniform population of $100$ people in each city. In Fig.~\ref{Figure:Benchmark:static-even-d-NMI-binsizevsnm}, we plot the NMI of algorithmic partitions versus planted partitions for several values of the resolution parameter $\gamma$ using the NG null model and both spatial null models. In combination with Fig.~\ref{Figure:Benchmark:static-even-d-NMI-gammavsnm-tab} in the main text, which has $N=50$ cities, we observe no qualitative changes in NMI aside from an expected increase in variability when $N$ is small.

\section{Variation of Edge Density Parameter $\mu$} \label{Appendix:mu}

\begin{figure}[tbp]
	\centering
    \includegraphics[width=0.5\linewidth]{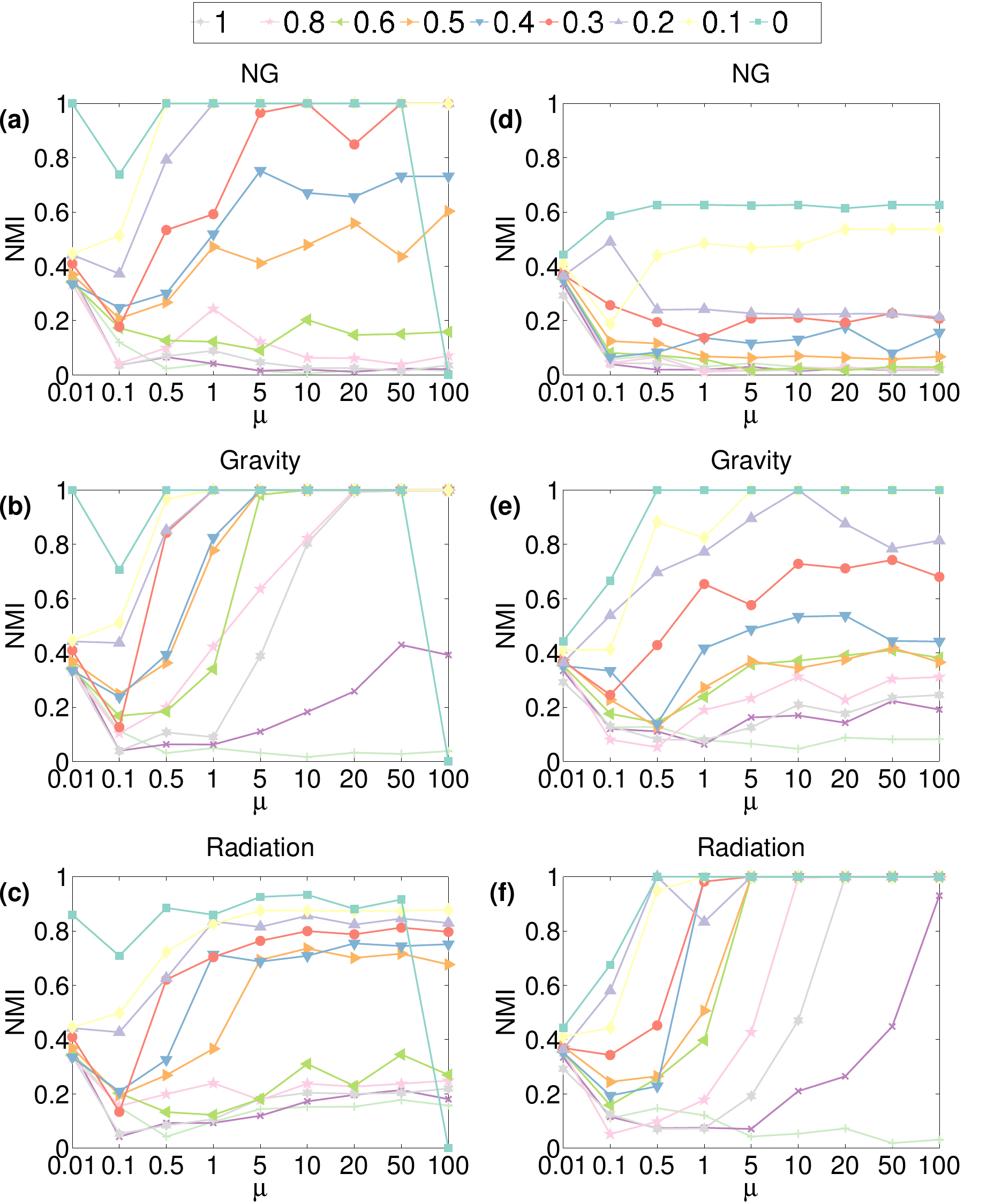}
\caption{NMI between algorithmically-detected partitions, which we obtain by maximizing modularity with $\gamma = 1$, and planted partitions for uniform population static spatial benchmarks with $N = 50$, a size parameter of $l = 10$, uniform city populations of $100$, and several values of inter-community connectivity $\lambda_d$.  
We plot the NMI scores as a function of the edge density parameter $\mu$ for (left) the distance benchmark and (right) the flux benchmark. 
\label{Figure:Benchmark:static-even-rovsnm}}
\end{figure}

\begin{figure}[tbp]
\centering
  \includegraphics[width=0.5\linewidth]{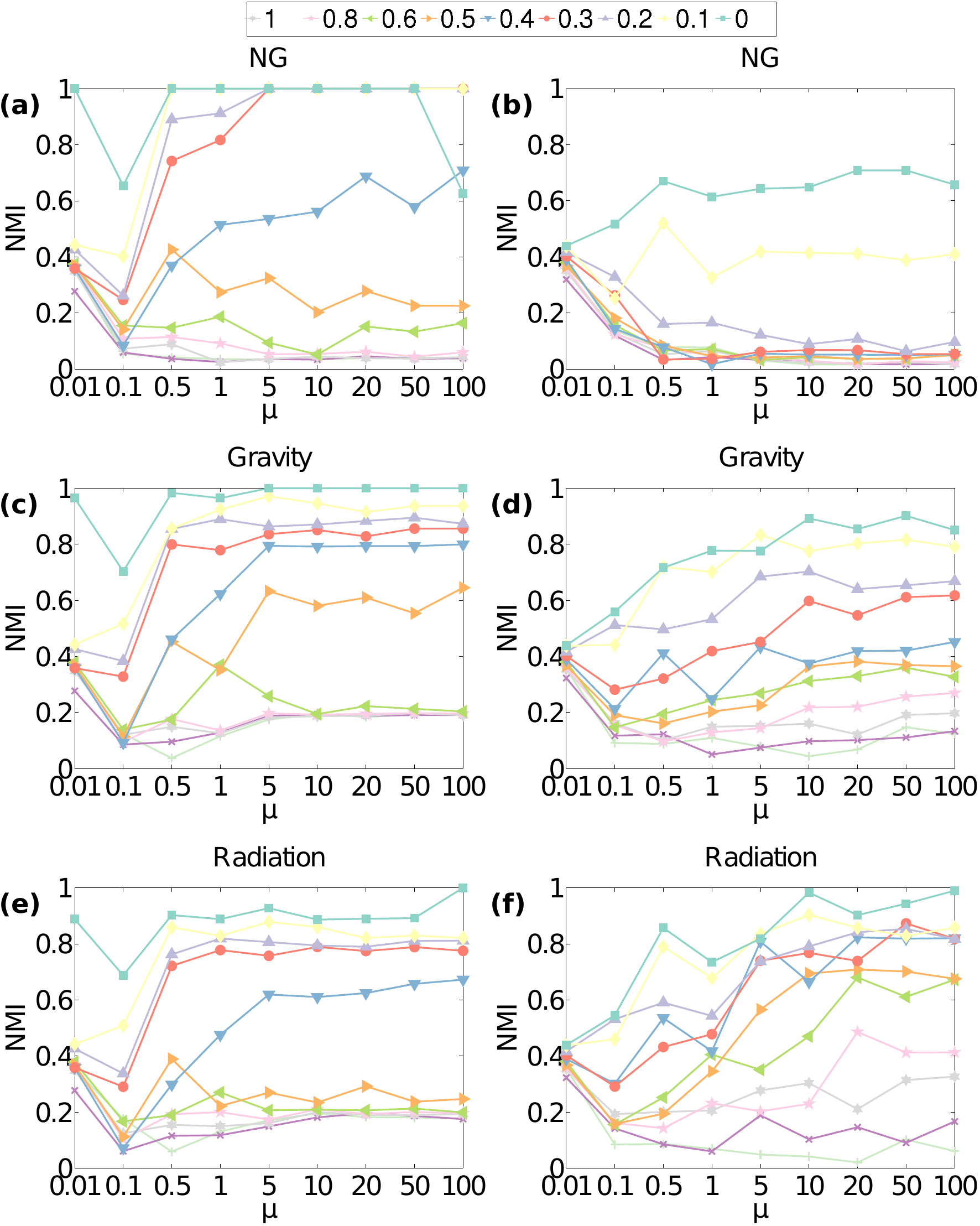}
\caption{NMI between algorithmically-detected partitions, which we obtain by maximizing modularity with $\gamma = 1$,  and planted partitions for random population static spatial benchmarks with $N = 50$, a size parameter of $l = 10$, city populations $n$ selected uniformly at random from $[0,100]$, and several values of inter-community connectivity $\lambda_d$.  We plot the NMI scores as a function of the edge density parameter $\mu$ for (left) the distance benchmark and (right) the flux benchmark. 
\label{Figure:Benchmark:static-varyPop-rovsnm}}
\end{figure}

We present the results of varying the edge density parameter $\mu$ in static benchmarks. The edge density has a strong effect on the ability of the modularity-maximization methods to detect communities. For $\mu \lessapprox 5$, we obtain smaller NMI scores than the maximum attained for each particular $\lambda_d$ for larger $\mu$ values. 
(See Figs.~\ref{Figure:Benchmark:static-even-rovsnm} and \ref{Figure:Benchmark:static-varyPop-rovsnm}.)
We therefore focus on using a density parameter of $\mu =100$ in the main text to follow the choice that was used for the benchmarks networks in Ref.~\cite{Expert2011}.

\section{``Distance and Population'' Benchmark} \label{app:benchmark-distpop}

In this section, we construct a ``distance and population'' spatial benchmark. In Fig.~\ref{Figure:Benchmark:static-even-d-NMI-gammavsnm-tab} in the main text we observed that the gravity null model performs best on the uniform population distance benchmark, but the NG null model performs better than spatial null models on the random population distance benchmark because the edge placement in that benchmark does not include population information. Here, we study the effects of incorporating population into edge probabilities for the ``distance and population'' benchmark.
  
\begin{figure}
\centering		  \includegraphics[width=0.5\linewidth]{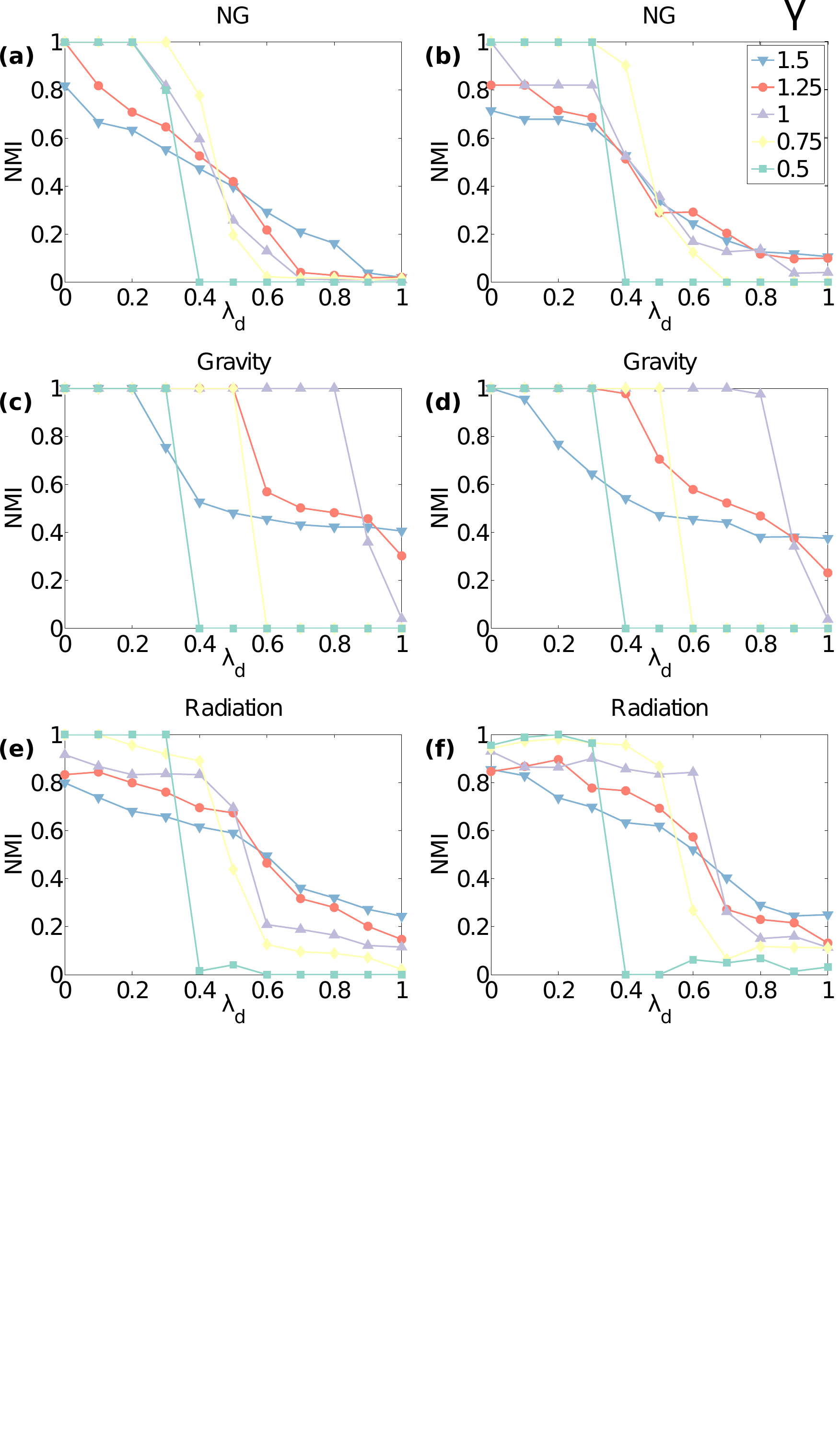}
\caption{NMI between algorithmically-detected community structure, which we obtain by maximizing modularity, and planted community structure in ``distance and population'' static spatial benchmarks with (left) uniform populations and (right) random populations. We use $N = 50$, $l = 10$, $m=10$, $\mu = 100$, and $\gamma=1$ for various values of $\omega$ (colored curves) as a function of $\lambda_d$. We detected communities by optimizing modularity using the (top) NG, (middle) gravity, and (bottom) radiation null models
\label{Figure:Benchmark:distpop}}
\end{figure}

We construct the new type of benchmark network in the same manner as the distance benchmark in Section \ref{Section:benchmarks}, but we now incorporate population into the edge-placement probability by taking $p^{\mathrm{distpop}}_{ij} = \frac{p_i p_j \lambda(c_i,c_j)}{Z_1 d_{ij}}$. As expected, this brings back the advantage that the gravity null model has for the uniform population distance benchmark (compare Fig.~\ref{Figure:Benchmark:distpop} with Fig.~\ref{Figure:Benchmark:static-even-d-NMI-gammavsnm-tab} in the main text). The radiation null model has the second-best performance on this benchmark, with a better performance than on the random population distance benchmark. However, it does not do as well as it did on the random population flux benchmark (see Fig.~\ref{Figure:Benchmark:static-even-d-NMI-gammavsnm-tab}). 

\section{Community detection on random population multilayer spatial benchmarks}\label{Appendix:multi-random}

We now study the influence of the parameters $\gamma$ and $\omega$ on the community structure for random-population multilayer temporally-stable benchmarks. We first compare the results to our findings from static benchmarks by varying $\gamma$ and $\lambda_d$ for fixed values of $\omega$. 
We study the performance of modularity maximization with the NG, gravity, and radiation null models on random population benchmarks (see Fig.~\ref{Figure:Benchmark:multi-gammavsnm}) with parameter values of $N = 50$, $l = 10$, and $m = 10$ layers using $\gamma \in \{0.5,0.75,1,1.25,1.5\}$ and $\omega \in \{0.1,0.25,0.5,0.75,1\}$. We only show plots for $\omega=0.1$, as the values of $\omega$ do not noticeably influence the results.

We obtain results that are similar to our results for the corresponding static benchmarks inn Fig.~\ref{Figure:Benchmark:static-even-d-NMI-gammavsnm-tab}. 

\begin{figure}
	\centering
    \includegraphics[width=0.5\linewidth]{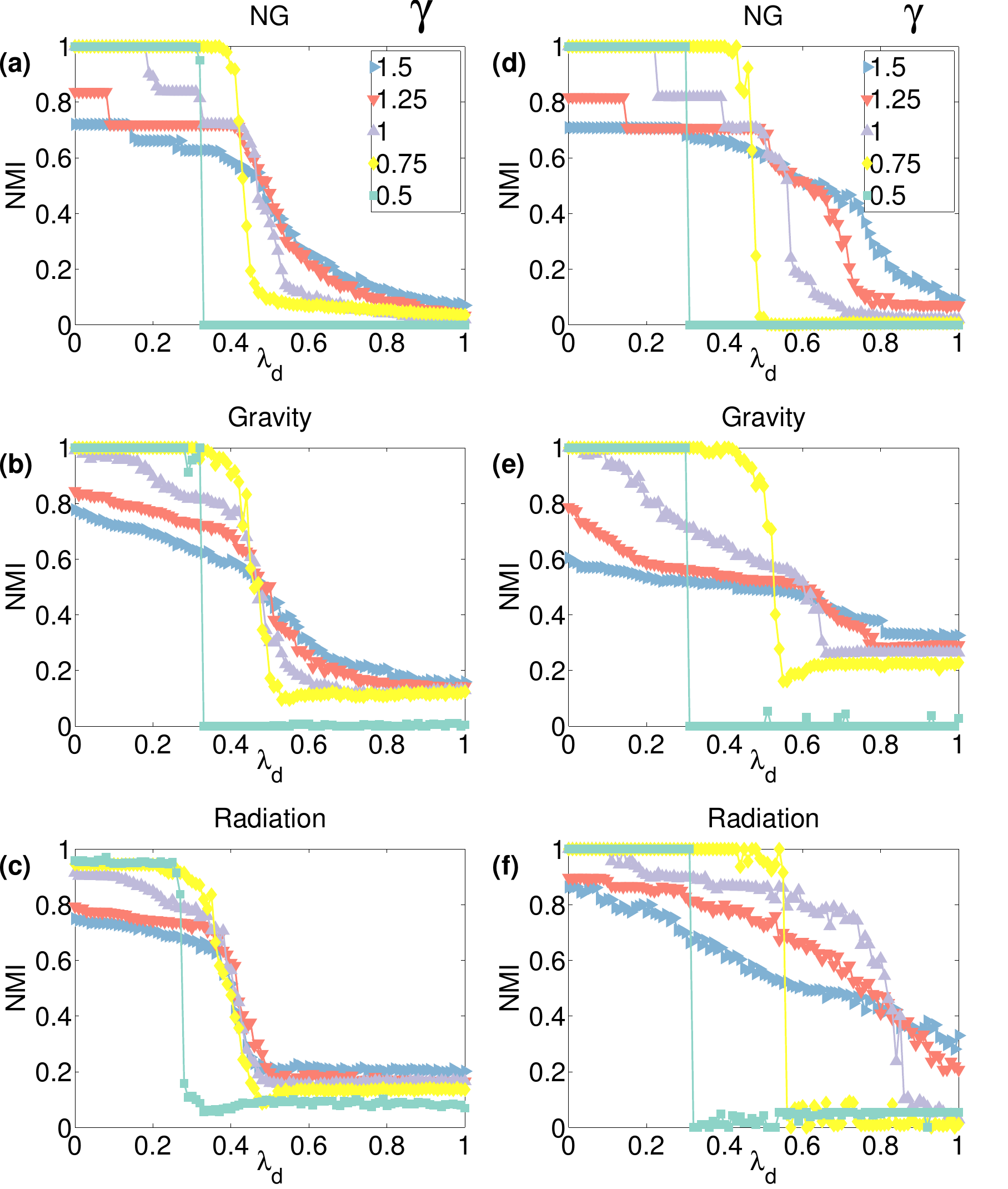}
\caption{NMI between algorithmically-detected community structure, which we obtain by maximizing modularity, and planted community structure in random-population, temporally-stable multilayer spatial benchmarks.  We choose the population of each of the $N = 50$ cities uniformly at random from the set $\{1,\ldots,100\}$. We consider various values of the resolution parameter $\gamma$, and the other parameter values are $l = 10$, $m=10$, $\mu = 100$, and $\omega=0.1$. We plot NMI as a function of $\lambda_d$ for (left) the distance benchmark and (right) the flux benchmark using the (top) NG, (middle) gravity, and (bottom) radiation null models. 
 \label{Figure:Benchmark:multi-rand-gammavsnm}}
\end{figure}

Once again, we find that the choice of $\gamma$ has a large influence on the quality of algorithmic partitions, and (as with our findings for static benchmarks) that $\gamma = 1$ seems to yield the best performance (i.e., the largest NMI scores) for low values of $\lambda_d$, whereas larger values of $\gamma$ perform better for larger $\lambda_d$ (see Fig.~\ref{Figure:Benchmark:multi-rand-gammavsnm}).  
The effect of varying $\gamma$ is most pronounced for the radiation null model on flux benchmarks.

\begin{figure}
    \includegraphics[width=0.5\linewidth]{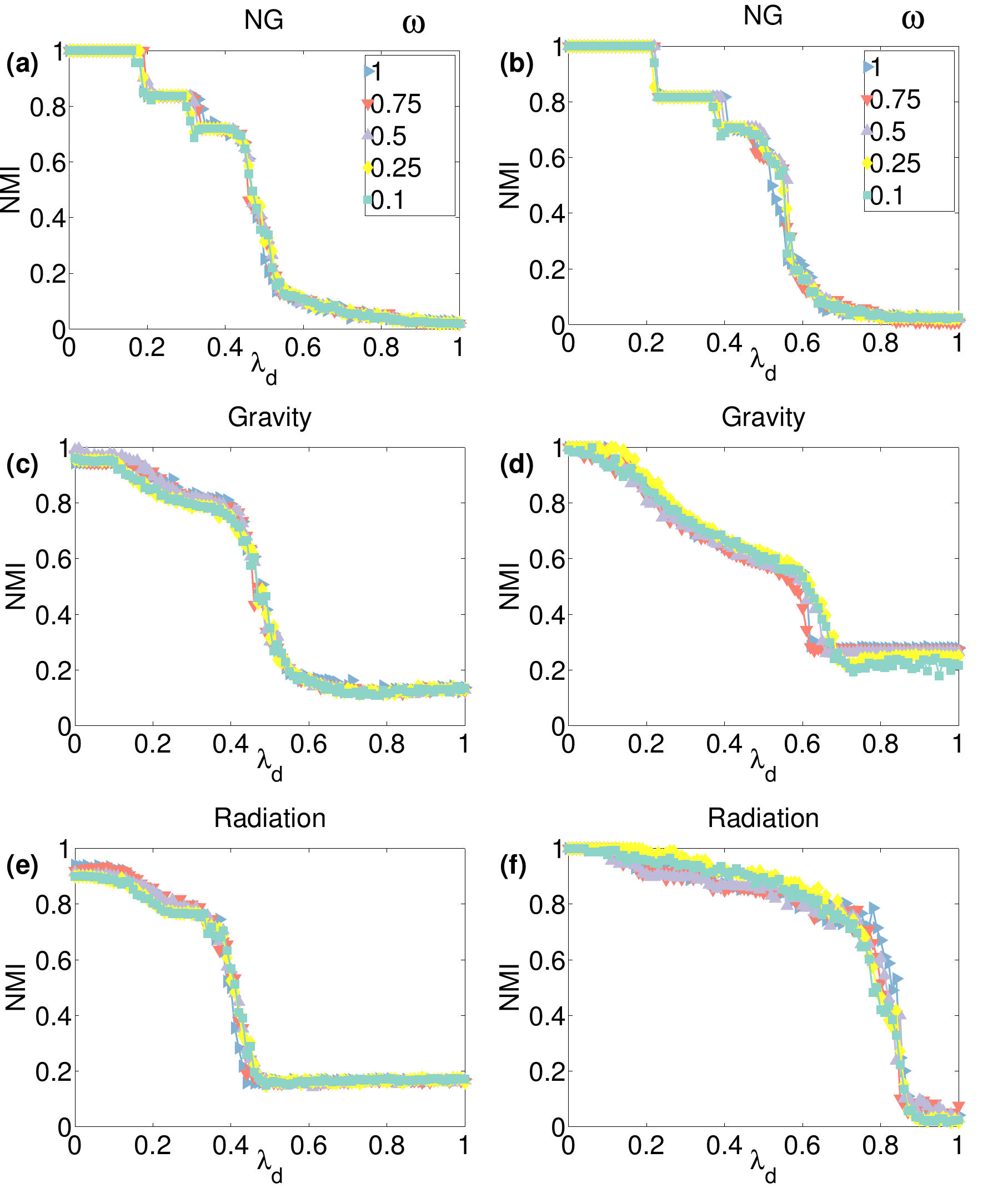}
\caption{NMI between algorithmically-detected community structure, which we obtain by maximizing modularity, and planted community structure in random-population, temporally-stable multilayer spatial benchmarks. We choose the population of each of the $N = 50$ cities uniformly at random from the set $\{1,\ldots,100\}$. We consider various values of the parameter $\omega$, and the other parameter values are $l = 10$, $m=10$, $\mu = 100$, and $\gamma = 1$. We plot NMI as a function of $\lambda_d$ for (left) the distance benchmark and (right) the flux benchmark using the (top) NG, (middle) gravity, and (bottom) radiation null models.
\label{Figure:Benchmark:multi-rand-omegavsnm}}
\end{figure}

We now examine the NMI of algorithmic versus planted partitions in temporally-stable multilayer benchmarks for fixed $\gamma = 1$ while varying $\omega$ and $\lambda_d$. As we show in Fig.~\ref{Figure:Benchmark:multi-rand-omegavsnm}, we find that the value of $\omega$ usually has very little effect on our ability to detect the planted communities via modularity maximization --- the same as for the uniform population temporally stable multilayer benchmarks (see Fig.~\ref{Figure:Benchmark:multi-omegavsnm}). The parameter $\omega$ becomes important for the random-population, temporally-evolving multilayer benchmarks in the same manner as what we observed in the main text for uniform population benchmark networks (not shown; see Fig.~\ref{Figure:Benchmark:multi-t-gammavsnm} in the main text for the uniform population results).


\section{Province-level Communities for Multilayer Benchmarks} \label{Appendix:bench-regions}

\begin{figure}[tbp]
  \includegraphics[width=0.5\linewidth]{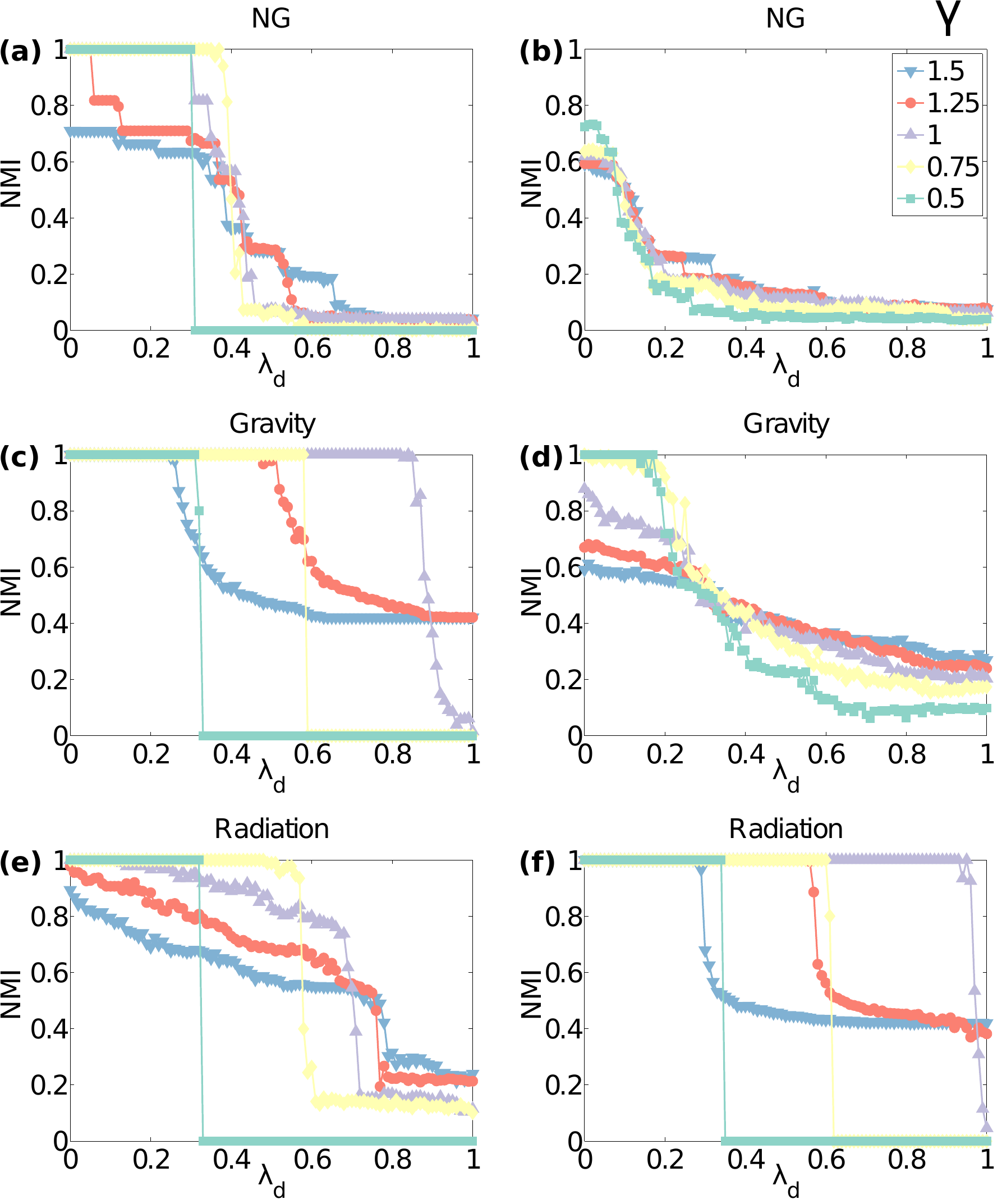}
\caption{NMI between algorithmically-detected province-level community structures, which we obtain by maximizing modularity, for uniform population ($n_i=100$ for all $i$) temporally stable multilayer spatial benchmarks with $m=10$ layers. Each layer has a single-layer planted partition with $N = 50$ cities, a size parameter of $l = 10$, and a density parameter of $\mu = 100$.  We use $\omega = 0.1$ and consider various values of the resolution parameter $\gamma$, and we plot NMI as a function of the inter-community connectivity $\lambda_d$ for (left) the distance benchmark and (right) the flux benchmark. 
   \label{Figure:Benchmark:multi-nodelevel-u-gammavsnm}}
\end{figure}

In Fig.~\ref{Figure:Benchmark:multi-nodelevel-u-gammavsnm}, we present our results for province-level community detection on uniform population temporally stable multilayer benchmarks. As one can see by comparing these results to those in Fig.~ \ref{Figure:Benchmark:multi-gammavsnm}, we obtain similar NMI scores for the performance of community detection for province-level communities as we did for the ordinary community detection in multilayer networks.

\section{Community Detection on Aggregated dengue fever Data}
\label{Appendix:regions}

\begin{figure*}[tbp]
\hfill(a)\includegraphics[width=0.27\linewidth]{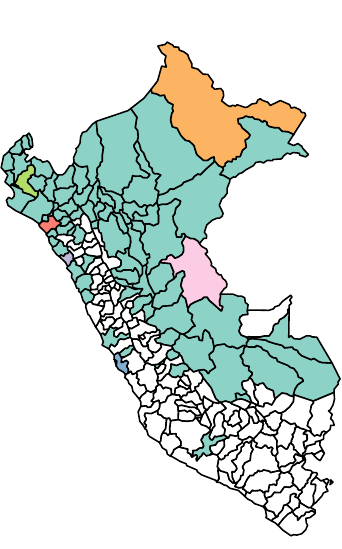}
\hfill(b)\includegraphics[width=0.27\linewidth]{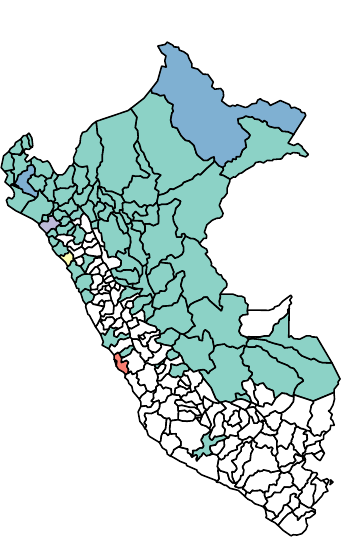}
\hfill(c)\includegraphics[width=0.27\linewidth]{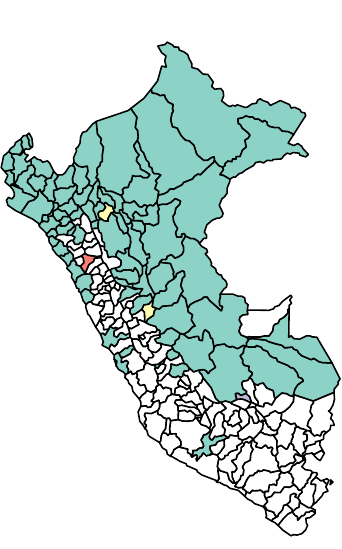}

\caption{Algorithmically-detected community structures, which we obtain using modularity maximization, for static dengue fever correlation networks that we construct using the entire set of time series (i.e., we use $\tau = 1$ and $\Delta=779$) using (a) the gravity null model, (b) the radiation null model, and (c) the correlation null model for a resolution-parameter value of $\gamma=1$. We color provinces on a map of Peru according to their community assignments. White provinces are ones in which our data does not include any reported cases of dengue fever in the indicated time window. 
   \label{Figure:Dengue-fullyaggregated}}
\end{figure*}

In Fig.~\ref{Figure:Dengue-fullyaggregated}, we show additional results of community detection on fully aggregated networks (i.e., we use $\tau=1$ and $\Delta=779$) from the dengue fever times series. In Section~\ref{dengue-region-NG} of the main text, in Fig.~\ref{Figure:Dengue-regionlevel-map}(a) we showed the results of modularity maximization using the NG null model. 
We now also show similar results for the gravity, radiation, and correlation null models.
The gravity, radiation, and correlation null models find one large community and a few small communities. Because of the aggregation, we have lost the rich set of information that we were able to study using multilayer community detection.

\end{document}